\documentclass[10pt, twocolumn, notitlepage, superscriptaddress, prx, longbibliography]{revtex4-1}

\usepackage{here}
\usepackage{amsmath}         
\usepackage{graphicx}
\usepackage{braket}         
 \usepackage{lettrine}
 \usepackage{color}
 \usepackage{times}
\usepackage{hhline}
\usepackage{tabularx}
\usepackage{calc}

\begin{document}

\title{
Theory of Current-Induced Angular Momentum Transfer Dynamics in Spin-Orbit Coupled Systems
\looseness=-1}    

\author{Dongwook Go}
\email{d.go@fz-juelich.de}
\affiliation{Peter Gr\"unberg Institut and Institute for Advanced Simulation, Forschungszentrum J\"ulich and JARA, 52425 J\"ulich, Germany \looseness=-1}
\affiliation{Institute of Physics, Johannes Gutenberg University Mainz, 55099 Mainz, Germany}
\affiliation{Department of Physics, Pohang University of Science and Technology, Pohang 37673, Korea}
\affiliation{Basic Science Research Institute, Pohang University of Science and Technology, Pohang 37673, Korea}

\author{Frank Freimuth}
\affiliation{Peter Gr\"unberg Institut and Institute for Advanced Simulation, Forschungszentrum J\"ulich and JARA, 52425 J\"ulich, Germany \looseness=-1}

\author{Jan-Philipp Hanke}
\affiliation{Peter Gr\"unberg Institut and Institute for Advanced Simulation, Forschungszentrum J\"ulich and JARA, 52425 J\"ulich, Germany \looseness=-1}

\author{Fei Xue}
\affiliation{Physical Measurement Laboratory, National Institute of Standards and Technology, Gaithersburg, MD 20899, USA}
\affiliation{Institute for Research in Electronics and Applied Physics \& Maryland Nanocenter, University of Maryland, College Park, MD 20742}

\author{Olena Gomonay}
\affiliation{Institute of Physics, Johannes Gutenberg University Mainz, 55099 Mainz, Germany}

\author{\\ Kyung-Jin Lee}
\affiliation{Department of Materials Science and Engineering, Korea University, Seoul 02841, Korea}
\affiliation{KU-KIST Graduate School of Converging Science and Technology, Korea University, Seoul 02841, Korea}

\author{Stefan Bl\"ugel}
\affiliation{Peter Gr\"unberg Institut and Institute for Advanced Simulation, Forschungszentrum J\"ulich and JARA, 52425 J\"ulich, Germany \looseness=-1}

\author{Paul M. Haney}
\email{paul.haney@nist.gov}
\affiliation{Physical Measurement Laboratory, National Institute of Standards and Technology, Gaithersburg, MD 20899, USA}

\author{Hyun-Woo Lee}
\affiliation{Department of Physics, Pohang University of Science and Technology, Pohang 37673, Korea}

\author{Yuriy Mokrousov}
\email{y.mokrousov@fz-juelich.de}
\affiliation{Peter Gr\"unberg Institut and Institute for Advanced Simulation, Forschungszentrum J\"ulich and JARA, 52425 J\"ulich, Germany \looseness=-1}
\affiliation{Institute of Physics, Johannes Gutenberg University Mainz, 55099 Mainz, Germany}

\begin{abstract}
Motivated by the importance of understanding various competing mechanisms to the current-induced spin-orbit torque on magnetization in complex magnets, we develop a unified theory of current-induced spin-orbital coupled dynamics in magnetic heterostructures. The theory describes angular momentum transfer between different degrees of freedom in solids, {\it e.g.}, the electron orbital and spin, the crystal lattice, and the magnetic order parameter. Based on the continuity equations for the spin and orbital angular momenta, we derive equations of motion that relate spin and orbital current fluxes and torques describing the transfer of angular momentum between different degrees of freedom, achieved in a steady state under an applied external electric field. We then propose a classification scheme for the mechanisms of the current-induced torque in magnetic bilayers. Based on our first-principles implementation within the density functional theory, we apply our formalism to two different magnetic bilayers, Fe/W(110) and Ni/W(110), which are chosen such that the orbital and spin Hall effects in W have opposite sign and the resulting spin- and orbital-mediated torques can compete with each other. We find that while the spin torque arising from the spin Hall effect of W is the dominant mechanism of the current-induced torque in Fe/W(110), the dominant mechanism in Ni/W(110) is the orbital torque originating in the orbital Hall effect of the non-magnetic substrate. It leads to negative and positive \emph{effective} spin Hall angles, respectively, which can be directly identified in experiments. This clearly demonstrates that our formalism is ideal for studying the angular momentum transfer dynamics in spin-orbit coupled systems as it goes beyond the ``spin current picture'' by naturally incorporating the spin and orbital degrees of freedom on an equal footing. Our calculations reveal that, in addition to the spin and orbital torque, other contributions such as the interfacial torque and self-induced anomalous torque within the ferromagnet are not negligible in both material systems.  
\end{abstract}

\date{\today}                 
\maketitle

\section{Introduction}


Spin-orbit coupling plays a central role in a plethora of phenomena occurring in magnetic multilayers \cite{Hellman2017}. Current-induced spin-orbit torque is one of the most important examples, and is a workhorse in the field of spintronics \cite{Gambardella2011, Manchon2019}.  In contrast to spin-transfer torque in spin valve structures, a device utilizing spin-orbit torque does not require an extra ferromagnetic layer to create spin polarized current.  Instead, nonequilibrium spin currents and spin densities are generated in nonmagnetic materials due to spin-orbit coupling. The magnitude of spin-obit torque can be sufficient to induce magnetic switching, as demonstrated in magnetic bilayers consisting of a nonmagnet and a ferromagnet \cite{Miron2011b, Liu2012a, Liu2012b, Yu2014, Baumgartner2017}. Spin-orbit torque also enables fast current-induced magnetic domain wall motion \cite{Miron2011a, Ryu2013, Emori2013, Martinez2013}. Several microscopic mechanisms of current-induced spin-orbit torque have been proposed. However, quantification of the individual contributions is challenging both theoretically and experimentally. Moreover, our understanding of the phenomenon based on the properties of the electronic structure is rather unsatisfactory yet. 

\begin{figure*}[ht!]
\includegraphics[angle=0, width=0.65\textwidth]{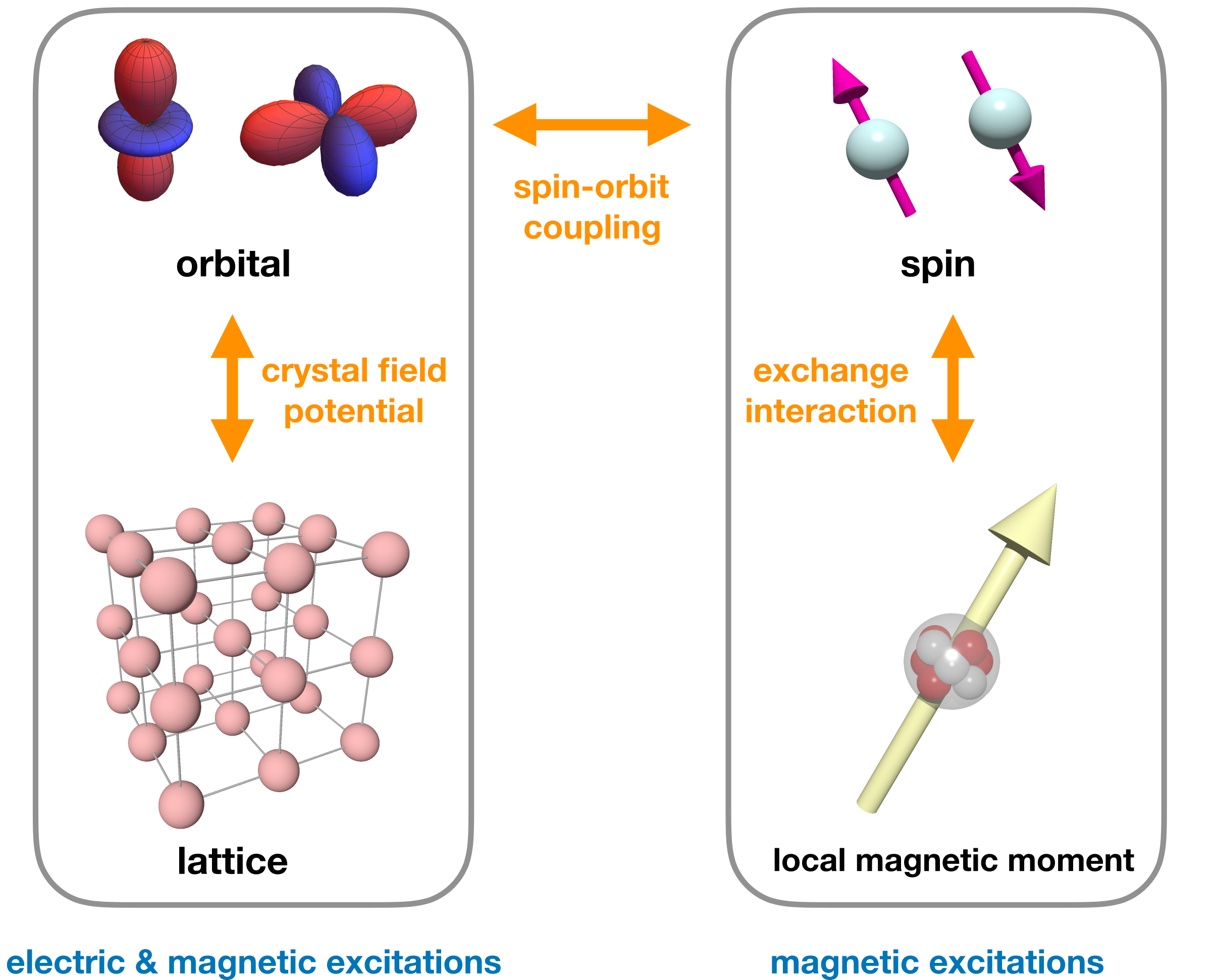}
\caption{
\label{fig:overview}
Interactions between angular-momentum-carrying degrees of freedom in solids: spin and orbital of the electron, the crystal lattice, and the local magnetic moment. Orange arrows indicate microscopic interactions by which angular momentum is exchanged: the spin-orbit coupling for interaction between the spin and orbital momenta of an electron, crystal field potential for the interaction between the lattice and the orbital angular momentum of the electron, and exchange interaction for the interaction between the local magnetic moment and the spin of the electron. The orbital and lattice in the left column can have both electric and magnetic excitations, while the spin and local magnetic moment in the right column have only magnetic excitations. This property marks the orbital degree of freedom as an essential element in describing magnetoelectric responses, such as the current-induced torque.
}
\end{figure*}

In this work, we examine the fundamental physical nature of spin-orbit torque in the view of angular momentum exchange between different degrees of freedom in solids. The possible channels for angular momentum transfer among these degrees of freedom are schematically shown in  Fig.~\ref{fig:overview}. It is conceptually important to separate (i) angular momentum carried by a conduction electron angular momentum encoded in its orbital and spin parts of the wave function, (ii) mechanical angular momentum of the lattice, and (iii) spin angular momentum encoded into the local magnetic moment emerging as a result of magnetic ordering. 
These degrees of freedom interact with each other and exchange angular momentum. For example, spin-orbit coupling mediates an angular momentum transfer between spin and orbital degrees of the electron, crystal field potential leads to an orbital angular momentum transfer between the electron and the lattice, and exchange interaction enables spin transfer between the conduction electron's spin and local magnetic moment.  In its most profound definition, the spin-orbit torque is understood as an angular momentum flow from the surrounding lattice harvested by the local magnetic moment $-$ a process which is mediated by spin-orbit entangled electrons. Here, taking this fundamental viewpoint as the foundation, we provide a unified and complete picture of possible scenarios on current-induced torque on the magnetization.

\begin{figure*}[t!]
	\includegraphics[angle=0, width=0.85\textwidth]{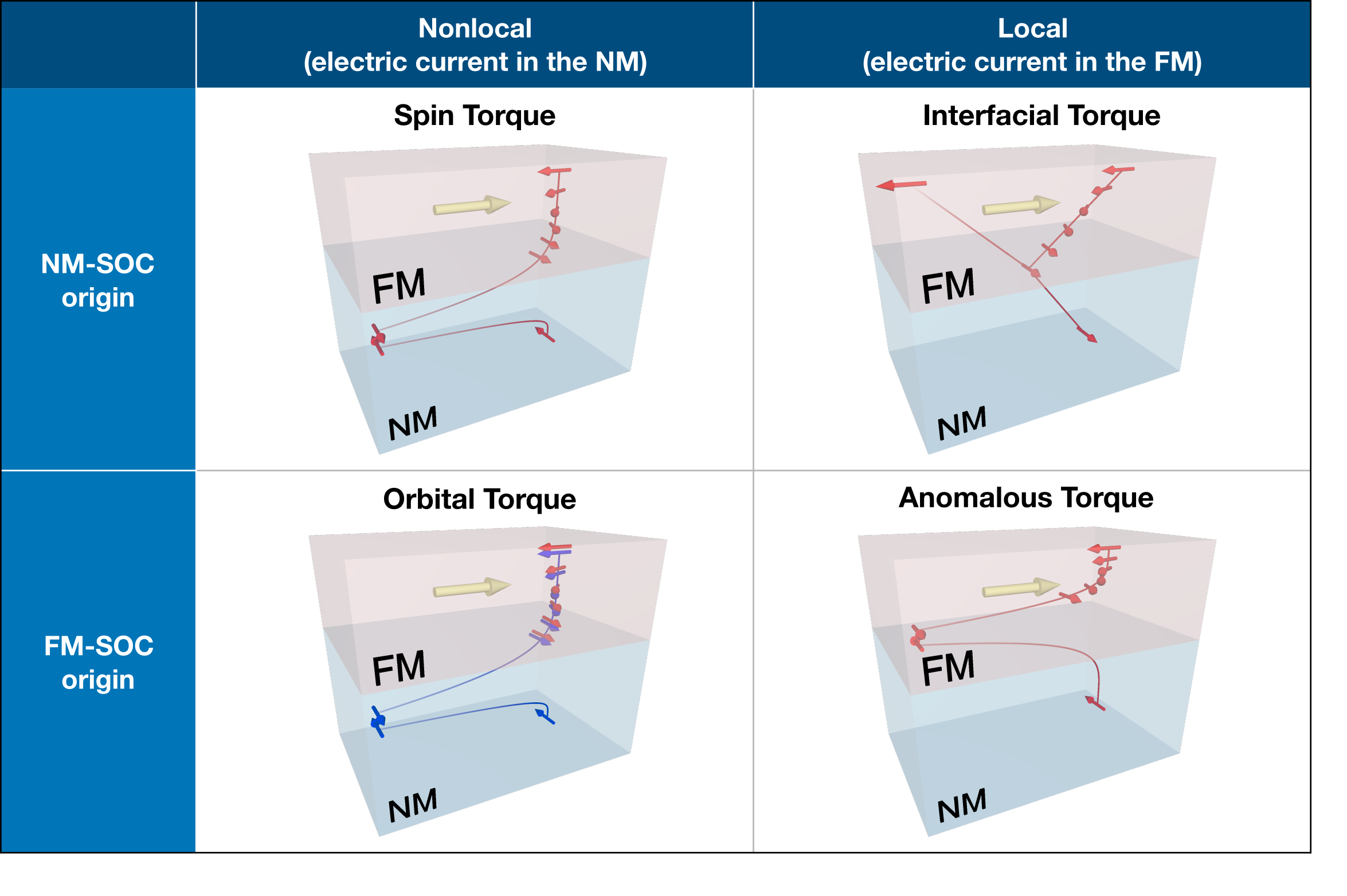}
	\caption{
		\label{fig:classification}
		Classification of the mechanisms of the current-induced torque. The row represents the origin of spin-orbit coupling in either the nonmagnet or in the ferromagnet. The column represents the locality of the torque: i.e.,  whether the torque acting on the ferromagnet originates from the electrical current flowing in the nonmagnet (nonlocal) or in the ferromagnet itself (local). The red arrows represent the spin, and the blue arrows represent the orbital angular momentum. The local magnetic moment is represented with a big yellow arrow. 
	}
\end{figure*}

Depending on the specifics of a particular angular momentum exchange transfer channel, which takes place in different parts of the solid e.g. in the bulk or at the interface, we can understand various competing mechanisms in non-uniform magnetic heterostructures in an unified manner. Here, we choose to consider a bilayer geometry $-$ comprising a nonmagnet adjecent to a ferromagnet $-$ which is most widely studied in experiments. Within our viewpoint, we  classify the mechanisms of the current-induced torque into four different scenarios, which are schematically illustrated in Fig.~\ref{fig:classification}. The classification is based on two independent criteria: (1) the spatial origin of the spin-orbit interaction, 
and (2) the spatial origin of the current responsible for the angular momentum generation, which is absorbed by the magnetization. 
This classification is discussed in detail below, and demonstration of its relevance and completeness is the main goal of this work. 

In magnetic bilayers consisting of a nonmagnet and a ferromagnet, the spin Hall effect arising from the nonmagnet is considered to be one of the main mechanisms for generating a torque on the magnetization of the ferromagnet \cite{Liu2012a, Liu2012b}. That is, an electrical current in the nonmagnet induces a transverse spin current, which is injected into the ferromagnet and results in a torque (upper left panel in Fig. \ref{fig:classification}). In this picture, the spin Hall conductivity of the nonmagnet is assumed to be a bulk property, and the spin injection and resulting torque generation on the local magnetic moment is explained by the theory of the spin-transfer torque \cite{Stiles2002, Ralph2008}. We denote such contribution due to spin injection from the nonmagnet as a spin torque. This analysis considers the spin-orbit coupling only in the nonmagnet and neglects the spin-orbit coupling at the nonmagnet/ferromagnet interface and in the ferromagnet. Moreover, current-induced effects from the ferromagnet are neglected. 
The spin-orbit coupling effect at the nonmagnet/ferromagnet interface has been considered to be another dominant mechanism and intensively investigated \cite{Manchon2009, Garate2009, Kim2012, Haney2013a, Haney2013b, Amin2016a, Amin2016b, Kim2017, Amin2018}. Since the Rashba-type interfacial states are formed at the nonmagnet/ferromagnet interface due to the broken inversion symmetry \cite{Rashba1960, Manchon2015, Bercioux2015}, scattering of electrons from the interface leads to finite spin density and current \cite{Kim2017, Amin2018}, which interacts with and exerts a torque on the local magnetic moments of the ferromagnet (upper right panel in Fig. \ref{fig:classification}). We denote this contribution as interfacial torque. 

While the role of spin-orbit coupling in the ferromagnet has been considered to be negligible as compared to that of the spin-orbit coupling in the nonmagnet, which usually comprises heavy atomic species, it has been found that spin-orbit coupling in the ferromagnet can induce a sizable amount of self-induced torque by the generation of the intrinsic spin current, e.g., via the spin Hall effect \cite{Amin2019, Wang2019, Davison2020}. The corresponding torque contribution is called the anomalous torque in analogy to the anomalous Hall effect in the ferromagnet \cite{Wang2019}. When inversion symmetry is present in a stand-alone ferromagnet, the net anomalous torque amounts to zero. However, in the nonmagnet/ferromagnet bilayer, where the inversion symmetry is broken at the interface, the anomalous torque may exert a finite torque (lower right panel in Fig. \ref{fig:classification}), comparable to the spin torque and interfacial torque. The above mechanisms (spin torque, interfacial torque, and anomalous torque) arise from spin-dependent scattering in the bulk or at the interface, and rely on the concept of spin current or spin density. 

Recently, a mechanism of the torque generation based on the orbital angular momentum injection has been proposed~\cite{Go2020}. This mechanism is fundamentally different from the other mechanisms in that it requires the consideration of the orbital part of the electron's angular momentum, rather than its spin. 
Called the orbital torque, it relies on two processes as described in lower left panel of Fig. \ref{fig:classification}. First, the orbital angular momentum or its current is generated, which can be achieved for instance by the orbital Hall effect \cite{Tanaka2008, Kontani2009, Go2018, Jo2018}. Second, the orbital angular momentum is injected into the ferromagnet and transfers its angular momentum to the local magnetic moment. In this process, the injected orbital angular momentum should couple to the spin of the conduction electron, which interacts with the local magnetic moment via the exchange interaction. Thus, it requires the spin-orbit coupling within the ferromagnet. Since the orbital Hall conductivity can be truly gigantic, exceeding that of the spin Hall conductivity of heavy elements \cite{Tanaka2008, Kontani2009} by an order of magnitude, the orbital torque contribution to the current-induced torque can be substantial (note that in the remainder of the paper, we use the terms current-induced spin-orbit torque and current-induced torque interchangeably). Moreover, since the orbital Hall effect does not require the spin-orbit coupling, which is in contrast to the spin Hall effect, the orbital Hall conductivity is gigantic even in light elements \cite{Go2018, Jo2018}.

In nonmagnet/ferromagnet bilayers, the orbital Hall effect and spin Hall effect coexist in the nonmagnet, especially when the nonmagnet consists of heavy elements. Thus, depending on the material combinations, the orbital torque and spin torque may add up or cancel each other \cite{Go2020}. To enhance the torque efficiency of the device, it is favorable to have the same sign of the orbital torque and spin torque. On the other hand, the case when the sign of the orbital torque and spin torque are opposite is of interest as well, because when the magnitude of the orbital torque is larger than that of the spin torque, the sign of the measured \emph{effective} spin Hall angle in the nonmagnet/ferromagnet bilayer will be opposite to the sign promoted by the spin Hall conductivity of the nonmagnet. Considering that the sign is a more robust quantity than the magnitude in torque measurements, such a sign change can serve as the first hint of an active orbital torque mechanism.

It turns out that all of the above mentioned mechanisms (spin torque, interfacial torque, anomalous torque, and orbital torque) contribute to both fieldlike torque and and dampinglike torque, often with comparable magnitudes. The former gives rise to a precessional motion of the magnetization with respect to the spin accumulation direction, and the latter leads the magnetization to point away/toward an effective field direction. This complicates the analysis of the experiments. Since previous theoretical models have been developed assuming a restricted setup and evaluated only specific contributions \cite{Kim2017, Amin2019}, e.g., when the spin-orbit coupling exists only at the interface, it is hard to compare magnitudes of different contributions directly. On the other hand, first-principles approaches often evaluate the total torque from linear response theory \cite{Freimuth2014, Freimuth2015, Geraton2015, Mahfouzi2018, Belashchenko2019, Guimaraes2020}, which makes it  difficult to assess contributions by different mechanisms quantitatively.

Thus, it is necessary to develop a unified theory within which different mechanisms of the current-induced torque are classified and can be separately evaluated for a given system. This would bridge the gap between the theoretical pictures set up by models and first-principles calculations of real materials. The main difficulty here lies in the nonlocality of magnetoelectric coupling \cite{Valenzuela2006, Kimura2007} and different sources of the spin-orbit coupling. The orbital torque mechanism \cite{Go2020} is highly nonlocal in nature, with the orbital current converted into the spin current in the ferromagnet. In view of the existing analysis based on the spin current, the orbital torque mechanism appears abnormal as the spin current seems to emerge out of nowhere, while in fact it originates in the orbital current. This implies that tracing only the spin current inevitably fails to describe the orbital torque. In general, the spin is not conserved in the presence of spin-orbit coupling, and the spin current does not directly correspond to the spin accumulation or torque on the local magnetic moment \cite{Chen2018}. However, it is important to realize that the angular momentum of the spin is not simply lost. Instead, it is transferred to other degrees of freedom. Therefore, in our theory, we  track not only the flow of  spin but also the flow of orbital angular momentum, as well as their interactions with other degrees of freedom in solids, such as crystal lattice and local magnetic moment. Detailed analysis of the transfer of angular momentum between these channels provides a long-sought insight into the microscopic nature of different competing mechanisms of current-induced torque.

Recent theories imply that the current-induced dynamics and transport of the spin in the presence of spin-orbit coupling originate in the orbital degrees of freedom \cite{Kontani2009, Go2018}. For example, while the orbital Hall effect occurs regardless of the spin-orbit coupling, the spin Hall effect is a consequence of the orbital Hall effect by virtue of the spin-orbit coupling \cite{Go2018}. Depending on the correlation (or relative orientation) between the spin and orbital angular momentum, the relative sign of the orbital Hall effect and spin Hall effect may be the same or opposite, following Hund's rule behavior \cite{Kontani2009, Go2018}. In this sense, the orbital Hall effect can be considered as a precursor to the spin Hall effect. Another example is a Rashba-type state, which is responsible for the interfacial torque generation. It is well known that the Rashba state originates in a chiral orbital angular momentum texture \cite{Park2013a, Park2013b, Go2017}. Such an \emph{orbital} Rashba effect persists even in the absence of spin-orbit coupling, which induces current-induced orbital dynamics and transport \cite{Salemi2019, Canonico2020}. Through spin-orbit coupling, the orbital Rashba state couples to the spin and the spin texture emerges, thus leading to spin dynamics. In general, such a hierarchy is expected to be a rather universal feature. The reason is the following: in the microscopic Hamiltonian of the electrons in solids, the spin cannot interact with an external electric field unless the spin-orbit coupling is present. On the other hand, the orbital degree of freedom, originated in the real-space behavior of the wave functions and distribution of charge, directly couples to an external electric field (see Fig.~\ref{fig:overview}). Hence, under the perturbation by an external electric field, the orbital dynamics is expected to occur prior to the spin dynamics and regardless of the spin-orbit coupling, and the spin dynamics becomes correlated with the orbital dynamics due to the spin-orbit coupling. Therefore, the orbital degree of freedom should be explicitly incorporated into a theoretical formulation to properly describe the current-induced torque, or magnetoelectric coupling phenomena in general. This will help to achieve clarity in resolving various contributions to the current-induced torques.

In this paper, we develop a theoretical formalism that can track the flow and transfer of the angular momentum between spin and orbital degrees of freedom of electrons, the crystal lattice, and the local magnetic moment in the presence of an external electric field. Following the continuity equations for the spin and orbital angular momentum of the electron, which was outlined in Ref. \cite{Haney2010}, we clarify every channel for the angular momentum transfer: between spin-orbital, orbital-lattice, and spin-local magnetic moment. Then we derive equations of motion which hold in the steady state in the presence of an external electric field. For the angular momentum transfer between electron's spin and local magnetic moment, which is directly related to the current-induced torque, we propose criteria for classifying different microscopic mechanisms based on physical properties: whether the magnetoelectric coupling is of local or nonlocal nature and whether it originates in the atomic spin-orbit coupling of the nonmagnet or the ferromagnet. In this way, we classify the mechanism of the current-induced torque as spin torque, orbital torque, interfacial torque, and anomalous torque, and separately evaluate them for a given system.

As a proof of principle, we implement our formalism in the density functional theory framework, and perform first-principles calculations for two real material systems: Fe/W(110) and Ni/W(110), which are carefully chosen with the expectation that the spin torque and orbital torque have an opposite sign in these bilayers. We show that the current-induced torque in Fe/W(110) is dominated by the spin torque contribution, that is, the spin current flux in Fe equals the torque acting on the local magnetic moment. As a result, the effective spin Hall angle is negative, as it is well known for W. On the other hand, we find that the orbital torque is dominant over the spin torque in Ni/W(110). As a result, it leads to a positive sign of the effective spin Hall angle, which is opposite to the sign of the spin Hall conductivity in W. This peculiar result is due to a positive sign of the orbital Hall conductivity in W. In Ni/W(110) it is found that angular momentum transfer from the orbital to the spin channel is pronounced in the ferromagnet, which is a crucial requirement for the orbital torque mechanism. We attribute the different behavior of Fe/W(110) and Ni/W(110) to the difference in the electronic structure, where the correlation between the spin and orbital angular momenta in the ferromagnet is more pronounced in Ni/W(110) than Fe/W(110) near the Fermi energy. In addition, we find that the interfacial torque and anomalous torque are not negligible in both Fe/W(110) and Ni/W(110). These results clearly demonstrate the advantages of our theoretical formalism tracking the flow and transfer of the angular momentum through various degrees of freedom. Moreover, a different sign of the effective spin Hall angle in two different systems can be readily measured in experiment.

The  paper is organized as follows. In Sec. \ref{sec:theoretical_formalism}, we develop a theoretical formalism that describes angular momentum transfer between the spin and the orbital angular momentum of the electron, lattice, and local magnetic moment in the steady state under an external electric field. We propose a classification scheme for the mechanisms of current-induced torque and provide definitions of spin torque, orbital torque, interfacial torque, and anomalous torque. In Sec. \ref{sec:first-principles_calculation}, we apply this formalism to perform a first principles study of current induced torques in Fe/W(110) and Ni/W(110) bilayers. In Sec. \ref{sec:discussion}, we further discuss the disentangling of the various mechanisms of current-induced torque and comment on several issues of orbital transport and dynamics. This includes similarity and difference between the orbital current and spin current, and implications on experiments. Finally, Sec. \ref{sec:conclusion} summarizes and concludes the paper.

\section{Theoretical Formalism}
\label{sec:theoretical_formalism}

\subsection{Overview}
\label{subsec:overview}

In this section, we develop a theoretical formalism that describes angular momentum transfer between different degrees of freedom to identify competing mechanisms of the current-induced torque separately. Before presenting detailed equations, we provide a motivation and an overview of the formalism that we aim to derive. Figure \ref{fig:overview} shows interactions between spin and orbital momenta of the electron, lattice, and local magnetic moment, each of which carry angular momentum in solids. Considering microscopic interactions, the electron's spin interacts with the local magnetic moment via the exchange interaction, the electron's orbital moment interacts with the lattice via the crystal field potential, and the electron's spin and orbital momenta are coupled by the spin-orbit coupling. It is important to note that the local magnetic moment and the electron's spin on the right column of Fig. \ref{fig:overview} are related to magnetic excitations, i.e., in the absence of spin-orbit coupling they do not respond to an electric field. On the other hand, the electron's orbital and the crystal lattice, in the left column of Fig. \ref{fig:overview}, react to an application of an external electric field, and their orbital dynamics couples to a magnetic field. Therefore, the electronic orbital degree of freedom is a core element in describing magnetoelectric coupling, {\it e.g.}, the current-induced torque. Note that a charge excitation of the ions in the lattice is efficiently screened by the electrons in metals, which are our main interest in this paper. Moreover, we assume that the lattice degrees of freedom are frozen (absence of a phonon excitation) and we neglect a coupling between the ions and an external electric field. Therefore, according to this physical picture, the current-induced torque arises as follows: An external electric field excites the orbital dynamics, with which the spin dynamics is entangled by the spin-orbit coupling. The resulting spin dynamics alters the local magnetic moment by the exchange interaction.

An exception to this picture is a noncollinear magnet, where the orbital angular momentum is associated with the scalar spin chirality \cite{Hoffmann2015, Hanke2017} or Skyrmion charge \cite{dosSantosDias2016,Lux2018}. Here, spin and orbital momenta may interact even without relativistic spin-orbit coupling~\cite{Grytsiuk2020}. Although such topological orbital angular momentum exhibits exotic dynamic phenomena associated with complex spin structures \cite{Zhang2019}, we leave this case to future work.

In the rest of this section, we first define different mechanisms of the current-induced torque in Sec. \ref{subsec:classification}, which we aim to disentangle for a given magnetic bilayer system. To achieve this, we start from the effective single-particle Hamiltonian to separately define the spin-orbit coupling, the crystal field potential, and the exchange interaction, which is adapted for the density functional theory framework (Sec. \ref{subsec:effective_Hamiltonian}). Then we derive the continuity equations for the spin and orbital angular momentum in Sec. \ref{subsec:continuity_equations}. In the continuity equations, rates for the changes of the spin and orbital angular momentum are captured by the influxes of the spin and orbital angular momentum as well as torques describing the angular momentum transfer between different degrees of freedom. To evaluate individual contributions appearing in the continuity equations under an external electric field, we consider interband and intraband contributions within the Kubo formula (Sec. \ref{subsec:Kubo_formula}). However, we point out that the interband contribution does not satisfy the stationary condition in the steady state (Sec. \ref{subsec:stationary_condition}). To resolve this problem, we propose a balance-type equation that describe a relation between the interband and intraband contributions in the steady state, which we call the interband-intraband correspondence. The application of the interband-intraband correspondence to the continuity equations of the spin and orbital angular momentum leads to the equations of motion (Sec. \ref{subsec:steady_state_eom}), which is the main result of this section. Meanwhile, the intraband contribution satisfies the stationary condition by itself, for which we derive the equations of motion as well.

\subsection{Classifying Mechanisms of the Current-Induced Torque}
\label{subsec:classification}

We aim to identify and disentangle various competing mechanisms for the current-induced torque with our formalism. Before presenting the detailed formalism, we define various mechanisms of the current-induced torque more precisely. We consider two independent criteria: (1) whether it is an effect due to spin-orbit coupling in the nonmagnet or the ferromagnet, and (2) whether it is due to electrical current flowing in the nonmagnet or in the ferromagnet. Figure \ref{fig:classification} presents a table of the mechanisms of the current-induced torque, where the row classifies whether the spin-orbit coupling originates in the nonmagnet or in the ferromagnet, and the column classifies whether the nature of the torque response is nonlocal or local. We define the nonlocal and local nature of the torque as the response arising in the ferromagnet from the electrical current flowing in the nonmagnet and the torque arising in the ferromagnet from the electrical current flowing in the nonmagnet, respectively. Thus, we classify microscopic mechanisms of the current-induced torque as follows:
\begin{itemize}
	
	\item Spin torque (nonlocal, spin-orbit coupling from nonmagnet): Electric current flowing in the nonmagnet generates a transverse \emph{spin} current via the spin Hall effect. The \emph{spin} current is injected to the ferromagnet and transferred to the local magnetic moment.
	
	\item Orbital torque (nonlocal, spin-orbit coupling from ferromagnet): Electric current flowing in the nonmagnet generates a transverse \emph{orbital} current via the orbital Hall effect. The \emph{orbital} current is injected into the ferromagnet and couples the spin in the ferromagnet via spin-orbit coupling. The converted spin or spin current generate a torque on the local magnetic moment.
	
	\item Interfacial torque (local, spin-orbit coupling from nonmagnet): Electric current flowing in the ferromagnet scatters from the nonmagnet/ferromagnet interface. By the spin-orbit coupling of the nonmagnet, the interfacial scattering may alter the direction of the spin, i.e., by spin-orbit filtering or spin-orbit precession \cite{Amin2018}. The reflected spin exerts torque on the local magnetic moment.
	
	\item Anomalous torque (local, spin-orbit coupling from ferromagnet): Electric current flowing in the ferromagnet induces transverse spin current via the spin Hall effect. As the inversion symmetry is broken by the nonmagnet/ferromagnet interface, spin accumulation at the top and at the bottom of the ferromagnet become asymmetric, leading to a finite torque on the local magnetic moment.
	
\end{itemize}
We remark that the our definition of the interfacial torque is restricted rather than general. For example, our definition neglect an effect of the current flowing in the nonmagnet in the proximity of the interface. The spin Hall or orbital Hall current in the nonmagnet may be enhanced near the interface, but we include this effect into the definition of the spin torque or the orbital torque, respectively. Thus, the definition of the interfacial torque agrees with the picture that spin-orbit effects in the ferromagnet originate in the proximity-induced spin-orbit coupling from the nonmagnet. Meanwhile, we emphasize that not only the orbital torque but also all the other mechanisms involve an excitation of the orbital angular momentum or its current, because electric response of the spin follows the orbital response via spin-orbit coupling. 


\subsection{Effective Single-Particle Hamiltonian}
\label{subsec:effective_Hamiltonian}

Within the effective single-particle description, such as the Kohn-Sham treatment within the density functional theory, the general electronic Hamiltonian in a solid is formally written as
\begin{eqnarray}
\mathcal{H} 
= 
\int d^3 r
{\Psi}^\dagger (\mathbf{r})
\left[
\frac{\mathbf{p}^2}{2m}
+
V_\mathrm{eff} (\mathbf{r})
\right]
{\Psi} (\mathbf{r}),
\label{eq:Hamiltonian_total}
\end{eqnarray}
where $\Psi (\mathbf{r})$ and $\Psi^\dagger (\mathbf{r})$ are electron annihilation and creation field operators in the second quantization representation, respectively. Here, $\mathbf{p}=-i\hbar \boldsymbol{\nabla}_\mathbf{r}$ is the momentum operator, $\hbar$ is the reduced Plank constant, and $m$ is the electron mass. The effective single-particle potential $V_\mathrm{eff} (\mathbf{r})$ can be divided into the spin-orbit coupling $V_\mathrm{SO} (\mathbf{r})$, the exchange interaction $V_\mathrm{XC} (\mathbf{r})$, and the crystal field potential $V_\mathrm{CF} (\mathbf{r})$:
\begin{eqnarray}
V_\mathrm{eff} (\mathbf{r})
=
V_\mathrm{SO} (\mathbf{r})
+
V_\mathrm{XC} (\mathbf{r})
+
V_\mathrm{CF} (\mathbf{r}).
\label{eq:potential_total}
\end{eqnarray}
We define $V_\mathrm{CF} (\mathbf{r})$ such that it is independent of the spin. The spin-orbit coupling and exchange interaction are explicitly written as
\begin{eqnarray}
V_\mathrm{SO} (\mathbf{r})
&=&
\beta \boldsymbol{\sigma} \cdot \boldsymbol{\nabla}_\mathbf{r} V_\mathrm{CF} (\mathbf{r}) \times \mathbf{p},
\label{eq:spin_orbit_interaction}
\\
V_\mathrm{XC} (\mathbf{r})
&=&
\mu_B \boldsymbol{\Omega}_\mathrm{XC} (\mathbf{r}) \cdot \boldsymbol{\sigma},
\label{eq:exchange_interaction}
\end{eqnarray}
respectively. Here, $\boldsymbol{\sigma}$ is the vector of the Pauli matrices representing the spin, $\beta = \hbar/4m^2c^2$ with the speed of light $c$, $\mu_\mathrm{B}$ is the Bohr magneton, and $\boldsymbol{\Omega}_\mathrm{XC} (\mathbf{r})$ is an effective magnetic field caused by the exchange interaction. We construct $V_\mathrm{SO} (\mathbf{r})$ by neglecting $V_\mathrm{XC} (\mathbf{r})$ as an approximation. Note that the degrees of freedom of the lattice and the local magnetic moment are \emph{implicitly} included in this description, entering as coordinates in the respective potentials $V_\mathrm{XC} (\mathbf{r})$ and $V_\mathrm{CF} (\mathbf{r})$. In the evaluation of operators we use symmetrized representations such that the hermiticity is kept in the numerical implementation. However, we present non-symmetrized forms throughout the paper for notational brevity.

\subsection{Continuity Equations for Spin and Orbital Angular Momenta}
\label{subsec:continuity_equations}

The continuity equations for spin and orbital angular momentum have been introduced by Haney and Stiles in Ref. \cite{Haney2010}. Here, we derive the expression adapted for the first-principles calculation based on the density functional theory, starting from the general single particle Hamiltonian [Eqs. \eqref{eq:Hamiltonian_total} and \eqref{eq:potential_total}]. In the Heisenberg picture (indicated by the hat symbol below), we define the orbital angular momentum and spin density operators as
\begin{subequations}
\label{eq:spin_orbital_density_definition}
\begin{eqnarray}
\hat{\boldsymbol{l}} (\mathbf{r},t)
&=&
\hat{\Psi}^\dagger (\mathbf{r},t) \mathbf{L} \hat{\Psi} (\mathbf{r},t)
\label{eq:orbital_density_definition},
\\
\hat{\mathbf{s}} (\mathbf{r},t)
&=&
\hat{\Psi}^\dagger (\mathbf{r},t) \mathbf{S} \hat{\Psi} (\mathbf{r},t)
\label{eq:spin_density_definition}.
\end{eqnarray}
\end{subequations}
While the spin $\mathbf{S}$ is represented by the vector of the Pauli matrices $\mathbf{S}=(\hbar/2)\boldsymbol{\sigma}$, evaluation of the orbital angular momentum is nontrivial in periodic solids because the position $\mathbf{r}$ is ill-defined under periodic boundary conditions. Nonetheless, we can calculate the orbital angular momentum with respect to the atomic spheres called muffin tins centered at the positions of the atoms:
\begin{subequations}
\begin{gather}
\mathbf{L}
=
\sum_\mu \mathbf{L}_\mu,
\\
\mathbf{L}_\mu
=
\Theta ( R_\mu - r_\mu )
\left(\mathbf{r}_\mu
\times \mathbf{p}
\right).
\end{gather}
\label{eq:orbital_angular_momentum_definition}
\end{subequations}
Here, $\Theta (x)$ is the Heaviside step function, $\mu$ is the index of an atom in the unit cell whose center is located at $\boldsymbol{\tau}_\mu$, $\mathbf{r}_\mu = \mathbf{r} - \boldsymbol{\tau}_\mu$ is the displacement from the atom center, and $R_\mu$ is the radius of the muffin tin. 
This method is called atom-centered approximation, and it gives a reliable result when orbital currents are associated with partially occupied $d$ or $f$ shells, which are localized around atomic centers. Thus, the usage of the atom-centered approximation is justified in magnetic bilayers consisting of transition metal elements, Fe/W(110) and Ni/W(110), which are in the focus of our study. Under the atom-centered approximation, the size of the region in real space which gives rise to the orbital angular moment is smaller than that of a wave packet, thus the orbital can be treated as an \emph{internal} degree of freedom, similar to the spin (see Sec. \ref{subsec:orbital_current_vs_spin_current} for the discussion). However, the atom-centered approximation neglects contributions from nonlocal currents, e.g., in Chern insulators and noncollinear magnets \cite{Hanke2016}, and ultimately one should resort to the modern theory of orbital magnetization \cite{Thonhauser2005, Ceresoli2006, Shi2007}.

For the orbital angular momentum and spin densities defined in Eq. \eqref{eq:spin_orbital_density_definition}, we can derive continuity equations from the Heisenberg equations of motion. These are formally written as
\begin{subequations}
\label{eq:continuity_equations}
\begin{eqnarray}
\frac{\partial \hat{l}_
\alpha(\mathbf{r},t)}{\partial t}
&=&
\frac{1}{i\hbar}
\left[
\hat{l}_\alpha (\mathbf{r},t), \hat{\mathcal{H}}(t)
\right]
\nonumber
\\
&=&
-
\boldsymbol{\nabla}_\mathbf{r} 
\cdot
\hat{\mathbf{Q}}^{L_\alpha} (\mathbf{r},t)
+
\hat{{T}}^{L_\alpha} (\mathbf{r},t),
\label{eq:continuity_equation_orbital}
\\
\frac{\partial \hat{s}_\alpha (\mathbf{r},t)}{\partial t}
&=&
\frac{1}{i\hbar}
\left[
\hat{s}_\alpha (\mathbf{r},t), \hat{\mathcal{H}}(t)
\right]
\nonumber
\\
&=&
-
\boldsymbol{\nabla}_\mathbf{r} 
\cdot
\hat{\mathbf{Q}}^{S_\alpha} (\mathbf{r},t)
+
\hat{T}^{S_\alpha} (\mathbf{r},t),
\label{eq:continuity_equation_spin}
\end{eqnarray}
\end{subequations}
where $\alpha=x,y,z$. Here,
\begin{subequations}
\begin{eqnarray}
\hat{\mathbf{Q}}^{L_\alpha} (\mathbf{r},t)
&=&
\frac{1}{2}
\hat{\Psi}^\dagger (\mathbf{r},t)
\left\{
L_\alpha, \mathbf{v}
\right\}
\hat{\Psi} (\mathbf{r},t),
\label{eq:orbital_current}
\\
\hat{\mathbf{Q}}^{S_\alpha} (\mathbf{r},t)
&=&
\frac{1}{2}
\hat{\Psi}^\dagger (\mathbf{r},t)
\left\{
S_\alpha, \mathbf{v}
\right\}
\hat{\Psi} (\mathbf{r},t),
\label{eq:spin_current}
\end{eqnarray}
\end{subequations}
are orbital and spin current operators, respectively, where 
\begin{eqnarray}
\mathbf{v} = \frac{i\hbar}{2m}
\left(
{\boldsymbol{\nabla}}_\mathbf{r}^\mathrm{L}
-
{\boldsymbol{\nabla}}_\mathbf{r}^\mathrm{R}
\right)
+ \beta \boldsymbol{\sigma} \times \boldsymbol{\nabla}_\mathbf{r} V_\mathrm{CF} (\mathbf{r})
\end{eqnarray}
is the velocity operator (${\boldsymbol{\nabla}}_\mathbf{r}^\mathrm{L}$ and ${\boldsymbol{\nabla}}_\mathbf{r}^\mathrm{R}$ act on the left and on the right, respectively), and 
\begin{subequations}
\begin{eqnarray}
\hat{T}^\mathbf{L} (\mathbf{r},t)
&=&
\frac{1}{i\hbar}
\hat{\Psi}^\dagger (\mathbf{r},t)
[
\mathbf{L}, V_\mathrm{eff} (\mathbf{r})
]
\hat{\Psi} (\mathbf{r},t),
\\
\hat{T}^\mathbf{S} (\mathbf{r},t)
&=&
\frac{1}{i\hbar}
\hat{\Psi}^\dagger (\mathbf{r},t)
[
\mathbf{S}, V_\mathrm{eff} (\mathbf{r})
]
\hat{\Psi} (\mathbf{r},t)
\end{eqnarray}
\end{subequations}
are torque operators for the orbital angular momentum and spin, respectively.

The appearance of the torques in Eq. \eqref{eq:continuity_equations} signals the fact that the orbital angular momentum and spin are not conserved. This implies that the angular momentum is transferred from the electron to other degrees of freedom as described in Fig.~\ref{fig:overview}. The electrons exchange orbital angular momentum with the lattice and with the electron's spin via the crystal field potential $V_\mathrm{CF} (\mathbf{r})$ and spin-orbit potential $V_\mathrm{SO} (\mathbf{r})$, respectively. Thus, the torque acting on the orbital angular momentum of the electron is decomposed as
\begin{eqnarray}
\hat{T}^\mathbf{L} (\mathbf{r},t)
&=&
\hat{T}^\mathbf{L}_\mathrm{CF} (\mathbf{r},t)
+
\hat{T}^\mathbf{L}_\mathrm{SO} (\mathbf{r},t),
\end{eqnarray}
where 
\begin{eqnarray}
\hat{T}_\mathrm{CF}^\mathbf{L} (\mathbf{r},t)
&=&
\frac{1}{i\hbar}
\hat{\Psi}^\dagger (\mathbf{r},t)
[
\mathbf{L}, V_\mathrm{CF} (\mathbf{r}) + V_\mathrm{XC} (\mathbf{r})
]
\hat{\Psi} (\mathbf{r},t),
\ \ \ 
\label{eq:CF_torque}
\\
\hat{T}_\mathrm{SO}^\mathbf{L} (\mathbf{r},t)
&=&
\frac{1}{i\hbar}
\hat{\Psi}^\dagger (\mathbf{r},t)
[
\mathbf{L}, V_\mathrm{SO} (\mathbf{r})
]
\hat{\Psi} (\mathbf{r},t).
\label{eq:SO_torque_orbital}
\end{eqnarray}
We denote $\hat{T}_\mathrm{CF}^\mathbf{L} (\mathbf{r},t)$ as the {\it crystal field torque} and $\hat{T}_\mathrm{SO}^\mathbf{L} (\mathbf{r},t)$ as the {\it spin-orbital torque}. Note that we included the effect of $V_\mathrm{XC}(\mathbf{r})$ in the definition of the crystal field torque, as it contains non-spherical component in general. On the other hand, the electron exchanges the spin angular momentum with the local magnetic moment and the electron's orbital angular momentum via $V_\mathrm{XC} (\mathbf{r})$ and $V_\mathrm{SO} (\mathbf{r})$, respectively. Thus, the torque acting on the electron's spin can be decomposed as
\begin{eqnarray}
\hat{T}^\mathbf{S} (\mathbf{r},t)
&=&
\hat{T}_\mathrm{XC}^\mathbf{S} (\mathbf{r},t)
+
\hat{T}_\mathrm{SO}^\mathbf{S} (\mathbf{r},t),
\end{eqnarray}
where 
\begin{eqnarray}
\hat{T}_\mathrm{XC}^\mathbf{S} (\mathbf{r},t)
&=&
\frac{1}{i\hbar}
\hat{\Psi}^\dagger (\mathbf{r},t)
[
\mathbf{S}, V_\mathrm{XC} (\mathbf{r})
]
\hat{\Psi} (\mathbf{r},t),
\label{eq:exchange_torque}
\\
\hat{T}^\mathbf{S}_\mathrm{SO} (\mathbf{r},t)
&=&
\frac{1}{i\hbar}
\hat{\Psi}^\dagger (\mathbf{r},t)
[
\mathbf{S}, V_\mathrm{SO} (\mathbf{r})
]
\hat{\Psi} (\mathbf{r},t).
\label{eq:spin-orbital_torque_spin}
\end{eqnarray}
We denote $\hat{T}_\mathrm{XC}^\mathbf{S} (\mathbf{r},t)$ as the {\it exchange torque} and  $\hat{T}^\mathbf{S}_\mathrm{SO} (\mathbf{r},t)$ as the {\it spin-orbital torque}. Note that $\hat{T}^\mathbf{L}_\mathrm{SO} (\mathbf{r},t)$ and $\hat{T}^\mathbf{S}_\mathrm{SO} (\mathbf{r},t)$ differ, and we specify them as the spin-orbital torques acting on the orbital and spin, respectively.

We have a few remarks on the different torques and their definitions. In the absence of the spin-orbit coupling, the spin-orbital torques vanish. Thus in a steady state, where $\langle{\partial \hat{s}_\alpha (\mathbf{r},t)}/{\partial t}\rangle=0$, Eq. \eqref{eq:continuity_equation_spin} becomes $\langle {T}_\mathrm{XC}^{S_\alpha} (\mathbf{r})\rangle = \boldsymbol{\nabla}_\mathbf{r} \cdot \langle {\mathbf{Q}}^{S_\alpha} (\mathbf{r}) \rangle$. Here, $\langle \cdots \rangle$ represents expectation value in the steady state. This implies that the spin current divergence is absorbed by the local magnetic moment. Thus, this corresponds to the spin-transfer torque in the absence of the spin-orbit coupling. If we consider the opposite situation where the spin current flux is absent, occurring e.g. in atomically thin magnetic films, where the spin current effect can be neglected along the perpendicular direction to the film plane, Eq. \eqref{eq:continuity_equation_spin} becomes $\langle {T}_\mathrm{XC}^{S_\alpha} (\mathbf{r}) \rangle = - \langle {T}_\mathrm{SO}^{S_\alpha} (\mathbf{r}) \rangle$. Thus, the exchange torque amounts to the spin-orbital torque. This is related to the widely used terminology, \emph{spin-orbit} torque \cite{Manchon2009}. However, in our terminology, the net torque acting on the local magnetic moment is the exchange torque, which may differ from the \emph{spin-orbital} torque due to the presence of the spin current flux. In general, both the spin current flux and spin-orbital torque contribute to the exchange torque.

We obtain additional insight from explicitly evaluating the torques in a simplified situation. Let us first consider the exchange torque. By using Eqs. \eqref{eq:exchange_interaction} and \eqref{eq:exchange_torque}, the exchange torque can be written as  
\begin{eqnarray}
\hat{T}_\mathrm{XC}^\mathbf{S} (\mathbf{r},t)
=
\mu_\mathrm{B}
\hat{\Psi}^\dagger (\mathbf{r},t)
\left[
\boldsymbol{\sigma} \times \boldsymbol{\Omega}_\mathrm{XC} (\mathbf{r})
\right]
\hat{\Psi} (\mathbf{r},t)
\label{eq:XC_torque_explicit}
\end{eqnarray}
in general. Thus, it describes a precession of the spin with respect to the direction of the exchange field. On the other hand, by using Eqs. \eqref{eq:spin_orbit_interaction} and \eqref{eq:spin-orbital_torque_spin}, the spin-orbital torque acting on the spin is formally written as 
\begin{eqnarray}
\hat{T}_\mathrm{SO}^\mathbf{S} (\mathbf{r},t)
=
\beta 
\hat{\Psi}^\dagger (\mathbf{r},t)
\left[
\boldsymbol{\sigma} \times 
\left\{
\boldsymbol{\nabla}_\mathbf{r} V_\mathrm{CF} (\mathbf{r}) \times \mathbf{p}
\right\}
\right]
\hat{\Psi} (\mathbf{r},t).
\nonumber
\\
\end{eqnarray}
Since it depends on the spatial gradient of $V_\mathrm{CF} (\mathbf{r})$, the dominant contribution to it is concentrated near the atom centers, where $V_\mathrm{CF} (\mathbf{r})$ is almost spherical. Thus, within the muffin tins, we can approximately write  $\boldsymbol{\nabla}_\mathbf{r} V_\mathrm{CF} (\mathbf{r})  \approx \sum_\mu \Theta \left( R_\mu - {r}_\mu \right) [\partial V_\mathrm{CF} (r_\mu)/\partial r_\mu]$. Within this approximation
\begin{eqnarray}
V_\mathrm{SO} (\mathbf{r})
\approx
\sum_\mu  
\hat{\Psi}^\dagger (\mathbf{r},t)
\left[
\xi_\mu (r_\mu) 
\mathbf{L}_\mu \cdot \boldsymbol{\sigma}
\right]
\hat{\Psi} (\mathbf{r},t).
\end{eqnarray}
Thus, the spin-orbital torque becomes
\begin{eqnarray}
\hat{T}_\mathrm{SO}^\mathbf{S} (\mathbf{r},t)
\approx
\sum_\mu 
\xi_\mu (r_\mu) 
\left(
\mathbf{L}_\mu \times \boldsymbol{\sigma}
\right),
\label{eq:spin_orbital_torque_approx}
\end{eqnarray}
where 
\begin{eqnarray}
\xi_\mu (r_\mu) 
=
\frac{\beta}{r_\mu} \frac{d V_\mathrm{CF} (r_\mu)}{d r_\mu}
\label{eq:spin-orbit coupling_strength}
\end{eqnarray}
is the strength of the spin-orbit coupling for the $\mu$-th atom. Therefore, Eq. \eqref{eq:spin_orbital_torque_approx} indicates that the spin-orbital torque describes a mutual precession between the orbital angular momentum and the spin. That is,
\begin{eqnarray}
\hat{T}_\mathrm{SO}^\mathbf{S} (\mathbf{r},t) 
\approx
- \hat{T}_\mathrm{SO}^\mathbf{L} (\mathbf{r},t).
\end{eqnarray}
While it is approximately true in most systems, we keep superscripts $\mathbf{S}$ and $\mathbf{L}$ separately, because $\hat{T}_\mathrm{SO}^\mathbf{S} (\mathbf{r},t)$ and $-\hat{T}_\mathrm{SO}^\mathbf{L} (\mathbf{r},t)$ differ in general due to nonspherical contributions to the $V_\mathrm{SO} (\mathbf{r})$ although it is small.

Meanwhile, the crystal field torque cannot be expressed in simple terms. In general, it describes an angular momentum transfer between the lattice and the electronic orbital angular momentum. It originates due to the breaking of the continuous rotation symmetry by the crystal field, which differentiates specific directions depending on the structure of the crystal, and leads to various anisotropic effects.

\subsection{Kubo Formula: Interband and Intraband Responses}
\label{subsec:Kubo_formula}

The current-induced torque corresponds to the response of the exchange torque to an electric field, [Eqs. \eqref{eq:exchange_torque} and \eqref{eq:XC_torque_explicit}]. One of the most widely used approaches for its calculation is the linear response theory, where often interband and intraband contributions are evaluated separately. The interband contribution originates in the change of a given state by a coherent superposition of the eigenstates for a given $\mathbf{k}$: in response to an external electric field $\boldsymbol{\mathcal{E}} = \mathcal{E}_x \hat{\mathbf{x}}$ the periodic part of the Bloch state $\ket{u_{n\mathbf{k}}}$ changes as
\begin{eqnarray}
\ket{u_{n\mathbf{k}}}
\rightarrow
\ket{u_{n\mathbf{k}}} + \ket{\delta u_{n\mathbf{k}}},
\end{eqnarray}
where
\begin{eqnarray}
\ket{\delta u_{n\mathbf{k}}}
=
i\hbar e \mathcal{E}_x
\sum_{m\neq n}
\frac{
\ket{u_{m\mathbf{k}}}
\bra{u_{m\mathbf{k}}}
v_x (\mathbf{k})
\ket{u_{n\mathbf{k}}}
}{
\left(
E_{n\mathbf{k}} - E_{m\mathbf{k}} + i\eta
\right)^2
}.
\label{eq:state_change}
\end{eqnarray}
Here, $e>0$ is the absolute value of the charge of the electron, $\mathbf{k}$ is the crystal momentum, $E_{n\mathbf{k}}$ is the energy eigenvalue for the periodic part of the $n$-th Bloch state $\ket{u_{n\mathbf{k}}}$. The infinitesimally small number $\eta >0$ arises from the causality relation. That is, in describing time-evolution of the state, the electric field is adiabatically turned on from $t=-\infty$ to $t=0$ by the vector potential $\mathbf{A} (t) =  - te^{\eta t/\hbar} \mathcal{E}_x \hat{\mathbf{x}}$ such that $\boldsymbol{\mathcal{E}} = -\partial\mathbf{A}(t)/\partial t$. As a result, the interband response of an observable $\mathcal{O}$ is given by
\begin{eqnarray}
\label{eq:interband_raw}
\left\langle \mathcal{O} \right\rangle^\mathrm{inter}
=
2
\sum_{n\mathbf{k}}
f_{n\mathbf{k}}
\mathrm{Re}
\left[
\bra{u_{n\mathbf{k}}}
\mathcal{O} (\mathbf{k})
\ket{\delta u_{n\mathbf{k}}}
\right],
\end{eqnarray}
where $f_{n\mathbf{k}}$ is the Fermi-Dirac distribution function for the state $\ket{u_{n\mathbf{k}}}$. By combining Eqs. \eqref{eq:state_change} and \eqref{eq:interband_raw} and manipulating the dummy indices $n$ and $m$, we arrive at
\begin{eqnarray}
\label{eq:interband_final}
\left\langle {\mathcal{O}} \right\rangle^\mathrm{inter}
&=&
e\hbar \mathcal{E}_x
\sum_{n\neq m}
\sum_{\mathbf{k}}
\left(
f_{n\mathbf{k}} - f_{m\mathbf{k}}
\right)
\\
& &
\times
\mathrm{Im}
\left[
\frac{
\bra{u_{n\mathbf{k}}}
\mathcal{O} (\mathbf{k})
\ket{u_{m\mathbf{k}}}
\bra{u_{m\mathbf{k}}}
v_x (\mathbf{k})
\ket{u_{n\mathbf{k}}}
}{(E_{n\mathbf{k}} - E_{m\mathbf{k}} + i\eta)^2}
\right].
\nonumber
\end{eqnarray}
Here, we define $\mathcal{O} (\mathbf{k}) = e^{-i\mathbf{k}\cdot\mathbf{r}}\mathcal{O} e^{i\mathbf{k}\cdot\mathbf{r}}$ in $\mathbf{k}$-space. The interband contribution in Eq. \eqref{eq:interband_final} is also known as the  {\it intrinsic contribution} since it depends only on the electronic structure, the eigenstates and their energy eigenvalues in the ground state. 

On the other hand, the intraband response arises due to a shift of the Fermi surface by disorder scattering. The leading contribution arises from the change of the occupation function:
\begin{eqnarray}
\left\langle {\mathcal{O}} \right\rangle^\mathrm{intra}
=
\sum_{n\mathbf{k}}
(f_{n\mathbf{k}+\Delta\mathbf{k}} - f_{n\mathbf{k}})
\bra{u_{n\mathbf{k}}}
\mathcal{O} (\mathbf{k})
\ket{u_{n\mathbf{k}}},
\nonumber
\\
\label{eq:Kubo_intra}
\end{eqnarray}
which is also referred to as {\it  Boltzmann-like contribution}. Here, $\Delta k_x = -{e\mathcal{E}_x \tau}/{\hbar}$ is the shift of the Fermi surface caused by the electric field $\boldsymbol{\mathcal{E}} = \mathcal{E}_x \hat{\mathbf{x}}$, and $\tau$ is the momentum relaxation time. Up to linear order in $\Delta\mathbf{k}$, 
\begin{eqnarray}
f_{n\mathbf{k}+\Delta\mathbf{k}} - f_{n\mathbf{k}}
&\approx &
\hbar \Delta\mathbf{k}
f'_{n\mathbf{k}} 
\bra{u_{n\mathbf{k}}}
v_x (\mathbf{k})
\ket{u_{n\mathbf{k}}},
\end{eqnarray}
where $f'_{n\mathbf{k}} = \partial f_{n\mathbf{k}}/\partial E_{n\mathbf{k}}$. Thus, the intraband contribution is written as 
\begin{eqnarray}
\left\langle {\mathcal{O}} \right\rangle^\mathrm{intra}
=
-e\mathcal{E}_x \tau
\sum_{n\mathbf{k}}
f'_{n\mathbf{k}}
& &
\bra{u_{n\mathbf{k}}}
\mathcal{O} (\mathbf{k})
\ket{u_{n\mathbf{k}}}
\nonumber
\\
\times
& &
\bra{u_{n\mathbf{k}}}
v_x (\mathbf{k})
\ket{u_{n\mathbf{k}}}.
\label{eq:intraband_final}
\end{eqnarray}
Note that it is described by a single phenomenological parameter $\tau$, which is assumed to be state-independent. As $\tau$ increases, i.e., as the resistivity decreases, the intraband contribution linearly increases. In general, the momentum relaxation time depends on the particular state in the electronic structure. In ferromagnets, for example, it is known that the momentum relaxation times of the majority and minority electrons are different, which plays an important role in understanding various magnetotransport effects \cite{Mathon2001}. However, within the approach that we pursue here, as given by Eq. \eqref{eq:intraband_final}, we do not consider these effects.

\subsection{Stationary Condition in the Steady State}
\label{subsec:stationary_condition}

A serious problem of the linear response described by Eqs. \eqref{eq:interband_final} and \eqref{eq:intraband_final} is that the stationary condition is not satisfied. That is,
\begin{eqnarray}
\left\langle 
\frac{d\mathcal{O}}{dt}
\right\rangle^\mathrm{intra}
+
\left\langle 
\frac{d\mathcal{O}}{dt}
\right\rangle^\mathrm{inter}
\neq 0,
\end{eqnarray}
where $d\mathcal{O}/dt = [\mathcal{O},\mathcal{H}]/i\hbar$. Thus, the continuity equations~\eqref{eq:continuity_equations} are not satisfied if one naively evaluates the sum of the interband and intraband contributions. This discrepancy is due to the inconsistent treatment of disorder scattering, which is only taken into account by the Fermi surface shift within the relaxation time approximation. In general, the effect of disorder scattering enters the equation via the self-energy correction and vertex correction. It is known that a consistent treatment of the self-energy and vertex corrections up to the same order as the perturbation (which is a disorder potential in this case) makes the continuity equation satisfied. This is known as the Ward identity~\cite{Mahan2000}. However, such treatment is computationally demanding, and it requires us to assume a specific model of the disorder potential.

Instead, we propose a remedy by finding a nontrivial relation between the interband and intraband contributions. This allows us to evaluate the response functions given by Eqs. \eqref{eq:interband_final} and \eqref{eq:intraband_final} and retain the stationary condition. We find that the following relation holds:
\begin{eqnarray}
\frac{1}{\tau}
\left\langle 
\mathcal{O}
\right\rangle^\mathrm{intra}
=
\left\langle 
\frac{d\mathcal{O}}{dt}
\right\rangle^\mathrm{inter}
\label{eq:interband-intraband_correspondence}
\end{eqnarray}
as long as the operator $\mathcal{O}(\mathbf{k})$ does not have $\mathbf{k}$-dependence. The proof is presented in Appendix \ref{app:interband-intraband_correspondence}. A physical interpretation of Eq. \eqref{eq:interband-intraband_correspondence} is the following. The right hand side of the equation describes intrinsic \emph{pumping} of $\mathcal{O}$, which depends only on the electronic structure. The left hand side of the equation is related to a relaxation process, which tend to suppress deviations from the equilibrium value of $\mathcal{O}$. In the steady state, the intrinsic pumping and the relaxation rates are equal, thus $\langle \mathcal{O} \rangle^\mathrm{intra}$ is determined by the relaxation rate $\tau$. Therefore, Eq. \eqref{eq:interband-intraband_correspondence} describes a balance between a tendency to increase $\mathcal{O}$ by the intrinsic process and a relaxation rate by the extrinsic process. For the spin operator, Eq. \eqref{eq:interband-intraband_correspondence} holds precisely since it does not have $\mathbf{k}$-dependence. On the other hand, the orbital angular momentum operator [Eq. \eqref{eq:orbital_angular_momentum_definition}] depends on $\mathbf{k}$ since it contains momentum operator $\mathbf{p}$, which turns into $e^{-i\mathbf{k}\cdot\mathbf{r}} \mathbf{p} e^{i\mathbf{k}\cdot\mathbf{r}}=\mathbf{p}+\hbar\mathbf{k}$ in $\mathbf{k}$-space representation. However, the $\mathbf{k}$-dependence of the local orbital momentum is usually very small within the atom-centered approximation as it is usually dominated by a $\mathbf{k}$-independent contribution, i.e., $\mathbf{L}(\mathbf{k}) \approx \mathbf{L}(0)$. In Secs.~\ref{subsec:Fe/W(110)} and \ref{subsec:Ni/W(110)}, we verify that Eq.~\eqref{eq:interband-intraband_correspondence} is satisfied for the orbital angular momentum with high precision, which implies that $\mathbf{k}=0$ contribution in $\mathbf{L}(\mathbf{k})$ dominates and determines overall behavior of the orbital angular momentum operator within the atom-centered approximation.


%
Meanwhile, the intraband contribution alone satisfies the steady state condition:
\begin{eqnarray}
\left\langle 
\frac{d\mathcal{O}}{dt}
\right\rangle^\mathrm{intra}
=0.
\label{eq:intraband_stationary}
\end{eqnarray}
A proof of the stationary condition for the intraband contribution is given in Appendix \ref{app:stationary_condition_intra}. Note that for the intraband contribution, the stationary condition does not rely on $\mathbf{k}$-dependence of $\mathcal{O}(\mathbf{k})$, which is in contrast to the interband-intraband correspondence [Eq. \eqref{eq:interband-intraband_correspondence}]. Equations \eqref{eq:interband-intraband_correspondence} and \eqref{eq:intraband_stationary} are used to derive the equations of motion below.

\subsection{Steady State Equations of Motion for Spin and Orbital Angular Momenta}
\label{subsec:steady_state_eom}

By applying the interband-intraband correspondence [Eq. \eqref{eq:interband-intraband_correspondence}] to the continuity equations [Eq. \eqref{eq:continuity_equations}], we arrive at the following equations:
\begin{widetext}
\begin{subequations}
\label{eq:equations_of_motion_inter}
\begin{eqnarray}
\label{eq:equation_of_motion_orbital_inter}
\frac{1}{\tau} 
\left\langle 
l_\alpha (\mathbf{r}) 
\right\rangle^\mathrm{intra}
&=&
-\boldsymbol{\nabla}_{\mathbf{r}} \cdot 
\left\langle 
{\mathbf{Q}}^{L_\alpha} (\mathbf{r})
\right\rangle^\mathrm{inter}
+
\left\langle 
T_\mathrm{CF}^{L_\alpha} (\mathbf{r})
\right\rangle^\mathrm{inter}
+
\left\langle 
T_\mathrm{SO}^{L_\alpha} (\mathbf{r})
\right\rangle^\mathrm{inter},
\\
\label{eq:equation_of_motion_spin_inter}
\frac{1}{\tau} 
\left\langle 
s_\alpha (\mathbf{r}) 
\right\rangle^\mathrm{intra}
&=&
-\boldsymbol{\nabla}_\mathbf{r} \cdot 
\left\langle 
{\mathbf{Q}}^{S_\alpha} (\mathbf{r})
\right\rangle^\mathrm{inter}
+
\left\langle 
T_\mathrm{XC}^{S_\alpha} (\mathbf{r})
\right\rangle^\mathrm{inter}
+
\left\langle 
T_\mathrm{SO}^{S_\alpha} (\mathbf{r})
\right\rangle^\mathrm{inter}.
\end{eqnarray}
\end{subequations}
\end{widetext}
Note that that the time dependence no longer appears since the equations describe the steady state. Also, the hat symbol for the Heisenberg picture is removed. Equation~\eqref{eq:equations_of_motion_inter} relates the current fluxes and torques of the intrinsic origin to the intraband accumulation of the orbital angular momentum and spin. Application of Eq. \eqref{eq:intraband_stationary} leads to constraints between intraband contributions for the current fluxes and torques of the orbital angular momentum and the spin:
\begin{widetext}
\begin{subequations}
\label{eq:equations_of_motion_intra}
\begin{eqnarray}
\label{eq:equation_of_motion_orbital_intra}
& &
-\boldsymbol{\nabla}_{\mathbf{r}} \cdot 
\left\langle 
{\mathbf{Q}}^{L_\beta} (\mathbf{r})
\right\rangle^\mathrm{intra}
+
\left\langle 
T_\mathrm{CF}^{L_\beta} (\mathbf{r})
\right\rangle^\mathrm{intra}
+
\left\langle 
T_\mathrm{SO}^{L_\beta} (\mathbf{r})
\right\rangle^\mathrm{intra}
=
0,
\\
\label{eq:equation_of_motion_spin_intra}
& &
-\boldsymbol{\nabla}_\mathbf{r} \cdot 
\left\langle 
{\mathbf{Q}}^{S_\beta} (\mathbf{r})
\right\rangle^\mathrm{intra}
+
\left\langle 
T_\mathrm{XC}^{S_\beta} (\mathbf{r})
\right\rangle^\mathrm{intra}
+
\left\langle 
T_\mathrm{SO}^{S_\beta} (\mathbf{r})
\right\rangle^\mathrm{intra}
=
0.
\end{eqnarray}
\end{subequations}
\end{widetext}
The above equations constitute equations of motion for the spin and orbital angular momenta, which are coupled by the spin-orbit coupling, in the steady state reached after an external electric field has been applied. This is one of the main results of our work. Previous theories on the current-induced torque have focused on evaluating linear response of the exchange torque [Eq. \eqref{eq:exchange_torque}] \cite{Freimuth2014, Freimuth2015, Geraton2015, Ghosh2018, Mahfouzi2018, Belashchenko2019, Manchon2020}. In contrast, Eqs. \eqref{eq:equations_of_motion_inter} and \eqref{eq:equations_of_motion_intra} enable one to identify individual microscopic mechanisms responsible for current-induced torque, as we illustrate next.

\section{First-principles calculations}
\label{sec:first-principles_calculation}

\begin{figure}[t!]
	\includegraphics[angle=0, width=0.43\textwidth]{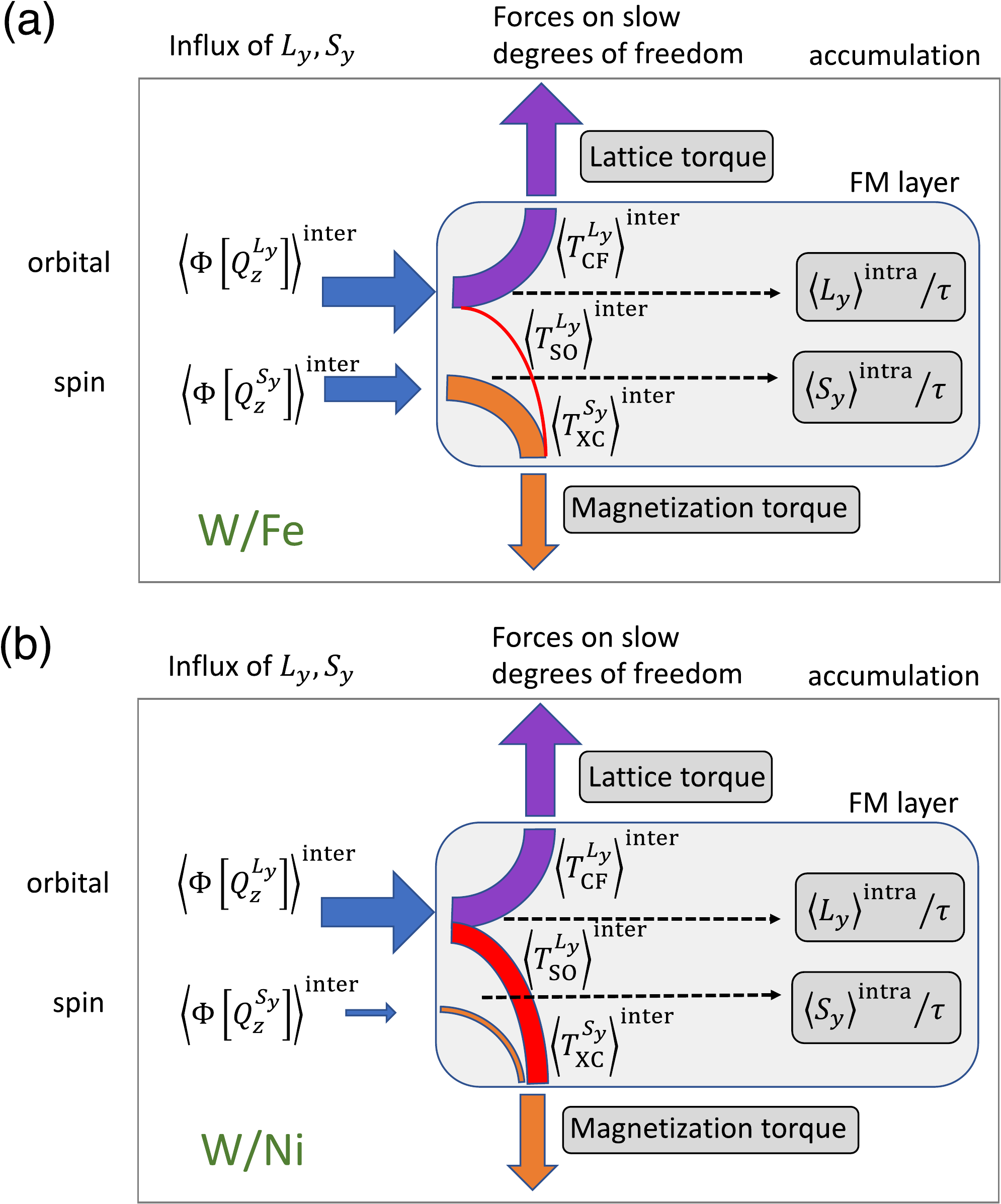}
	\caption{
		\label{fig:summary}
		Schematics of the angular momentum flow in (a) W/Fe and (b) W/Ni. We note that (a) in W/Fe a torque on the magnetization is mostly coming from the spin current influx. (b) On the other hand, in W/Ni, there is a significant contribution of the spin-orbital torque to the magnetization torque.
	}
\end{figure}

In this section we apply the formalism presented in the previous section to two specific systems: W/Fe and W/Ni bilayers.  Before presenting an in-depth analysis of these systems based on the formalism presented in the previous section, it is useful to begin with an overview of the systems' behavior. The angular momentum flows that we calculate for the two systems are illustrated schematically in Fig. \ref{fig:summary}. For the W/Fe system, the flux of orbital angular momentum into the ferromagnetic layer is mostly transferred to a torque on the lattice, while the flux of spin angular momentum is mostly transferred to a torque on the magnetization. This behavior is emblematic of the conventional spin Hall effect combined with spin transfer picture of spin-orbit torque in bilayer systems. The W/Ni system exhibits qualitatively different behavior: the orbital angular momentum flux entering the ferromagnetic layer contributes substantially to the torque on the magnetization, indeed a magnitude which exceeds the contribution from the spin current flux. In this case, the more prominent spin-orbit coupling in Ni enables a flow of angular momentum from orbital to spin degrees of freedom. The distinction between W/Fe and W/Ni is evident by a different sign of the current-induced torque on the magnetization in the two systems (equivalently, a different sign of the effective spin Hall effect).  In the following sections we begin with a description of the key differences in the electronic structure of the two systems which underlie the difference in their magnetic response.  We then briefly discuss the symmetry constraints on the system, and finally present an in-depth analysis of the terms entering the conservation of angular momentum in Eq.~\eqref{eq:equations_of_motion_inter}.   

\subsection{Motivation for Choice of Material Systems}

\begin{figure}[t!]
\includegraphics[angle=0, width=0.4\textwidth]{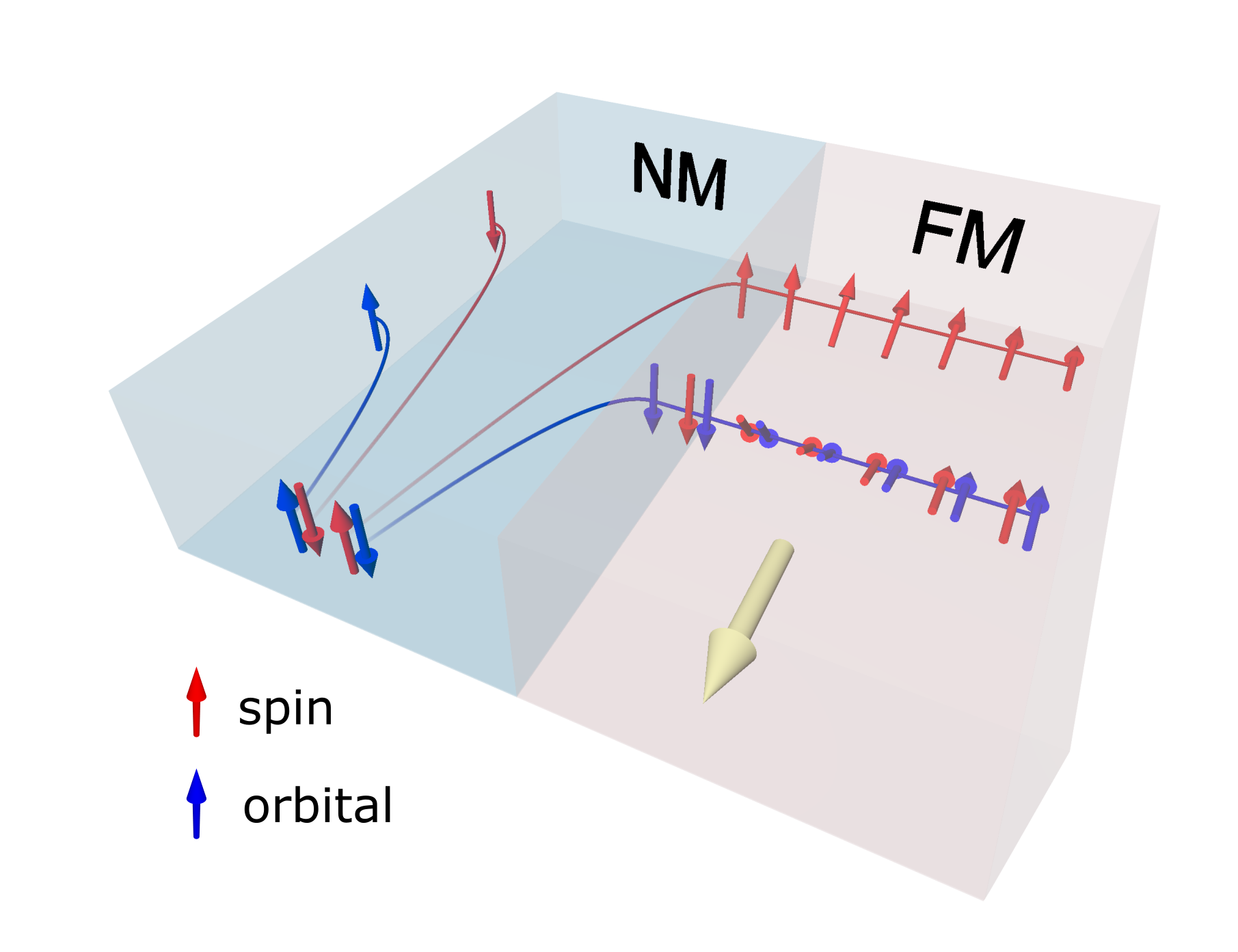}
\caption{
\label{fig:ST_vs_OT}
Competition between the orbital torque and the spin torque when the directions of the orbital Hall effect and spin Hall effect are opposite in the nonmagnet (NM). In the ferromagnet (FM), rotations of the angular momentum represent angular momentum transfer to the local magnetic moment by dephasing, whose directions are opposite for the spin injection and orbital injection.
}
\end{figure}

One of the main motivations in choosing a material system is to find a system with dominant orbital torque behavior, which has been elusive since the first theoretical prediction~\cite{Go2020}, and compare with a \emph{conventional} system where the spin torque is dominant. To do this, consider a case in which the signs of the orbital torque and spin torque are opposite. The sign of the net torque acting on the local magnetic moment will vary depending on whether the orbital torque is larger than the spin torque, or vice versa. This implies that when the orbital torque is dominant over the spin torque, the sign of the torque acting on the local moment can be opposite to that expected from the spin torque mechanism only. This situation can be realized either (1) when the spin Hall effect and orbital Hall effect in the nonmagnet have opposite signs and the spin-orbit correlation in the ferromagnet is positive or (2) when the spin Hall effect and orbital Hall effect in the nonmagnet have same sign and the spin-orbit correlation in the ferromagnet is negative. The spin-orbit correlation in the ferromagnet is important in the orbital torque mechanism because the injected orbital angular momentum in the ferromagnet first couples to the spin and then exerts a torque on the local magnetic moment. For typical $3d$ ferromagnets, such as Fe, Co, and Ni, the spin-orbit correlation is expected to be positive as $d$ shells are more than half-filled, which tends to align the orbital and spin angular momenta along the same direction. Thus, we aim to achieve the case (1), which is schematically illustrated in Fig. \ref{fig:ST_vs_OT}. As the directions orbital Hall effect and spin Hall effect are opposite, the angular momentum transfers by dephasing, which are represented as the rotation of the arrows in the ferromagnet in Fig. \ref{fig:ST_vs_OT}, are also opposite.

One of the key features of the orbital torque mechanism is that it relies on the spin-orbit coupling of the ferromagnet, thus the orbital torque depends on the choice of the ferromagnet. Although the spin-orbit coupling strength is similar for typical $3d$ ferromagnets such as Fe, Co, and Ni, the resulting effect of spin-orbit coupling depends on details of the electronic structure, such as the band structure, band  filling, magnitude of the exchange splitting, etc. This explains a noticeable difference of the spin Hall conductivities of Fe and Ni: $\sigma_\mathrm{SH}^\mathrm{Fe}=519\ (\hbar/e) (\Omega\mathrm{cm})^{-1}$ and $\sigma_\mathrm{SH}^\mathrm{Ni}=1688\ (\hbar/e) (\Omega\mathrm{cm})^{-1}$ \cite{Amin2019}. Thus, even among $3d$ ferromagnets the \emph{effective} spin-orbit coupling strength $-$ which incorporates not only the spin-orbit coupling itself but also electronic structure effects $-$ can vary significantly. We expect that the \emph{effective} spin-orbit coupling strength is much stronger in Ni than in Fe, and we show this by explicit calculations below.

\begin{figure*}[t!]
\includegraphics[angle=0, width=0.75\textwidth]{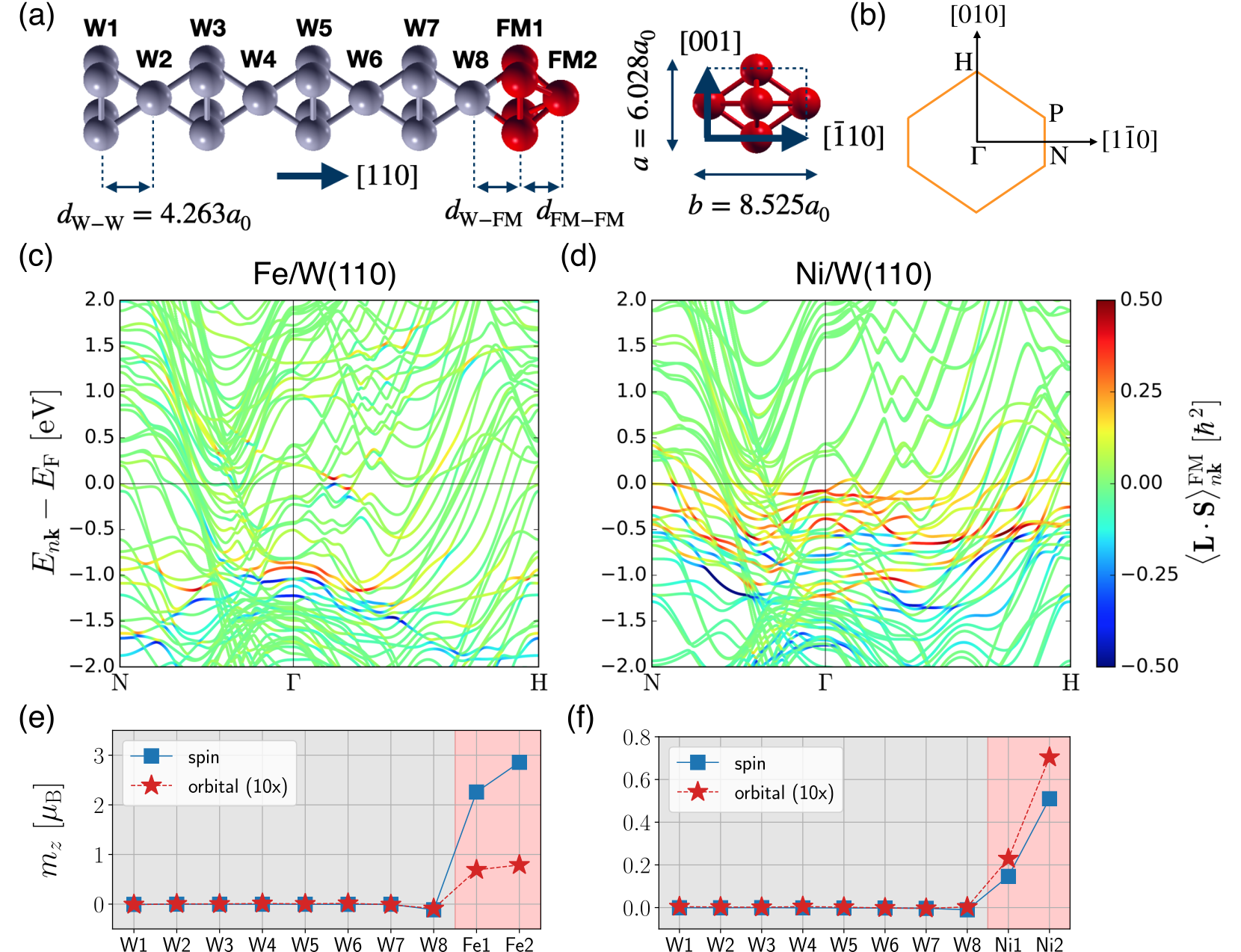}
\caption{
\label{fig:structure}
(a) Crystal structure of ferromagnet (FM)/W(110), where FM = Fe or Ni. Side and top views are displayed on the left and right, respectively. (b) First Brillouin zone and high symmetry points of bcc(110) film. Electronic energy dispersion $E_{n\mathbf{k}}$ and the spin-orbit correlation in the ferromagnet $\langle \mathbf{L}\cdot\mathbf{S} \rangle_{n\mathbf{k}}^\mathrm{FM}$ for (c) Fe/W(110) and (d) Ni/W(110), which are represented by the line and color map, respectively. Note that $\langle \mathbf{L}\cdot\mathbf{S} \rangle_{n\mathbf{k}}^\mathrm{FM}$ is much more pronounced in Ni compared to Fe near the Fermi energy $E_\mathrm{F}$. Layer-resolved plots of the spin (blue squares) and orbital (red stars) moments for (e) Fe/W(110) and (f) Ni/W(110). Comparing Fe/W(110) and Ni/W(110), the spin moment in Fe is much larger than that in Ni, but the relative ratio of the orbital moment over the spin moment is much larger in Ni. This implies that the orbital degree of freedom is not frozen in Ni/W(110), while it is quenched in Fe/W(110).}
\end{figure*}

Therefore, we consider nonmagnet/ferromagnet bilayers where the nonmagnet exhibits an opposite sign of the orbital Hall effect and spin Hall effect, while the ferromagnet is varied such that the strength of effective spin-orbit coupling is controlled. This leads us to the choice of  Fe/W and Ni/W bilayers $-$ two prototypical systems that satisfy these criteria. For W, the orbital Hall conductivity is by an order of magnitude larger than the spin Hall conductivity, with opposite sign \cite{Tanaka2008}. A reason for choosing Fe and Ni as ferromagnets comes from the expectation that the orbital-to-spin conversion efficiency of the orbital torque mechanism is much larger in Ni than it is in Fe. Moreover, both materials can be grown epitaxially along the $[110]$ direction of the body-centered cubic (bcc) structure. We denote these systems as Fe/W(110) and Ni/W(110), respectively. Meanwhile, Fe/W(110) has been previously studied for the anisotropic Dzyaloshinskii-Moriya interactions for stabilizing the anti-Skyrmion \cite{Hoffmann2017}. 

Figures \ref{fig:structure}(a) and \ref{fig:structure}(b) respectively display side and top views of the ferromagnet/W(110) structure, where ferromagnet = Fe or Ni. We consider 8 layers of W and 2 layers of the ferromagnet. We denote the magnetic atom closest to the interface as Fe1 and Ni1, while the magnetic atom at the surface of the slab is marked as Fe2 and Ni2. For the bcc(110) stack of the W layers, we assume that the film follows the bulk lattice parameters of the bcc W, whose lattice constant is $a=6.028 a_0$ in the cubic unit cell convention, where $a_0$ is the Bohr radius. As a result, the distance between the neighboring layers of W is $d_\mathrm{W-W}=a/\sqrt{2}=4.263 a_0$. The in-plane unit cell is of a rectangular shape, whose length along the $[001]$ and $[\bar{1}10]$ directions are $a=6.028 a_0$ and $b=\sqrt{2}a=8.525 a_0$, respectively. The layer distances between W-ferromagnet and ferromagnet-ferromagnet were optimized in order to minimize the total energy: $d_\mathrm{W-Fe}=3.825 a_0$ and $d_\mathrm{Fe-Fe}=3.296 a_0$ for Fe/W(110), and $d_\mathrm{W-Ni}=3.607 a_0$ and $d_\mathrm{Ni-Ni}=3.301 a_0$ for Ni/W(110). We assume that the local magnetic moment is oriented along the direction of $-\hat{\mathbf{z}}$, where $\hat{\mathbf{z}}$ is defined as the direction of $[110]$. The details of first-principles calculation are given in Appendix \ref{app:computational_method}.

\subsection{Spin-Orbit Correlation and Orbital Quenching}

The calculated electronic band structures of Fe/W(110) and Ni/W(110) are shown in Figs. \ref{fig:structure}(c) and \ref{fig:structure}(d), respectively. On top of each energy band $E_{n\mathbf{k}}$, the spin-orbit correlation in the ferromagnet $\langle \mathbf{L}\cdot\mathbf{S} \rangle^\mathrm{FM}_{n\mathbf{k}}$ is shown in color, which is defined as
\begin{eqnarray}
\left\langle 
\mathbf{L} \cdot \mathbf{S}
\right\rangle_{n\mathbf{k}}^\mathrm{FM}
=
\sum_{z \in \mathrm{FM}}
\bra{\psi_{n\mathbf{k}}}
P_z
\left(
\mathbf{L}
\cdot
\mathbf{S}
\right)
P_z
\ket{\psi_{n\mathbf{k}}}.
\end{eqnarray}
Here, $\ket{\psi_{n\mathbf{k}}}$ is the Bloch state of band $n$ at $k$-point $\mathbf{k}$, and $P_z$ is the projection operator onto a layer whose index is $z$. It can be seen that near the Fermi energy $E_\mathrm{F}$, the spin-orbit correlation is negligible in Fe/W(110). The hotspot of this quantity is located about 1.0 eV below the Fermi energy, whose effect is negligible in the steady state transport. On the other hand, in Ni/W(110) the spin-orbit correlation is much more pronounced for states near the Fermi energy. The positive sign of this correlation tends to align the orbital angular momentum and the spin in the same direction. 

The difference in the spin-orbit correlation directly affects the orbital moment of the ferromagnet in equilibrium. In Figs. \ref{fig:structure}(e) and \ref{fig:structure}(f), spin and orbital magnetic moments are plotted in each layer for Fe/W(110) and Ni/W(110), respectively. Blue square symbols and red star symbols respectively indicate the spin and orbital moments. For Fe/W(110) [Fig. \ref{fig:structure}(e)], the magnitude of the spin moment is large: $+2.259 \ \mu_\mathrm{B}$ and $+2.856\  \mu_\mathrm{B}$ for Fe1 and Fe2, respectively. On the other hand, the orbital moments of Fe1 and Fe2 are small: $+0.069\ \mu_\mathrm{B}$ and $+0.079\ \mu_\mathrm{B}$, respectively. The ratio of the orbital moment over the spin moment is 3.06 \% and 2.76 \% for Fe1 and Fe2, respectively, which is fairly small. Thus, the orbital magnetism is strongly quenched in Fe. This implies that even though the orbital angular momentum may be injected into Fe, i.e., by the orbital Hall effect of W, it is likely that most of the orbital angular momentum is relaxed to the lattice through the crystal field torque [Eq. \eqref{eq:CF_torque}] instead of being transferred to the angular momentum of the spin through the spin-orbital torque [Eq. \eqref{eq:SO_torque_orbital}]. Therefore, in Fe/W(110), it is expected that the orbital torque mechanism is not significant and the spin torque mechanism will be dominant, in accordance with common expectation. Meanwhile, we find proximity magnetism in W8 by the hybridization with Fe, where the spin and orbital moments are $-0.114\ \mu\mathrm{B}$ and $-0.009\ \mu_\mathrm{B}$, respectively.


In contrast to Fe/W(110), Ni atoms in Ni/W(110) exhibit much smaller spin moment but relatively large orbital moment. The spin moments are $+0.146\ \mu_\mathrm{B}$, $+0.510\ \mu_\mathrm{B}$ and the orbital moments are $+0.023\ \mu_\mathrm{B}$, $+0.070\ \mu_\mathrm{B}$ for Ni1 and Ni2, respectively. Remarkably, the ratio of the orbital moment over the spin moment is 15.64 \% and 13.80 \% for Ni1 and Ni2, respectively. Thus, the orbital moment is far from being quenched in Ni. Such electronic structure, which is prone to the formation of the orbital angular momentum, promotes the mechanism where an orbital Hall effect-induced orbital angular momentum can efficiently couple to the spin, resulting in the torque on the local magnetic moment. Therefore, at this point we expect that the orbital torque can be significantly larger than the spin torque in Ni/W(110), leading to the opposite effective spin Hall angle when compared to the Fe/W(110) bilayer. 

\begin{figure*}[ht!]
\includegraphics[angle=0, width=0.95\textwidth]{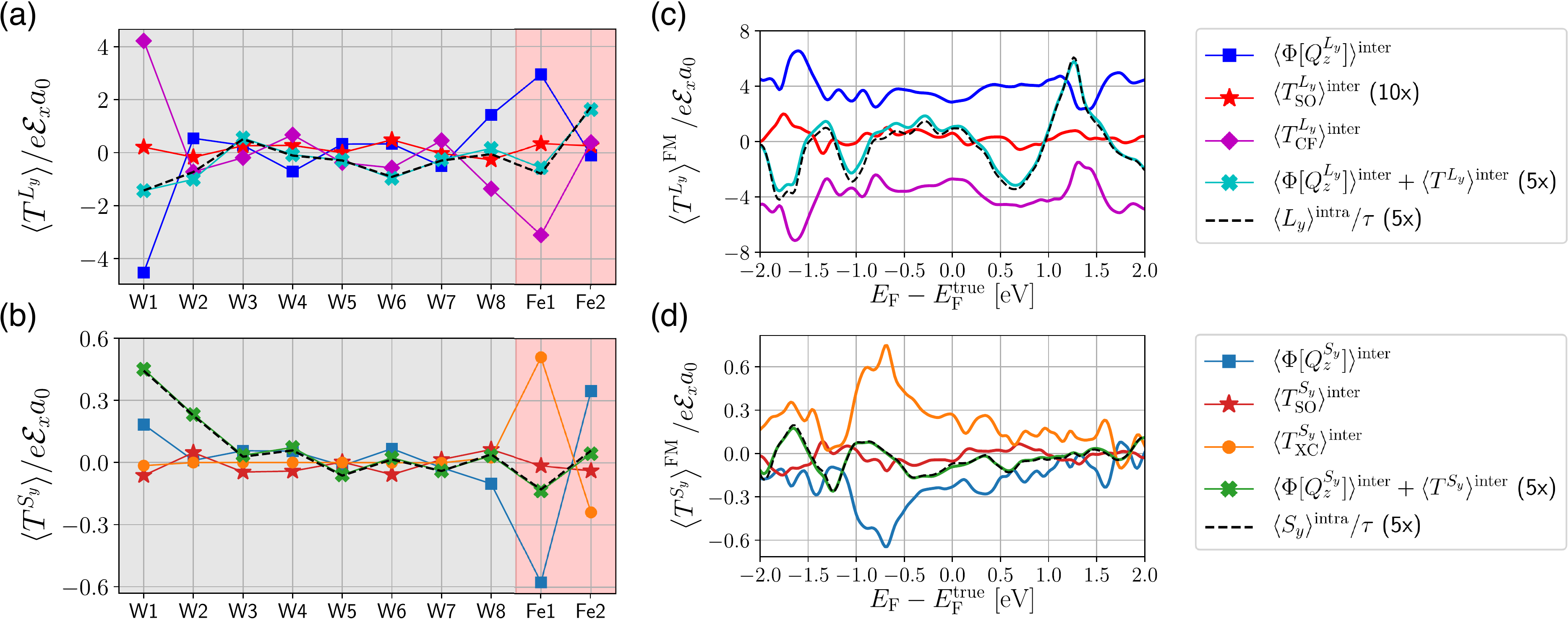}
\caption{
\label{fig:Fe_inter}
Electric response (per unit cell) of $L_y$ and $S_y$ current influxes $-$ $\Phi[Q_z^{L_y}]$ and $\Phi[Q_z^{S_y}]$ $-$ and various torques $-$ $T^{L_y}_\mathrm{SO}$, $T^{L_y}_\mathrm{CF}$, $T^{S_y}_\mathrm{SO}$, and $T^{S_y}_\mathrm{XC}$ $-$ arising from the interband processes and accumulation, and  arising from the intraband processes (divided by $\tau$) in Fe/W(110). Spatial profiles for (a) orbital and (b) spin quantities at the true Fermi energy $E_\mathrm{F} = E_\mathrm{F}^\mathrm{true}$. Fermi energy dependence for (c) orbital and (d) spin quantities, summed over the ferromagnet layers (Fe1 and Fe2). Note that the sum of the interband responses of the orbital/spin current influx and the total torque ($T^{L_y} = T^{L_y}_\mathrm{SO} + T^{L_y}_\mathrm{CF}$ and $T^{S_y} = T^{S_y}_\mathrm{SO} + T^{S_y}_\mathrm{XC}$ for orbital and spin, respectively) matches with the intraband response of the orbital/spin accumulation divided by $\tau$.
}
\end{figure*}

\subsection{Symmetry Constraints}
\label{subsec:symmetry_constraint}

Before presenting the results of first-principles calculations, we consider symmetry constraints on the electric response for quantities taking part in the equations of motion. We define $\hat{\mathbf{x}} \parallel [001]$, $\hat{\mathbf{y}}\parallel [1\bar{1}0]$, and $\hat{\mathbf{z}}\parallel [110]$, and apply an external electric field along the $\hat{\mathbf{x}}$ direction. We consider a situation when $\hat{\mathbf{m}} = -\hat{\mathbf{z}}$, for which the symmetry analysis reveals that only the $y$ component is nonzero in Eq. \eqref{eq:equations_of_motion_inter}. On the other hand, for the equations of motion of the intraband contribution [Eq. \eqref{eq:equations_of_motion_intra}], the $x$ component is the only non-zero component. Thus, we present the result for $\alpha=y$ and $\beta=x$ in Eqs. \eqref{eq:equations_of_motion_inter} and \eqref{eq:equations_of_motion_intra}, respectively. Details of the symmetry analysis are explained in Appendix \ref{app:symmetry_analysis}. The current-induced torque on the local magnetic moment is given by
\begin{subequations}
\begin{eqnarray}
T^\mathbf{m} 
&=&
-\left\langle T_\mathrm{XC}^{\mathbf{S}} \right\rangle^\mathrm{inter}
-
\left\langle T_\mathrm{XC}^{\mathbf{S}} \right\rangle^\mathrm{intra}
\\
&=&
-\hat{\mathbf{y}} \left\langle T_\mathrm{XC}^{S_y} \right\rangle^\mathrm{inter}  -\hat{\mathbf{x}} \left\langle T_\mathrm{XC}^{S_x} \right\rangle^\mathrm{intra} .
\label{eq:torque_m1}
\end{eqnarray}
\end{subequations}
We further decompose $T^\mathbf{m}$ into dampinglike ($T_\mathrm{DL}$) and fieldlike ($T_\mathrm{FL}$) components:
\begin{eqnarray}
T^\mathbf{m} 
&=&
T_\mathrm{DL} \hat{\mathbf{m}}\times (\hat{\mathbf{m}}\times\hat{\mathbf{y}})
+
T_\mathrm{FL} \hat{\mathbf{m}}\times\hat{\mathbf{y}}
\nonumber
\\
&=&
-T_\mathrm{DL} \hat{\mathbf{y}}
+T_\mathrm{FL} \hat{\mathbf{x}},
\label{eq:torque_m2}
\end{eqnarray}
By comparing Eqs. \eqref{eq:torque_m1} and \eqref{eq:torque_m2}, we have
\begin{subequations}
\begin{eqnarray}
T_\mathrm{DL} &=& \left\langle T_\mathrm{XC}^{S_y} \right\rangle^\mathrm{inter},
\\
T_\mathrm{FL} &=& -\left\langle T_\mathrm{XC}^{S_x} \right\rangle^\mathrm{intra}.
\end{eqnarray}
\end{subequations}
%
Below, we present the analysis for $L_y$ and $S_y$ components of quantities from Eqs. \eqref{eq:equation_of_motion_orbital_inter} and \eqref{eq:equation_of_motion_spin_inter}, respectively, which is closely related to that of the dampinglike torque. The analysis for $L_x$ and $S_x$ from Eqs. \eqref{eq:equation_of_motion_orbital_intra} and \eqref{eq:equation_of_motion_spin_intra} is presented in the Appendix \ref{app:intraband_response}. 
In order to perform the decomposition of the computed quantities into contributions coming from each atomic layer, we adopt the tight-binding representation of the equations of motion, as explained in detail in Appendix \ref{app:tight-binding_respresentation}. In the tight-binding representation, we denote orbital and spin current influxes, which correspond to the first terms in the right hand side of Eqs. \eqref{eq:continuity_equation_orbital} and \eqref{eq:continuity_equation_spin}, as $\Phi [Q_z^{L_\alpha}]$ and $\Phi [Q_z^{S_\alpha}]$.

\subsection{Fe/W(110)}
\label{subsec:Fe/W(110)}

In Fig. \ref{fig:Fe_inter}(a), spatial profiles of individual terms appearing in Eq. \eqref{eq:equation_of_motion_orbital_inter} are shown for $L_y$. Note that the current influx and torque have the same dimension, thus we omit the labels for the current influx in the $y$-axes. We find that $\langle \Phi [Q_z^{L_y}] \rangle^\mathrm{inter}$ (blue squares)  is negative near W1 and positive at W8, which corresponds to a positive sign of the orbital Hall conductivity. In concurrence with $\langle \Phi [Q_z^{L_y}] \rangle^\mathrm{inter}$, $\langle T_\mathrm{CF}^{L_y} \rangle^\mathrm{inter}$ (purple diamonds) appears in the opposite sign. However, $\langle T_\mathrm{SO}^{L_y} \rangle^\mathrm{inter}$ (red stars) is much smaller than $\langle \Phi [Q_z^{L_y}] \rangle^\mathrm{inter}$ and $\langle T_\mathrm{CF}^{L_y} \rangle^\mathrm{inter}$. This means that most of the the orbital current influx is absorbed by the lattice. Meanwhile, the sum of $\langle \Phi [Q_z^{L_y}] \rangle^\mathrm{inter}$ and the total torque $\langle T^{L_y} \rangle^\mathrm{inter} = \langle T^{L_y}_\mathrm{SO} \rangle^\mathrm{inter} + \langle T^{L_y}_\mathrm{CF} \rangle^\mathrm{inter}$ (cyan crosses), which corresponds to the right hand side of Eq. \eqref{eq:equation_of_motion_orbital_inter}, matches $\langle L_y \rangle^\mathrm{intra}/\tau$ (black dashed line), which corresponds to the left hand side of Eq. \eqref{eq:equation_of_motion_orbital_inter}. This confirms the validity of the equation of motion Eq. \eqref{eq:equation_of_motion_orbital_inter}. Slight deviations are due to a finite $\eta$ parameter assumed in the calculation of the interband responses by Eq. \eqref{eq:interband_final} (Appendix \ref{app:computational_method}) and $\mathbf{k}$-dependence of the orbital angular momentum operator (Appendix \ref{app:interband-intraband_correspondence}).

Analogously, spatial profiles of the individual terms appearing in Eq. \eqref{eq:equation_of_motion_spin_inter}, related to the spin degree of freedom, are displayed in Fig. \ref{fig:Fe_inter}(b). We remark that the responses related to spin are an order of magnitude smaller than those related to the orbital channel in Fig. \ref{fig:Fe_inter}(a). This is natural since the spin dynamics is caused by the orbital dynamics that occurs first. From the sign of $\langle \Phi [Q_z^{S_y}] \rangle^\mathrm{inter}$ (light blue squares), which is positive near W1 and negative near W8, we conclude that the sign of the spin Hall conductivity is negative. Only in Fe layers, $\langle T_\mathrm{XC}^{S_y} \rangle^\mathrm{inter}$ (orange circles) is sizable, where the exchange interaction is dominant. The overall positive sign of $\langle T_\mathrm{XC}^{S_y} \rangle^\mathrm{inter}$ in Fe layers corresponds to a negative sign of the \emph{effective} spin Hall angle. We observe a strong correlation between $\langle \Phi [Q_z^{S_y}] \rangle^\mathrm{inter}$ and $\langle T_\mathrm{XC}^{S_y} \rangle^\mathrm{inter}$. This implies that the spin current influx is mostly transferred to the local magnetic moment, which agrees with the spin torque mechanism. Meanwhile, $\langle T_\mathrm{SO}^{S_y} \rangle^\mathrm{inter}$ (dark red stars) is much smaller, but not negligible. The sum of $\langle \Phi [Q_z^{S_y}] \rangle^\mathrm{inter}$ and the total torque on the spin $\langle T^{S_y} \rangle^\mathrm{inter} = \langle T^{S_y}_\mathrm{SO} \rangle^\mathrm{inter} + \langle T^{S_y}_\mathrm{XC} \rangle^\mathrm{inter}$ (green crosses), the right hand side of Eq. \eqref{eq:equation_of_motion_spin_inter}, corresponds to $\langle S_y \rangle^\mathrm{intra}/\tau$ on the left hand side (black dashed line).

A pronounced value of $\langle \Phi [Q_z^{S_y}] \rangle^\mathrm{inter}$ near the Fe layers, compared to its value at W1, may seem anomalous [Fig. \ref{fig:Fe_inter}(b)]. However, it can be understood by looking at $\langle S_y \rangle^\mathrm{intra}$, which exhibits a much more pronounced magnitude in W1 and W2, as compared to its value in Fe1 and Fe2. That is, in Fe1 and Fe2, the spin current is efficiently absorbed by the ferromagnet instead of inducing the spin accumulation. The situation is opposite in W1 and W2, where such spin current absorption is not possible, and the spin current simply results in spin accumulation. A similar behavior, where the spin current is strongly enhanced near the ferromagnet interface, has been also predicted in Co/Pt \cite{Freimuth2015} and Py/Pt \cite{Wang2016}.

To understand the predicted behavior in terms of the electronic structure, we present the Fermi energy dependence of the computed quantities in Figs. \ref{fig:Fe_inter}(c) and \ref{fig:Fe_inter}(d) for spin and orbital channels, respectively, where a superscript FM means that it is summed over Fe1 and Fe2 layers. To arrive at these plots, we intentionally varied the Fermi energy $E_\mathrm{F}$ from $-2\ \mathrm{eV}$ to $+2\ \mathrm{eV}$ with respect to the true Fermi energy $E_\mathrm{F}^\mathrm{true}$, assuming that the potential [Eq. \eqref{eq:potential_total}] remains invariant when $E_\mathrm{F}$ changes. For the orbital channel [Eq. \eqref{eq:equation_of_motion_orbital_inter} and Fig. \ref{fig:Fe_inter}(c)], we observe that $\langle \Phi [Q_z^{L_y}] \rangle^\mathrm{inter}$ (blue solid line) and $\langle T_\mathrm{CF}^{L_y} \rangle^\mathrm{inter}$ (purple solid line) tend to cancel each other. Meanwhile, $\langle T_\mathrm{SO}^{L_y} \rangle^\mathrm{inter}$ (red solid line) is smaller than the rest of the contributions. Thus, most of the orbital angular momentum is transferred to the lattice instead of the spin. We find that the equation of motion [Eq. \eqref{eq:equation_of_motion_orbital_inter}] is valid over the whole range of $E_\mathrm{F}$, where the sum of $\langle \Phi [Q_z^{L_y}] \rangle^\mathrm{inter}$ and $\langle T^{L_y} \rangle^\mathrm{inter}$ (cyan solid line) corresponds to $\langle L_y \rangle^\mathrm{intra}/\tau$ (black dashed line).
The Fermi energy properties for the spin channel [Eq. \eqref{eq:equation_of_motion_spin_inter}] are shown in Fig. \ref{fig:Fe_inter}(d). Here, a strong correlation between $\langle \Phi [Q_z^{S_y}] \rangle^\mathrm{inter}$ (light blue solid line) and $\langle T^{S_y}_\mathrm{XC} \rangle^\mathrm{inter}$ (orange solid line) can be observed. We thus conclude that the spin torque mechanism is dominant over the whole range of $E_\mathrm{F}$. At the same time, $\langle T^{S_y}_\mathrm{SO} \rangle^\mathrm{inter}$ (dark red solid line) is suppressed, which implies that the contribution to the current-induced torque caused by the spin-orbit coupling in the ferromagnet, i.e., the orbital torque and anomalous torque mechanisms, is negligible.

\begin{figure}[t!]
\includegraphics[angle=0, width=0.43\textwidth]{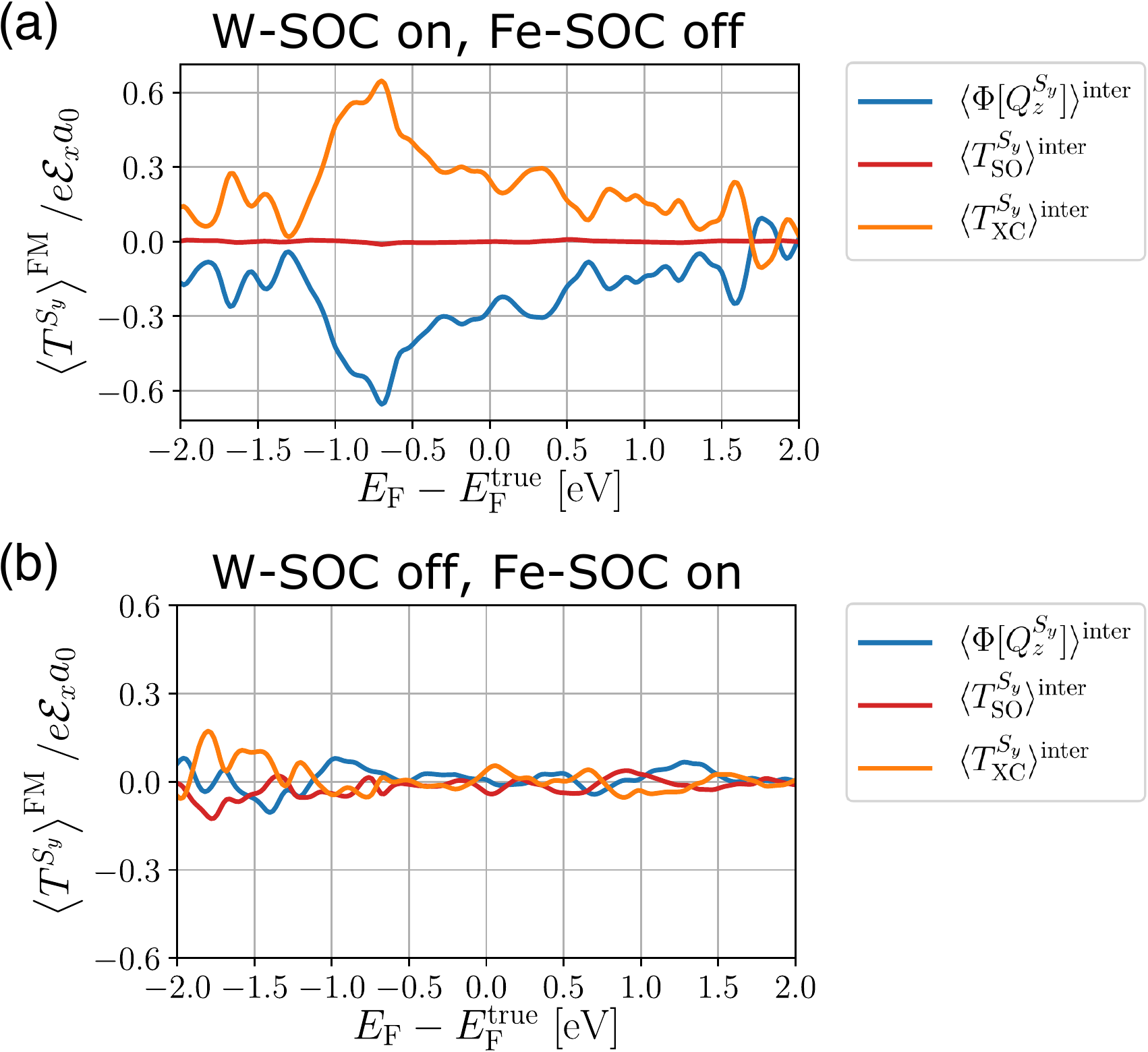}
\caption{
\label{fig:fermi_dep_Fe_SOC}
Fermi energy dependence of interband responses (per unit cell) of the spin current influx $\Phi[Q_z^{S_y}]$ (light blue solid line), spin-orbital torque $T_\mathrm{SO}^{S_y}$ (dark red solid line), and exchange torque $T_\mathrm{XC}^{S_y}$ (orange solid line), which are summed over the Fe layers in Fe/W(110). (a) The result when spin-orbit coupling is on in W and off in Fe, and (b) the result when spin-orbit coupling is off in W and on in Fe.}
\end{figure}

\begin{figure*}[t!]
\includegraphics[angle=0, width=0.95\textwidth]{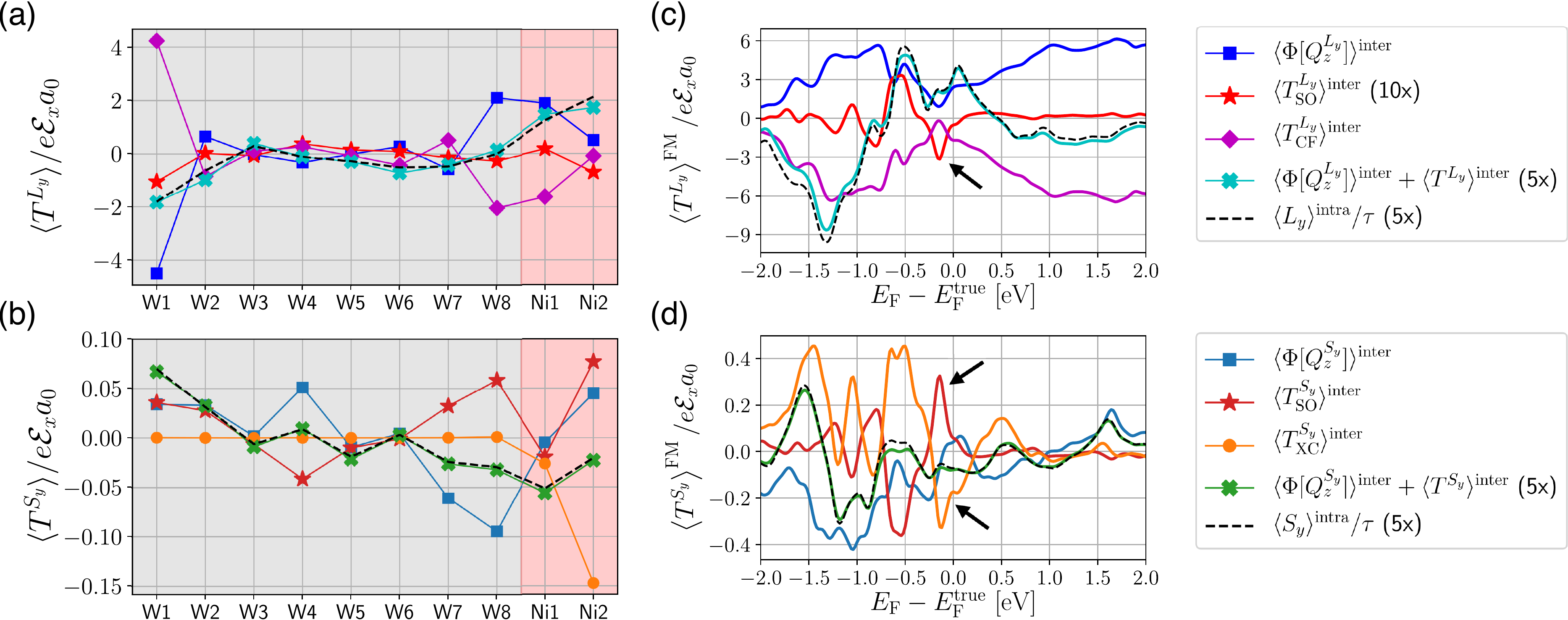}
\caption{
\label{fig:Ni_inter}
Electric response (per unit cell) of $L_y$ and $S_y$ current influxes $-$ $\Phi[Q_z^{L_y}]$ and $\Phi[Q_z^{S_y}]$ $-$ and various torques $-$ $T^{L_y}_\mathrm{SO}$, $T^{L_y}_\mathrm{CF}$, $T^{S_y}_\mathrm{SO}$, and $T^{S_y}_\mathrm{XC}$ $-$ arising from the interband processes and accumulation, and  arising from the intraband processes (divided by $\tau$) in Ni/W(110). Spatial profiles for (a) orbital and (b) spin quantities at the true Fermi energy $E_\mathrm{F} = E_\mathrm{F}^\mathrm{true}$. Fermi energy dependence for (c) orbital and (d) spin quantities, summed over the ferromagnet layers (Ni1 and Ni2). Note that the sum of the interband responses of the orbital/spin current influx and the total torque ($T^{L_y} = T^{L_y}_\mathrm{SO} + T^{L_y}_\mathrm{CF}$ and $T^{S_y} = T^{S_y}_\mathrm{SO} + T^{S_y}_\mathrm{XC}$ for orbital and spin, respectively) matches with the intraband response of the orbital/spin accumulation divided by $\tau$.
}
\end{figure*}

In order to clarify the microscopic mechanism of the current-induced torque better, we intentionally switch on and off the spin-orbit coupling in Fe or W atoms. When spin-orbit coupling is on in W and off in Fe [Fig. \ref{fig:fermi_dep_Fe_SOC}(a)], the Fermi energy dependence of $\langle \Phi [Q_z^{S_y}] \rangle^\mathrm{inter}$ (light blue solid line) perfectly matches  that of$\langle T^{S_y}_\mathrm{XC} \rangle^\mathrm{inter}$ with reversed sign (orange solid line), which supports the spin torque mechanism. On the other hand, $\langle T^{S_y}_\mathrm{SO} \rangle^\mathrm{inter}$ (dark red solid line) is essentially zero due to the absence of spin-orbit coupling in Fe. Meanwhile, when spin-orbit coupling is off in W and on in Fe [Fig. \ref{fig:fermi_dep_Fe_SOC}(b)], all the responses become very small. Thus, any contribution arising from the spin-orbit coupling of the ferromagnet (orbital torque or anomalous torque) is negligible.

\subsection{Ni/W(110)}
\label{subsec:Ni/W(110)}

In Figs. \ref{fig:Ni_inter}(a) and \ref{fig:Ni_inter}(b) we show the  plots of layer-resolved individual terms appearing in the equation of motion [Eq. \eqref{eq:equations_of_motion_inter}] for the $y$ component of the orbital and spin parts, respectively, in Ni/W(110). In Fig. \ref{fig:Ni_inter}(a), we find that the orbital Hall conductivity is positive in sign according to $\langle \Phi[Q_z^{L_y}] \rangle^\mathrm{inter}$ (blue squares). As in the case of Fe/W(110), $\langle \Phi[Q_z^{L_y}] \rangle^\mathrm{inter}$ and $\langle T_\mathrm{CF}^{L_y} \rangle^\mathrm{inter}$ (purple diamonds) are only different in sign, implying that the orbital angular momentum is transferred to the lattice. Thus, $\langle T_\mathrm{SO}^{L_y} \rangle^\mathrm{inter}$ (red stars) is much smaller. These features are similar to those we found in Fe/W(110). The interband-intraband correspondence between $\langle L_y \rangle^\mathrm{intra}/\tau$ (black dashed line) and the sum of $\langle \Phi[Q_z^{L_y}] \rangle^\mathrm{inter}$ and total torque $\langle T^{L_y} \rangle^\mathrm{inter}$ (cyan crosses) is also preserved.

On the other hand, as shown in Fig. \ref{fig:Ni_inter}(b), spatial profiles of spin quantitites are significantly different from those of Fe/W(110). First, we notice that $\langle \Phi[Q_z^{S_y}] \rangle^\mathrm{inter}$ (light blue squares) does not exhibit a close correlation with $\langle T^{S_y}_\mathrm{XC} \rangle^\mathrm{inter}$ (orange circles). Moreover, the sign of $\langle T^{S_y}_\mathrm{XC} \rangle^\mathrm{inter}$ is negative. This means \emph{positive} effective spin Hall angle in Ni/W(110), which is opposite to the negative sign of the spin Hall conductivity in W. This is in contrast to the common interpretation that the spin Hall angle is a property of the nonmagnet, regardless of the ferromagnet. Second, $\langle T^{S_y}_\mathrm{SO} \rangle^\mathrm{inter}$ (dark red stars) is comparable to the rest of the contributions, indicating the importance of spin-orbit coupling in Ni. Meanwhile, the interband-intraband correspondence stands with high precision (green crosses for the sum of $\langle \Phi[Q_z^{S_y}] \rangle^\mathrm{inter}$ and $\langle T^{S_y} \rangle^\mathrm{inter}$, and a black dashed line for $\langle S_y \rangle^\mathrm{intra}/\tau$).

The Fermi energy dependence of the computed quantities, shown in Figs. \ref{fig:Ni_inter}(c) and \ref{fig:Ni_inter}(d) for orbital and spin channels respectively, provides a detailed information on the overall trend. Although $\langle \Phi[Q_z^{L_y}] \rangle^\mathrm{inter}$ and $\langle T^{L_y}_\mathrm{CF} \rangle^\mathrm{inter}$ have opposite sign, their magnitudes differ we find that $\langle T^{L_y}_\mathrm{SO} \rangle^\mathrm{inter}$ is very pronounced near the Fermi energy, with corresponding peak indicated with a black arrow [Fig. \ref{fig:Ni_inter}(c)]. Since the response of the spin quantities is an order of magnitude smaller than that for the orbital channel, the pronounced spin-orbital torque, which is still much smaller than $\langle \Phi[Q_z^{L_y}] \rangle^\mathrm{inter}$ and $\langle T^{L_y}_\mathrm{CF} \rangle^\mathrm{inter}$, can have a significant effect on the dynamics of spin. In concurrence with the increase of $\langle T^{L_y}_\mathrm{SO} \rangle^\mathrm{inter}$, $\langle T^{L_y}_\mathrm{CF} \rangle^\mathrm{inter}$ is significantly decreased near the Fermi energy. This implies that a channel for the orbital angular momentum transfer to the lattice is suppressed.

As a result, the response of spin in Ni/W(110) exhibits a much more rich and complicated behavior when compared to Fe/W(110) [Fig. \ref{fig:Ni_inter}(d)]. We first notice that the correlation between $\langle \Phi[Q_z^{S_y}] \rangle^\mathrm{inter}$ (light blue solid line) and $\langle T^{S_y}_\mathrm{XC} \rangle^\mathrm{inter}$ (orange yellow solid line) is no longer present. Moreover, with the negative drop of $\langle T^{S_y}_\mathrm{XC} \rangle^\mathrm{inter}$, corresponding to the positive sign of the effective spin Hall angle, there is an associated positive peak from $\langle T^{S_y}_\mathrm{SO} \rangle^\mathrm{inter}$ (dark red solid line), which is indicated with a black arrow. This indicates that the spin is transferred from the orbital rather than spin current influx. Therefore, the orbital angular momentum is responsible for the current-induced torque in Ni/W(110). Meanwhile, the interband-intraband correspondence (green solid line for the sum of $\langle \Phi[Q_z^{S_y}] \rangle^\mathrm{inter}$ and $\langle T^{S_y} \rangle^\mathrm{inter}$ and black dashed line for $\langle S_y \rangle^\mathrm{intra}/\tau$) is satisfied.

\begin{figure}[t]
\includegraphics[angle=0, width=0.43\textwidth]{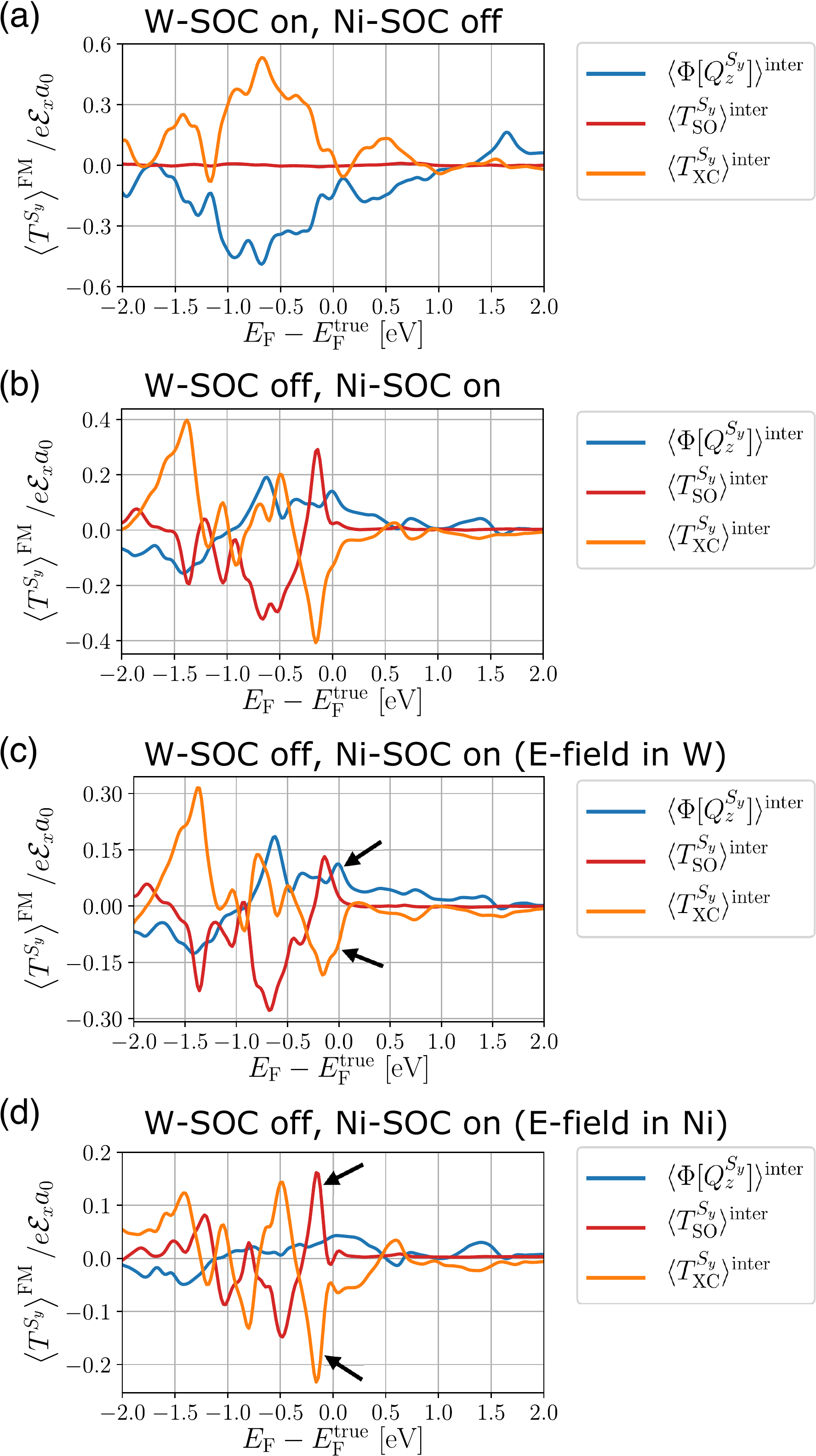}
\caption{
\label{fig:fermi_dep_Ni_SOC}
Fermi energy dependence of interband responses (per unit cell) of the spin current influx $\Phi[Q_z^{S_y}]$ (light blue solid line), spin-orbital torque $T_\mathrm{SO}^{S_y}$ (dark red solid line), and exchange torque $T_\mathrm{XC}^{S_y}$ (orange solid line), which are summed over the Ni layers in Ni/W(110), for the case when (a) the spin-orbit coupling is on in W and off in Ni, and (b)  the spin-orbit coupling is off in W and on in Ni. Both (c) and (d) show the results  when the spin-orbit coupling is off in W and on in Ni, and the external electric field is applied only in (c) W and (d) Ni layers.}
\end{figure}

As we have done for Fe/W(110), we switch on and off the spin-orbit coupling separately for W and Ni atoms in Ni/W(110) as well, showing the results in Fig. \ref{fig:fermi_dep_Ni_SOC}. In Fig. \ref{fig:fermi_dep_Ni_SOC}(a), the Fermi energy dependence of $\langle \Phi[Q_z^{S_y}] \rangle^\mathrm{inter}$, $\langle T^{S_y}_\mathrm{SO} \rangle^\mathrm{inter}$, and $\langle T^{S_y}_\mathrm{XC} \rangle^\mathrm{inter}$ is shown when the spin-orbit coupling of W is on and the spin-orbit coupling of Ni is off. First of all, we find that $\langle T^{S_y}_\mathrm{XC} \rangle^\mathrm{inter}$ is positive at the Fermi energy, which is opposite to the full-spin-orbit coupling case [Fig. \ref{fig:Ni_inter}(d)]. In this case, we find a strong correlation between $\langle \Phi[Q_z^{S_y}] \rangle^\mathrm{inter}$ and $\langle T^{S_y}_\mathrm{XC} \rangle^\mathrm{inter}$. Thus, the \emph{negative} sign of the effective spin Hall angle is caused by the spin injection from the spin Hall effect of W. However, such correlation is not as perfect as in the case of Fe/W(110) [Fig. \ref{fig:fermi_dep_Fe_SOC}(a)]. We attribute such difference to an interfacial mechanism, where the torque is generated regardless of the spin current. Meanwhile, $\langle T^{S_y}_\mathrm{SO} \rangle^\mathrm{inter}$ is negligible since the spin-orbit coupling of Ni is off.

When the spin-orbit coupling is off in W and on in Ni, nontrivial features show up in $\langle \Phi[Q_z^{S_y}] \rangle^\mathrm{inter}$, $\langle T^{S_y}_\mathrm{SO} \rangle^\mathrm{inter}$, and $\langle T^{S_y}_\mathrm{XC} \rangle^\mathrm{inter}$, which is in contrast to Fe/W(110) [Fig. \ref{fig:fermi_dep_Fe_SOC}(b)]. This is due to nontrivial spin-orbit correlation of Ni  shown in Fig. \ref{fig:structure}(d). Moreover, $\langle T^{S_y}_\mathrm{XC} \rangle^\mathrm{inter}$ is negative at the Fermi energy. We find that nontrivial peak features [black arrows in Fig. \ref{fig:Ni_inter}(d)] are reproduced in this calculation. Thus, we confirm that the latter peaks originate in the spin-orbit coupling of Ni. To further clarify the microscopic mechanisms, we apply the external electric field in W only [Fig. \ref{fig:fermi_dep_Ni_SOC}(c)] or Ni only [Fig. \ref{fig:fermi_dep_Ni_SOC}(d)] when the spin-orbit coupling of the W is off and the spin-orbit coupling of Ni is on, which correspond to the orbital torque and the anomalous torque contributions, respectively (more details can be found in the Appendix \ref{app:computational_method}).  In both cases, $\langle T^{S_y}_\mathrm{XC} \rangle^\mathrm{inter}$ exhibits a negative drop near $E_\mathrm{F}-E_\mathrm{F}^\mathrm{true}\approx 0.15\ \mathrm {eV}$, which is correlated with a positive peak of $\langle T^{S_y}_\mathrm{SO} \rangle^\mathrm{inter}$. This implies that for both cases the angular momentum transfer from the orbital channel to the spin channel is crucial. The difference is that for the orbital torque mechanism, Fig. \ref{fig:fermi_dep_Ni_SOC}(c), $\langle \Phi[Q_z^{S_y}] \rangle^\mathrm{inter}$ exhibits a positive peak at the Fermi energy (marked with a black arrow), which comes from the conversion of the orbital current into the spin current by the spin-orbit coupling of Ni. We find that it is correlated with a shoulder feature of $\langle T^{S_y}_\mathrm{XC} \rangle^\mathrm{inter}$ at the Fermi energy (marked with a black arrow). Such peak of $\langle \Phi[Q_z^{S_y}] \rangle^\mathrm{inter}$ implies that in the orbital torque mechanism, there are two different microscopic channels for the orbital-to-spin conversion: one for the spin converted from the orbital angular momentum via $\langle T^{S_y}_\mathrm{SO} \rangle^\mathrm{inter}$, and the other for the conversion of the orbital \emph{current} into the spin \emph{current} followed by the spin-transfer torque. Meanwhile, in Fig. \ref{fig:fermi_dep_Ni_SOC}(d), which corresponds to the anomalous torque mechanism, $\langle \Phi[Q_z^{S_y}] \rangle^\mathrm{inter}$ is not very pronounced, and only the peak of $\langle T^{S_y}_\mathrm{SO} \rangle^\mathrm{inter}$ is observed (indicated with a black arrow). The negative sign of $\langle T^{S_y}_\mathrm{XC} \rangle^\mathrm{inter}$ (positive sign of the effective spin Hall angle) is due to a positive sign of the spin Hall conductivity in Ni. We note that, as expected, for the anomalous torque mechanism, the orbital-to-spin conversion via $\langle T^{S_y}_\mathrm{SO} \rangle^\mathrm{inter}$ is crucial since it originates in the spin-orbit coupling of the ferromagnet. Therefore, we conclude that in Ni/W(110) the orbital torque and anomalous torque are the first and the second dominant mechanisms for the torque generation on the local magnetic moment.

\section{Discussion}
\label{sec:discussion}

\subsection{Disentangling Different Microscopic Mechanisms}

In Sec. \ref{sec:first-principles_calculation}, we found that the spin torque provides the dominant contribution to the current-induced torque in Fe/W(110) according to the correlation between the exchange torque and the spin current influx from W, which is reflected in the negative effective spin Hall angle [Fig. \ref{fig:fermi_dep_Fe_SOC}(a)]. In Ni/W(110), on the other hand, the orbital torque is found to be the most dominant contribution. The evidence for the orbital torque is provided by pronounced peaks in the spin-orbital torque and the spin current influx that suggests a positive effective spin Hall angle, associated with the exchange torque [Fig. \ref{fig:fermi_dep_Ni_SOC}(c)]. However, we also observed that the anomalous torque can be  associated with the spin-orbital torque [Fig. \ref{fig:fermi_dep_Ni_SOC}(d)] because the self-induced spin accumulation in the ferromagnet results from the current-induced orbital angular momentum. A crucial difference between the orbital torque and anomalous torque is that while the orbital torque is due to an electrical current flowing in the nonmagnet, the anomalous torque is due to an electrical current passing through the ferromagnet. In this respect, only the orbital torque is important for memory applications where the ferromagnetic layer must be patterned to form a physically separate memory cell, whereas both orbital torque and anomalous torque are important for applications based on magnetic textures (i.e., domain walls and Skyrmions) for which such patterning is not necessary.

We can disentangle each of the contributions in the current-induced torque of Fe/W(110) and Ni/W(110), according to the classification scheme outlined in Sec. \ref{subsec:classification}. The different contributions to the current-induced torque can be disentangled by modifying the system parameters ``by hand'' in the calculation.  To distinguish between local and nonlocal contributions to the torque, the electric field is selectively applied to only the ferromagnetic or nonmagnetic layer, respectively. We note, however, that this is an approximate measure since an electric current may flow in the ferromagnet(nonmagnet) although an electric field is applied only to the nonmagnet (ferromagnet) layer, as the electronic wave functions are delocalized across the film. For determining the spin-orbit coupling origin (nonmagnet versus ferromagnet), we do not simply turn on and off the spin-orbit coupling because it causes significant change of the band structure. Instead, we change the sign of the the spin-orbit coupling in the relevant layer, which changes the sign of its contribution. For example, we rely on the property that the sign of the orbital torque and anomalous torque should become opposite after flipping the sign of spin-orbit coupling in the ferromagnet, while the spin torque and interfacial torque remain invariant. By computing the torque under different system configurations, the four contributions to the current-induced torque can be determined, as illustrated in Sec. \ref{subsec:classification} and described in detail in Appendix \ref{app:disentanglement}. By this way, the sum of spin torque, orbital torque, interfacial torque, and anomalous torque equals the net torque when the electric field applied to the whole layers with actual spin-orbit coupling strength of each atom. Although this classification scheme relies on computational handles with no experimental counterpart, it provides a systematic basis for physically interpreting the results of calculations, which in turn enables the development of intuition about materials and system designs.

\begin{figure}[t!]
\includegraphics[angle=0, width=0.45\textwidth]{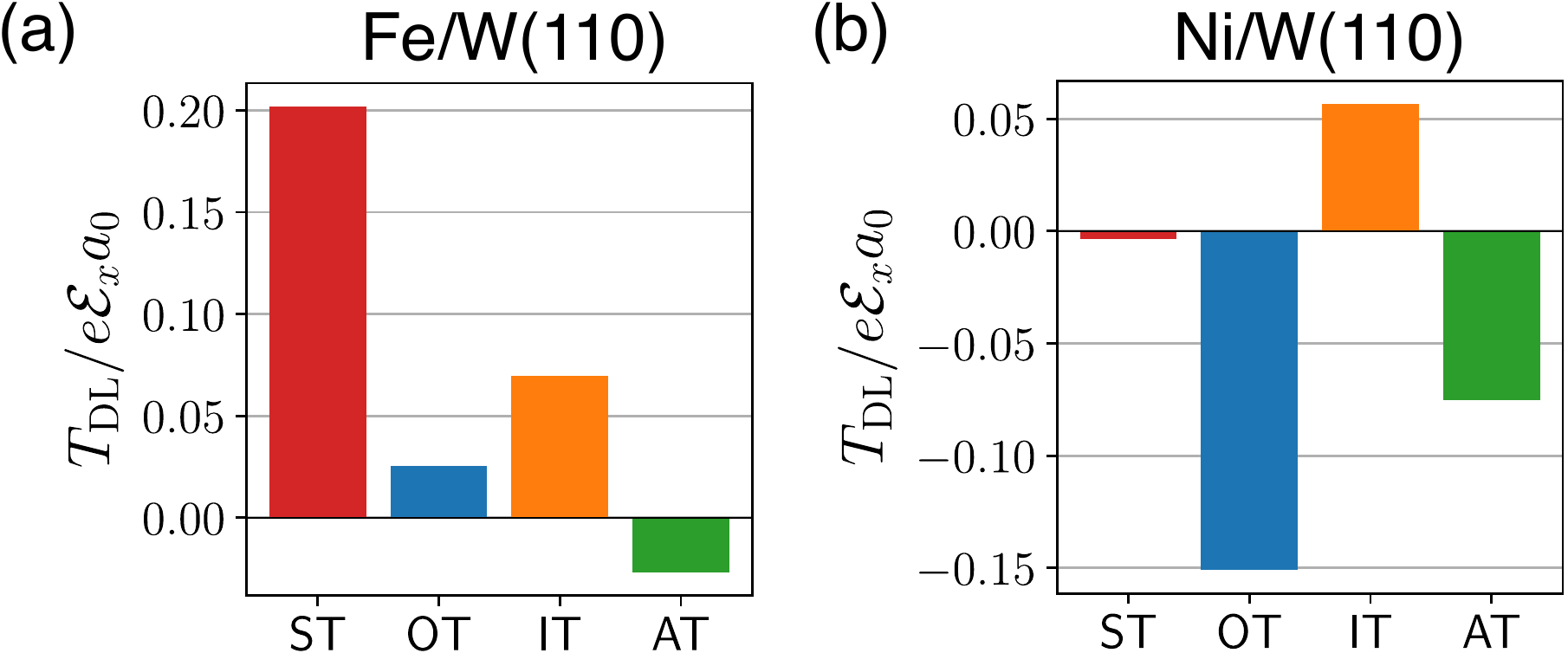}
\caption{
\label{fig:disentanglement}
Disentanglement of the dampinglike torque into the spin torque, orbital torque, interfacial torque, and anomalous torque in (a) Fe/W(11) and (b) Ni/W(110). Note that the spin torque and orbital torque are the most dominant mechanisms in Fe/W(110) and Ni/W(110), respectively. We note that the interfacial torque and anomalous torque are not negligible neither in  Fe/W(110) nor in Ni/W(110).
}
\end{figure}

In Figs. \ref{fig:disentanglement}(a) and \ref{fig:disentanglement}(b) we show the decomposition of the total dampinglike torque in Fe/W(110) and Ni/W(110), respectively, into separate contributions. In Fe/W(110), the spin torque is the most dominant contribution. However, our analysis reveals that the interfacial torque is not negligible, accounting for about 35 \% of the spin torque. Overall, the spin torque and interfacial torque are larger than the orbital torque and anomalous torque, implying that the spin-orbit coupling in W is more important than that in Fe. In Ni/W(110), on the other hand, the orbital torque is the most dominant contribution. The second largest contribution is the anomalous torque, which is comparable to a half of the orbital torque. The magnitude of the interfacial torque is not much smaller, reaching as much as 37 \% of the magnitude of the orbital torque. Overall, the orbital torque and anomalous torque are dominant over the spin torque and interfacial torque in Ni/W(110). This suggests that the spin-orbit coupling in Ni is more important than the spin-orbit coupling in W in this system, in contrast to an intuitive expectation that spin-orbit coupling in 3$d$ ferromagnets plays a minor role as compared to the spin-orbit coupling of the heavy element. These results are consistent with our analysis of the results presented in Figs. \ref{fig:fermi_dep_Fe_SOC} and \ref{fig:fermi_dep_Ni_SOC}.

\subsection{Orbital Current versus Spin Current}
\label{subsec:orbital_current_vs_spin_current}

Although the orbital current [Eq. \eqref{eq:orbital_current}] and the spin current [Eq. \eqref{eq:spin_current}] are defined in a similar way, there are conceptual differences. While the spin and its current can be locally defined everywhere in space, the orbital angular momentum is nonzero only inside the muffin-tin within the atom-centered approximation. Thus, the atom-centered approximation does not properly describe the interstitial region between muffin-tins, where the orbital information is supposed to be transported. Nonetheless, orbital current influx to a muffin-tin can be defined even within the atom-centered approximation, which is the reason why we evaluate the influx instead of the current itself throughout the manuscript. Heuristically, the orbital angular momentum is encoded in a vorticity of the phase of a wave function, which is exists not only in the muffin-tin but also in the interstitial region. It is the vorticity of the wave function that is transported through the interstitial region. The wave function is properly described in our calculation, so that we can reliably compute the flux of vorticity into the muffin-tin.


As the atom-centered approximation neglects the contribution from interstitial region, the crystal field torque in our calculation [Eq. \eqref{eq:CF_torque}] only describes angular momentum transfer from the orbital to the lattice within the muffin-tin, which is mostly concentrated near the surfaces and the interface [Figs. \ref{fig:Fe_inter}(a) and \ref{fig:Ni_inter}(a)]. In general, we expect that nonspherical component of the potential is more pronounced in the interstitial region, which provides another channel for angular momentum transfer from the electronic orbital to the lattice. However, as the $d$ character electronic wave function of a transition metal is localized inside the muffin-tin, we expect that additional contribution to the crystal field torque from the interstitial region is small.

\subsection{Experiments and Materials}

Although the effective spin Hall angle measured in experiments is the sum of all contributions to the torque on the local magnetic moment, it has been assumed that it is a property of the nonmagnet in nonmagnet/ferromagnet bilayers, which can be incorrect. For example, we have shown that the current-induced torque depends on the choice of the ferromagnet in ferromagnet/W(110), where ferromagnet is Fe or Ni. In this case, it is due to an opposite sign of the orbital Hall effect and spin Hall effect in W, and the resulting orbital-to-spin conversion efficiencies are different for Fe and Ni. As a result, even the sign of the effective spin Hall angle changes: from negative for Fe/W(110) to positive for Ni/W(110). We believe that such change-of-sign  behavior can be directly measured in experiment. More concretely, we suggest performing a spin-orbit torque experiment on an FeNi alloy in order to observe change of the effective spin Hall angle as the alloying ratio varies, with the effective spin Hall angle turning to zero at a certain critical concentration. 

We speculate that this behavior would be observed in other systems where the orbital Hall effect competes with the spin Hall effect. For example, among $5d$ elements, Hf, Ta, and Re exhibit gigantic orbital Hall conductivity, whose sign is opposite to that of the spin Hall conductivity \cite{Kontani2009}. Such behavior holds in general for groups 4-7 among transition metals. For $3d$ elements, such as Ti, V, Cr, and Mn, the spin Hall conductivity is much smaller than that of $5d$ elements, while the orbital Hall conductivity is almost as large as in $5d$ elements \cite{Jo2018}. Thus, the orbital torque contribution is expected to be more pronounced than the spin torque contribution when the nominally nonmagnetic substrate is made of $3d$ elements, as compared to the systems where the nonmagnet is made of $5d$ elements. Therefore, alloying not only the ferromagnet but also the nonmagnet provides a useful knob for observing competing mechanisms of the current-induced torque.

Layer thickness dependence of the spin-orbit torque has been measured in Ta/CoFeB/MgO \cite{Kim2013} and Hf/CoFeB/MgO \cite{Ramaswamy2016}, where the sign of the current-induced torque was found to change when the thickness of Ta or Hf was as small as $\approx 1\ \mathrm{nm}$ to $2\ \mathrm{nm}$. The origin of the sign change has been attributed to the competition between the bulk and interfacial mechanisms, which correspond to the spin torque and interfacial torque mechanisms in our terminology. Recently, such behavior has also been observed in a similar system Zr/CoFeB/MgO \cite{Zheng2020}, where a $4d$ element Zr was used instead of a $5d$ element. Due to a negligible spin Hall conductivity of Zr as compared to the orbital Hall conductivity, it has been proposed that the sign change occurs due to a competition between the spin torque and orbital torque \cite{Zheng2020}, instead of the competition between the spin torque and interfacial torque. Detailed investigation of these systems by our method may reveal the origin of the sign change.

Another widely-studied system in spintronics is a Pt-based magnetic heterostructure. Due to a large spin Hall conductivity of Pt \cite{Guo2008}, the spin torque is assumed to be the most dominant mechanism of the torque in Pt-based systems \cite{Liu2012a}. In Co/Pt, however, theoretical analysis revealed that the interfacial spin-orbit coupling contributes significantly to the fieldlike torque \cite{Haney2013b, Mahfouzi2020}. On the other hand, the dampinglike torque is attributed to the spin torque mechanism \cite{Freimuth2014, Mahfouzi2020}, which is also supported by experiments \cite{Fan2014}. Hiroki \emph{et al.} compared Ni/Pt and Fe/Pt bilayers, finding that the current-induced torque strongly depends on the choice of the ferromagnet \cite{Hiroki2020}. According to their interpretation, while the bulk effect is dominant in Ni/Pt, a pronounced interface effect in Fe/Pt not only leads to fieldlike torque but also suppresses the spin current injection from Pt, which leads to a distinct ferromagnet dependence of the torque \cite{Hiroki2020}. A similar conclusion has also been drawn in an experiment by Zhu \emph{et al.}, where the interfacial spin-orbit coupling has been varied by choosing different samples and annealing conditions \cite{Zhu2019}. Further investigation of the exact mechanism in these systems by theory is required.


For the study of the interplay between the spin and orbital degrees of freedom transition metal oxides may present a very fruitful playground. In transition metal oxides, a strong entaglement of the spin, orbital, and charge degrees of freedom has been intensively studied in the past \cite{Tokura2000, Mochizuki2004, Andrzej2012}. For example, magnetic properties of transition metal oxides are heavily affected by the orbital physics not only via the effect of spin-orbit coupling but also because of the anisotropic exchange interactions caused by the shape of participating orbitals \cite{Tokura2000}. However, most studies on the transition metal oxides have focused on their ground state properties, such as various competing magnetic phases. We expect that the investigation of the spin-orbital entangled dynamics would provide crucial insights into understanding the complex physics of transition metal oxides.

\section{Conclusion}
\label{sec:conclusion}

Motivated by various proposed mechanisms of the current-induced torques, which are challenging to disentangle both theoretically and experimentally, we developed a theory of current-induced spin-orbital coupled dynamics in magnetic heterostructures, which tracks the transfer of the angular momentum between different degrees of freedom in solids: spin and orbital of the electron, lattice, and local magnetic moment. By adopting the continuity equations for the orbital and spin angular momentum [Eq. \eqref{eq:continuity_equations}], we derived equations for the angular momentum dynamics in the steady state reached when an external electric field is applied, which provide relations between interband and intraband contributions to the current influx, torques, and accumulation of the spin and orbital angular momentum [Eqs. \eqref{eq:equations_of_motion_inter} and \eqref{eq:equations_of_motion_intra}]. 

This formalism is particularly useful for the detailed study of the microscopic mechanisms of the current-induced torque and we used its first principles implementation to investigate the spin-orbit torque origins in  Fe/W(110) and Ni/W(110) bilayers. In Fe/W(110), we observe a strong correlation between the spin current influx and the exchange torque, which is a key characteristic of the spin torque mechanism. On the other hand, such correlation is not observed in Ni/W(110). Instead, we observe a pronounced correlation between the exchange torque and the spin-orbital torque, indicating the transfer of angular momentum from the orbital to the spin channel. Moreover, the spin current influx exhibits a sign opposite to that of the spin Hall effect in W. This leads us to a conclusion that the orbital torque is dominant in Ni/W(110).

We further proposed a classification scheme of the different mechanisms of current-induced torque based on the criteria of whether the scattering source is in the nonmagnet-spin-orbit coupling or the ferromagnet-spin-orbit coupling, and whether the torque response is of local or nonlocal nature (Fig. \ref{fig:classification}). This analysis also confirms that the spin torque and orbital torque are the most dominant mechanisms in Fe/W(110) and Ni/W(110), respectively. However, we also find that the other contributions, interfacial torque and anomalous torque, are  not negligible as well. Our formalism enables an analysis of the angular momentum transport and transfer dynamics in detail, which clearly goes beyond the ``spin current picture''. Since it treats the spin and orbital degrees of freedom on an equal footing, it is ideal for systematically studying the spin-orbital coupled dynamics in complex magnetic heterostructures.


\begin{acknowledgements}
D.G. thanks insightful comments from discussions with Gustav Bihlmayer, Filipe Souza Mendes Guimar\~{a}es, Mathias Kl\"aui, OukJae Lee, Kyung-Whan Kim, and Daegeun Jo. K.-J. L., J.-P. H., and Y. M. acknowledge discussions with Mark D. Stiles. J.-P. H., and Y. M. additionally acknowledge discussions with Jairo Sinova. We thank Mark D. Stiles and Matthew Pufall for carefully reading the manuscript and providing insightful comments. We gratefully acknowledge the J\"ulich Supercomputing Centre for providing computational resources under project jiff40. D.G. and H.-W. Lee were supported by SSTF (Grant   No.BA-1501-07). F.X. acknowledges support under the Cooperative Research Agreement between the University of Maryland and the National Institute of Standards and Technology Physical Measurement Laboratory, Award 70NANB14H209, through the University of Maryland. We also acknowledge  funding  under SPP 2137 ``Skyrmionics" (project  MO  1731/7-1) and TRR 173 $-$ 268565370 (project A11) of the Deutsche Forschungsgemeinschaft (DFG, German Research Foundation).
\end{acknowledgements}

\appendix

\section{Interband-Intraband Correspondence}
\label{app:interband-intraband_correspondence}

Here we provide a proof of Eq. \eqref{eq:interband-intraband_correspondence}. We assume that the operator $\mathcal{O}$ does not have position dependence, which leads to $\mathcal{O}(\mathbf{k}) = e^{-i\mathbf{k}\cdot\mathbf{r}} \mathcal{O} e^{i\mathbf{k}\cdot\mathbf{r}} = \mathcal{O}$. From Eqs. \eqref{eq:intraband_final}, the left hand side of Eq. \eqref{eq:interband-intraband_correspondence} is written as
\begin{eqnarray}
\label{eq:intra_partial_1}
\frac{1}{\tau}
\left\langle \mathcal{O} \right\rangle^\mathrm{intra}
=
-\frac{e\mathcal{E}_x}{\hbar}
\sum_{n\mathbf{k}}
\partial_{k_x} f_{n\mathbf{k}}
\bra{u_{n\mathbf{k}}}
\mathcal{O}
\ket{u_{n\mathbf{k}}},
\end{eqnarray}
where we used
\begin{eqnarray}
\frac{\partial f_{n\mathbf{k}}}{\partial k_x}
=
\frac{\partial f_{n\mathbf{k}}}{\partial E_{n\mathbf{k}}}
\frac{\partial E_{n\mathbf{k}}}{\partial k_x}
=
f'_{n\mathbf{k}} \hbar 
\bra{u_{n\mathbf{k}}} v_x (\mathbf{k}) \ket{u_{n\mathbf{k}}}.
\end{eqnarray}
Application of integration by parts to the first term in Eq. \eqref{eq:intra_partial_1} leads to
\begin{eqnarray}
\frac{1}{\tau}
\left\langle \mathcal{O} \right\rangle^\mathrm{intra}
=
\frac{e\mathcal{E}_x}{\hbar}
\sum_{n\mathbf{k}}
f_{n\mathbf{k}}
& &
\left[
\bra{\partial_{k_x} u_{n\mathbf{k}}}
\mathcal{O} 
\ket{u_{n\mathbf{k}}}
\right.
\nonumber
\\
+
& &
\left.
\bra{u_{n\mathbf{k}}}
\mathcal{O} 
\ket{\partial_{k_x} u_{n\mathbf{k}}}
\right].
\label{eq:intra_partial_2}
\end{eqnarray}
It can be rewritten as
\begin{eqnarray}
& &
\frac{1}{\tau}
\left\langle \mathcal{O} \right\rangle^\mathrm{intra}
=
\frac{e\mathcal{E}_x}{\hbar}
\sum_{n\neq m}
\sum_{\mathbf{k}}
(f_{n\mathbf{k}} - f_{m\mathbf{k}})
\nonumber
\\
& &
\ \ \ \ \ \ 
\times 
\mathrm{Re}
\left[
\braket{\partial_{k_x} u_{n\mathbf{k}} 
| u_{m\mathbf{k}}}
\bra{u_{m\mathbf{k}}} \mathcal{O} \ket{u_{n\mathbf{k}}}
\right].
\end{eqnarray}
By using identities
\begin{eqnarray}
\braket{\partial_{k_x} u_{n\mathbf{k}} 
| u_{m\mathbf{k}}}
=
\frac{
\hbar 
\bra{u_{n\mathbf{k}}}
v_x (\mathbf{k})
\ket{u_{m\mathbf{k}}}
}{E_{n\mathbf{k}} - E_{m\mathbf{k}}}
\end{eqnarray}
and
\begin{eqnarray}
\bra{u_{m\mathbf{k}}}
\mathcal{O} 
\ket{u_{n\mathbf{k}}}
=
\frac{i\hbar
\bra{u_{m\mathbf{k}}}
(1/i\hbar) [\mathcal{O}, \mathcal{H}(\mathbf{k})]
\ket{u_{n\mathbf{k}}}
}{E_{n\mathbf{k}} - E_{m\mathbf{k}}},
\nonumber
\\
\end{eqnarray}
for $n\neq m$, we have
\begin{widetext}
\begin{subequations}
\begin{eqnarray}
\frac{1}{\tau}
\left\langle \mathcal{O} \right\rangle^\mathrm{intra}
&=&
-e\hbar \mathcal{E}_x
\sum_{n\neq m}
\sum_{\mathbf{k}}
\left(
f_{n\mathbf{k}} - f_{m\mathbf{k}}
\right)
\mathrm{Im}
\left[
\frac{
\bra{u_{n\mathbf{k}}}
v_x (\mathbf{k})
\ket{u_{m\mathbf{k}}}
\bra{u_{m\mathbf{k}}}
(1/i\hbar)
\left[
\mathcal{O} ,
H (\mathbf{k})
\right]
\ket{u_{n\mathbf{k}}}
}{(E_{n\mathbf{k}} - E_{m\mathbf{k}})^2}
\right]
\\
&=&
e\hbar \mathcal{E}_x
\sum_{n\neq m}
\sum_{\mathbf{k}}
\left(
f_{n\mathbf{k}} - f_{m\mathbf{k}}
\right)
\mathrm{Im}
\left[
\frac{
\bra{u_{n\mathbf{k}}}
(1/i\hbar)
\left[
\mathcal{O},
\mathcal{H} (\mathbf{k})
\right]
\ket{u_{m\mathbf{k}}}
\bra{u_{m\mathbf{k}}}
v_x (\mathbf{k})
\ket{u_{n\mathbf{k}}}
}{(E_{n\mathbf{k}} - E_{m\mathbf{k}} + i\eta)^2}
\right]
\\
&=&
\left\langle 
\frac{d\mathcal{O}}{dt}
\right\rangle^\mathrm{inter}.
\end{eqnarray}
\end{subequations}
\end{widetext}
This proves Eq. \eqref{eq:interband-intraband_correspondence}. In case when $\mathcal{O}(\mathbf{k})$ is $\mathbf{k}$-dependent, the deviation is given by
\begin{eqnarray}
\frac{1}{\tau}
\left\langle \mathcal{O} \right\rangle^\mathrm{deviation}
=
-
\frac{e\mathcal{E}_x}{\hbar}
f_n
\bra{u_{n\mathbf{k}}}
\partial_{k_x}
\mathcal{O} (\mathbf{k})
\ket{u_{n\mathbf{k}}},
\end{eqnarray}
such that 
\begin{eqnarray}
\frac{1}{\tau}
\left\langle \mathcal{O} \right\rangle^\mathrm{intra}
+
\frac{1}{\tau}
\left\langle \mathcal{O} \right\rangle^\mathrm{deviation}
=
\left\langle 
\frac{d\mathcal{O}}{dt}
\right\rangle^\mathrm{inter}
\end{eqnarray}
holds even when $\mathcal{O}(\mathbf{k})$ is $\mathbf{k}$-dependent.

\section{Stationary Condition of the Intraband Contribution}
\label{app:stationary_condition_intra}

For a proof of Eq. \eqref{eq:intraband_stationary}, we apply Eq. \eqref{eq:intraband_final} to $d\mathcal{O}/dt$:
\begin{eqnarray}
& & 
\left\langle \frac{d\mathcal{O}}{dt} \right\rangle^\mathrm{intra}
=
\\
& & 
-\frac{e\mathcal{E}_x\tau}{i\hbar^2}
\sum_{n\mathbf{k}}
\left[
\partial_{k_x}
f_{n\mathbf{k}}
\bra{u_{n\mathbf{k}}}
[
\mathcal{O} (\mathbf{k}),
\mathcal{H} (\mathbf{k})
]
\ket{u_{n\mathbf{k}}}
\right].
\nonumber
\end{eqnarray}
Because
\begin{eqnarray}
\bra{u_{n\mathbf{k}}}
[\mathcal{O}(\mathbf{k}), H(\mathbf{k})]
\ket{u_{n\mathbf{k}}}
=
0
\end{eqnarray}
for any Hermitian operator $\mathcal{O}$, we have
\begin{eqnarray}
\left\langle
\frac{d\mathcal{O}}{dt}
\right\rangle^\mathrm{intra}
=0.
\end{eqnarray}

\section{Computational Method}
\label{app:computational_method}

First-principles calculation consists of three steps. The first step is calculation of the electronic structure from the density functional theory. In this step, we obtain Bloch states and their energy eigenvalues. The second step is to obtain maximally-localized Wannier functions (MLWFs) starting from the Bloch states obtained in the first step. Once the MLWFs are found, matrix elements of all relevant operators (Hamiltonian, position, spin, and orbital) are expressed within the basis set of the MLWFs. Thus, a tight-binding model is obtained. The last step is evaluation of the interband and intraband responses of the individual terms in the equations of motion [Eqs. \eqref{eq:equations_of_motion_inter} and \eqref{eq:equations_of_motion_intra}] by solving the tight-binding model obtained from the second step.

The electronic structure of ferromagnet/W(110) (ferromagnet=Fe or Ni), whose lattice structure is shown in Fig. \ref{fig:structure}, is calculated self-consistently in the film mode of the full-potential linearized augmented plane wave method \cite{Wimmer1981} from the code \texttt{FLEUR} \cite{FLEUR}. We use Perdew-Burke-Ernzerhof exchange-correlation functional within the generalized gradient approximation \cite{Perdew1996}. Muffin-tin radii of the ferromagnet and W atoms are set to $2.1 a_0$ and $2.5 a_0$, respectively, where $a_0$ is the Bohr radius. The plane wave cutoff is set to $3.8 a_0^{-1}$. The Monkhorst-Pack $\mathbf{k}$-mesh of $24\times 24$ are sampled from the first Brillouin zone. The spin-orbit coupling is treated self-consistently within the second variation scheme. The layer distances $d_\mathrm{FM-FM}$ and $d_\mathrm{W-FM}$ are optimized such that the total energy is minimized. The optimized values for Fe/W(110) are $d_\mathrm{W-Fe}=3.825 a_0$ and $d_\mathrm{Fe-Fe}=3.296 a_0$, and those for Ni/W(110) are $d_\mathrm{W-Ni}=3.607 a_0$ and $d_\mathrm{Ni-Ni}=3.301 a_0$.

In order to obtain MLWFs, we initially project the Bloch states onto $d_{xy}$, $d_{yz}$, $d_{zx}$, and $sp_3d_2$ trial orbitals for each atom, and minimize their spreads using the code \texttt{WANNIER90} \cite{Pizzi2020}. We obtain in total 180 MLWFs out of 360 Bloch states, that is, 18 MLWFs for each atom. For the disentanglement of the inner and outer spaces, we set the frozen window as 2 eV above the Fermi energy. The Hamiltonian, position, spin, and orbital operators, which are evaluated beforehand within the Bloch basis, are then transformed to the basis of MLWFs, and the tight-binding model is obtained.

Individual terms appearing in the equations of motion [Eqs. \eqref{eq:equations_of_motion_inter} and \eqref{eq:equations_of_motion_intra}] are evaluated using Eqs. \eqref{eq:interband_final} and \eqref{eq:intraband_final} for interband and intraband contributions, respectively. The integration is performed over interpolated $\mathbf{k}$-mesh of $240\times 240$. For the interband contributions, we set $\eta = 25\ \mathrm{meV}$ for convergence, which describes broadening of the spectral weight by disorders. In the intraband contribution, we set the momentum relaxation time as $\tau=\hbar/2\Gamma$ with $\Gamma= 25\ \mathrm{meV}$, which corresponds to $\tau = 1.26\times 10^{-14}\ \mathrm{s}$. We set the temperature in the Fermi-Dirac distribution function as room temperature $T=300\ \mathrm{K}$. For the application of an external electric field specifically onto ferromagnet or W layers, we replaced $v_x$ in Eq. \eqref{eq:interband_final} by
\begin{subequations}
\begin{eqnarray}
v_x^\mathrm{FM}
&=&
\sum_{z \in \mathrm{FM}}
P_z
v_x 
+
v_x P_z
,
\\
v_x^\mathrm{W}
&=&
\sum_{z \in \mathrm{W}}
P_z
v_x 
+
v_x P_z
,
\end{eqnarray}
\label{eq:vel_projection}
\end{subequations}
where $P_z$ is the projection onto the MLWFs located in a layer whose index is $z$. We confirm that the 18 MLWFs are well localized in each layer. Note that Eq. \eqref{eq:vel_projection} is defined such that
\begin{eqnarray}
v_x
=
v_x^\mathrm{FM} + v_x^\mathrm{W}.
\end{eqnarray}

\section{Symmetry Analysis}
\label{app:symmetry_analysis}

In Sec. \ref{subsec:symmetry_constraint}, we state that only $y$ and $x$ components are  nonzero in Eqs. \eqref{eq:equations_of_motion_inter} and \eqref{eq:equations_of_motion_intra}, respectively. Here, we prove this by symmetry argument. Two important symmetries present in ferromagnet/W(110), where the magnetization is pointing the $z$ direction, are $\mathcal{T} \mathcal{M}_x$ and $\mathcal{T} \mathcal{M}_y$ symmetries. Here, $\mathcal{T}$ is the time-reversal operator and $\mathcal{M}_{x(y)}$ is the mirror reflection operator along the direction of $x(y)$. Since all the terms appearing in the same equation should transform in the same way, we consider only the response of a torque operator
\begin{eqnarray}
T^\mathbf{J}
=
\frac{d\mathbf{J}}{dt}
\end{eqnarray}
for a general angular momentum operator $\mathbf{J}$, which can be either orbital and spin origin. To find symmetry constraints on the interband [Eq. \eqref{eq:interband_final}] and intraband [Eq. \eqref{eq:intraband_final}] responses, we first investigate how matrix elements of $v_x$ and $T^\mathbf{J}$ transform. We define $U_\mathcal{T}$ and $U_{\mathcal{M}_{x(y)}}$ as Hilbert space representations of $\mathcal{T}$ and $\mathcal{M}_{x(y)}$, respectively. Note that $\mathcal{T}$ transforms $v_x$ and $T^\mathbf{J}$ as
\begin{eqnarray}
U_\mathcal{T}^{-1} 
v_x
U_\mathcal{T}
=
-v_x,
\end{eqnarray}
and
\begin{eqnarray}
U_\mathcal{T}^{-1} 
T^\mathbf{J}
U_\mathcal{T}
=
+
T^\mathbf{J},
\end{eqnarray}
respectively. On the other hand, $\mathcal{M}_x$ and $\mathcal{M}_y$ symmetries transform $v_x$ and $T^\mathbf{J}$ as
\begin{subequations}
\begin{eqnarray}
U_{\mathcal{M}_x}^{-1}
v_x
U_{\mathcal{M}_x}
=
- v_x,
\\
U_{\mathcal{M}_y}^{-1}
v_x
U_{\mathcal{M}_y}
=
+ v_x,
\end{eqnarray}
\end{subequations}
and 
\begin{subequations}
\begin{eqnarray}
U_{\mathcal{M}_x}^{-1} 
T^{J_x}
U_{\mathcal{M}_x}
=
+
T^{J_x},
\\
U_{\mathcal{M}_x}^{-1}
T^{J_y}
U_{\mathcal{M}_x}
=
-
T^{J_y},
\\
U_{\mathcal{M}_x}^{-1}
T^{J_z}
U_{\mathcal{M}_x}
=
-
T^{J_z},
\\
U_{\mathcal{M}_y}^{-1}
T^{J_x}
U_{\mathcal{M}_y}
=
-
T^{J_x},
\\
U_{\mathcal{M}_y}^{-1}
T^{J_y}
U_{\mathcal{M}_y}
=
+
T^{J_y},
\\
U_{\mathcal{M}_y}^{-1}
T^{J_z}
U_{\mathcal{M}_y}
=
-
T^{J_z}.
\end{eqnarray}
\end{subequations}
As a result, $\mathcal{T}\mathcal{M}_x$ and $\mathcal{T}\mathcal{M}_y$ symmetries transform $v_x$ and $T^\mathbf{J}$ as
\begin{subequations}
\begin{eqnarray}
U_{\mathcal{T}\mathcal{M}_x}^{-1}
v_x
U_{\mathcal{T}\mathcal{M}_x}
=
+ v_x,
\\
U_{\mathcal{T}\mathcal{M}_y}^{-1}
v_x
U_{\mathcal{T}\mathcal{M}_y}
=
- v_x,
\end{eqnarray}
\label{eq:symmetry_velocity}
\end{subequations}
and
\begin{subequations}
\begin{eqnarray}
U_{\mathcal{T}\mathcal{M}_x}^{-1} 
T^{J_x}
U_{\mathcal{T}\mathcal{M}_x}
=
+
T^{J_x},
\\
U_{\mathcal{T}\mathcal{M}_x}^{-1}
T^{J_y}
U_{\mathcal{T}\mathcal{M}_x}
=
-
T^{J_y},
\\
U_{\mathcal{T}\mathcal{M}_x}^{-1}
T^{J_z}
U_{\mathcal{T}\mathcal{M}_x}
=
-
T^{J_z},
\\
U_{\mathcal{T}\mathcal{M}_y}^{-1}
T^{J_x}
U_{\mathcal{T}\mathcal{M}_y}
=
-
T^{J_x},
\\
U_{\mathcal{T}\mathcal{M}_y}^{-1}
T^{J_y}
U_{\mathcal{T}\mathcal{M}_y}
=
+
T^{J_y},
\\
U_{\mathcal{T}\mathcal{M}_y}^{-1}
T^{J_z}
U_{\mathcal{T}\mathcal{M}_y}
=
-
T^{J_z},
\end{eqnarray}
\label{eq:symmetry_torque}
\end{subequations}
where $U_{\mathcal{T}\mathcal{M}_{x(y)}} = U_\mathcal{T} U_{\mathcal{M}_{x(y)}}$. Note that $\mathcal{T}$ and $\mathcal{M}_{x(y)}$ commute each other.

We remark that $U_\mathcal{T}$ and $U_{\mathcal{M}_{x(y)}}$ are anti-unitary and unitary operators, respectively. Thus, $U_{\mathcal{T}\mathcal{M}_{x(y)}}$ is anti-unitary. For an arbitrary anti-unitary operator $\Theta$, a matrix element of an operator $\mathcal{O}$ satisfies
\begin{eqnarray}
\bra{\Theta \phi}
\mathcal{O}
\ket{\Theta \psi} 
=
\bra{\phi} 
\left(
\Theta^{-1} \mathcal{O} \Theta
\right)
\ket{\psi}
^*.
\\
\nonumber
\end{eqnarray}
Thus, combining this result with Eqs. \eqref{eq:symmetry_velocity} and \eqref{eq:symmetry_torque} provides constraints on the interband [Eq. \eqref{eq:interband_final}] and intraband [Eq. \eqref{eq:intraband_final}] contributions.

As an illustration, let us demonstrate that both interband and intraband contributions vanishes for $T^{J_z}$. We consider $\mathcal{T}\mathcal{M}_x$ symmetry at first. By this, matrix elements of $v_x$ and $T^{J_z}$ transform as
\begin{eqnarray}
\bra{U_{\mathcal{T}\mathcal{M}_x} \psi_{m\mathbf{k}}}
v_x
\ket{U_{\mathcal{T}\mathcal{M}_x} \psi_{n\mathbf{k}}}
=
+
\bra{\psi_{n\mathbf{k}'}}
v_x
\ket{\psi_{m\mathbf{k}'}},
\nonumber
\\
\end{eqnarray}
and
\begin{eqnarray}
\bra{U_{\mathcal{T}\mathcal{M}_x} \psi_{n\mathbf{k}}}
T^{J_z}
\ket{U_{\mathcal{T}\mathcal{M}_x} \psi_{m\mathbf{k}}}
=
-
\bra{\psi_{m\mathbf{k}'}}
T^{J_z}
\ket{\psi_{n\mathbf{k}'}},
\nonumber
\\
\end{eqnarray}
where $\mathbf{k}' = (+k_x, -k_y, -k_z)$. On the other hand, $\mathcal{T}\mathcal{M}_y$ symmetry gives 
\begin{eqnarray}
\bra{U_{\mathcal{T}\mathcal{M}_y} \psi_{m\mathbf{k}}}
v_x
\ket{U_{\mathcal{T}\mathcal{M}_y} \psi_{n\mathbf{k}}}
=
-
\bra{\psi_{n\mathbf{k}''}}
v_x
\ket{\psi_{m\mathbf{k}''}},
\nonumber
\\
\end{eqnarray}
and
\begin{eqnarray}
\bra{U_{\mathcal{T}\mathcal{M}_y} \psi_{n\mathbf{k}}}
T^{J_z}
\ket{U_{\mathcal{T}\mathcal{M}_y} \psi_{m\mathbf{k}}}
=
-
\bra{\psi_{m\mathbf{k}''}}
T^{J_z}
\ket{\psi_{n\mathbf{k}''}},
\nonumber
\\
\end{eqnarray}
where $\mathbf{k}'' = (-k_x, +k_y, -k_z)$.

\begin{figure*}[t!]
\includegraphics[angle=0, width=0.8\textwidth]{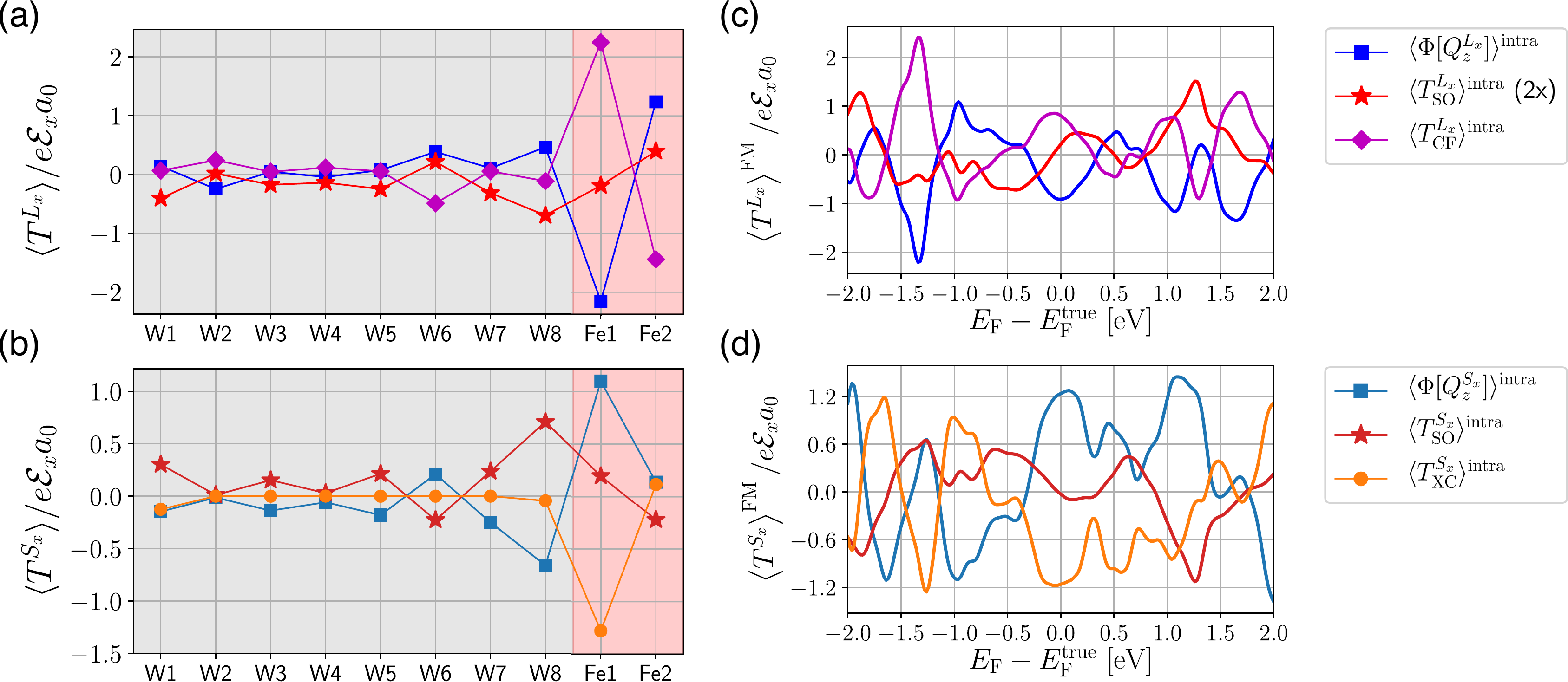}
\caption{
\label{fig:Fe_intra}
Electric response of current influxes $-$ $\Phi[Q_z^{L_y}]$ and $\Phi[Q_z^{S_y}]$ $-$ and various torques $-$ $T^{L_y}_\mathrm{SO}$, $T^{L_y}_\mathrm{CF}$, $T^{S_y}_\mathrm{SO}$, and $T^{S_y}_\mathrm{XC}$ $-$ arising from the intraband process in Fe/W(110). Spatial profiles for (a) the orbital and (b) the spin at true Fermi energy $E_\mathrm{F} = E_\mathrm{F}^\mathrm{true}$. Fermi energy dependences for (c) the orbital and (d) the spin, which are summed over the ferromagnet layers (Fe1 and Fe2).
}
\end{figure*}

\begin{figure*}[t!]
\includegraphics[angle=0, width=0.8\textwidth]{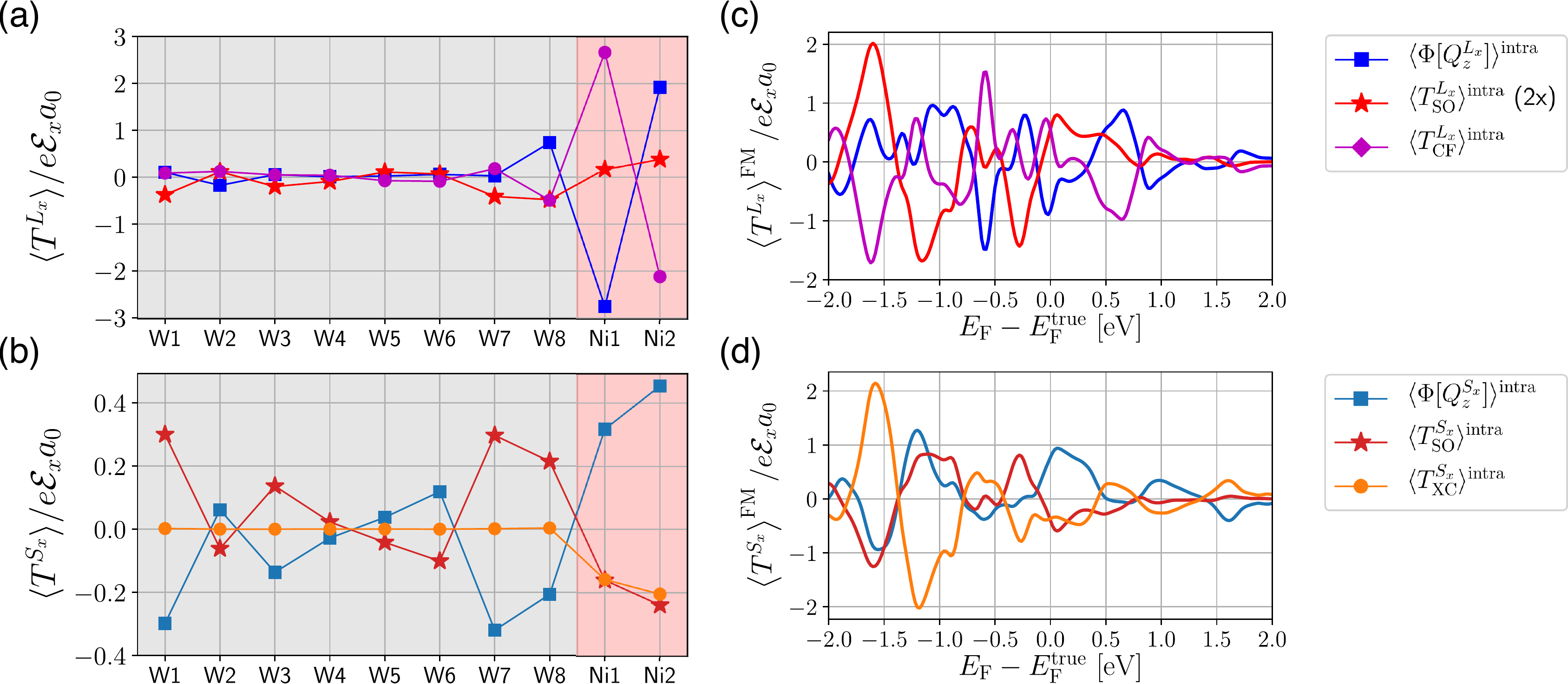}
\caption{
\label{fig:Ni_intra}
Electric response of current influxes $-$ $\Phi[Q_z^{L_y}]$ and $\Phi[Q_z^{S_y}]$ $-$ and various torques $-$ $T^{L_y}_\mathrm{SO}$, $T^{L_y}_\mathrm{CF}$, $T^{S_y}_\mathrm{SO}$, and $T^{S_y}_\mathrm{XC}$ $-$ arising from the intraband process in Ni/W(110). Spatial profiles for (a) the orbital and (b) the spin at true Fermi energy $E_\mathrm{F} = E_\mathrm{F}^\mathrm{true}$. Fermi energy dependences for (c) the orbital and (d) the spin, which are summed over the ferromagnet layers (Ni1 and Ni2).
}
\end{figure*}

A constraint for the interband contribution for $T^{J_z}$ [Eq. \eqref{eq:interband_final}] is given by $\mathcal{T}\mathcal{M}_y$ symmetry:
\begin{widetext}
\begin{subequations}
\begin{eqnarray}
\left\langle 
T^{J_z}
\right\rangle^\mathrm{inter}
\label{eq:symmetry_second_line}
&=&
e\hbar \mathcal{E}_x 
\sum_{n\neq m}
\sum_{\mathbf{k}}
(f_{n\mathbf{k}''} - f_{m\mathbf{k}''})
\mathrm{Im}
\left[
\frac{
\bra{U_{\mathcal{T}\mathcal{M}_y} \psi_{n\mathbf{k}}} 
T^{J_z} 
\ket{U_{\mathcal{T}\mathcal{M}_y} \psi_{m\mathbf{k}}}
\bra{U_{\mathcal{T}\mathcal{M}_y} \psi_{m\mathbf{k}}} 
v_x 
\ket{U_{\mathcal{T}\mathcal{M}_y} \psi_{n\mathbf{k}}}
}
{(E_{n\mathbf{k}''}- E_{m\mathbf{k}''} + i\eta)^2}
\right]
\\
&=&
e\hbar \mathcal{E}_x 
\sum_{n\neq m}
\sum_{\mathbf{k}}
(f_{n\mathbf{k}''} - f_{m\mathbf{k}''})
\mathrm{Im}
\left[
\frac{
\bra{\psi_{m\mathbf{k}''}} 
T^{J_z} 
\ket{\psi_{n\mathbf{k}''}}
\bra{\psi_{n\mathbf{k}''}} 
v_x 
\ket{\psi_{m\mathbf{k}''}}
}
{(E_{n\mathbf{k}''}- E_{m\mathbf{k}''} + i\eta)^2}
\right]
\\
&=&
e\hbar \mathcal{E}_x 
\sum_{n\neq m}
\sum_{\mathbf{k}}
(f_{m\mathbf{k}} - f_{n\mathbf{k}})
\mathrm{Im}
\left[
\frac{
\bra{\psi_{n\mathbf{k}}} 
T^{J_z} 
\ket{\psi_{m\mathbf{k}}}
\bra{\psi_{m\mathbf{k}}} 
v_x 
\ket{\psi_{n\mathbf{k}}}
}
{(E_{m\mathbf{k}}- E_{n\mathbf{k}} + i\eta)^2}
\right]
\\
&=&
-
\left\langle 
T^{J_x}
\right\rangle^\mathrm{inter}
\end{eqnarray}
\end{subequations}
\end{widetext}
in the limit $\eta \rightarrow 0+$. Thus, $\langle T^{J_z} \rangle^\mathrm{inter}$ is forbidden by $\mathcal{T}\mathcal{M}_y$ symmetry. In Eq. \eqref{eq:symmetry_second_line}, we used the fact that the linear response can also be written in terms of the transformed states. Note that we use the Bloch state representation instead of their periodic parts. For the intraband contribution, we have the following constraint by $\mathcal{T}\mathcal{M}_x$ symmetry:
\begin{widetext}
\begin{subequations}
\begin{eqnarray}
\left\langle 
T^{J_z}
\right\rangle^\mathrm{intra}
&=&
-\frac{e\mathcal{E}_x \tau}{\hbar}
\sum_{n\mathbf{k}}
\partial_{k_x'}
\left[
f_{n\mathbf{k}'}
\bra{U_{\mathcal{T}\mathcal{M}_x} \psi_{n\mathbf{k}}}
T^{J_z}
\ket{U_{\mathcal{T}\mathcal{M}_x} \psi_{n\mathbf{k}}}
\right]
\\
&=&
+\frac{e\mathcal{E}_x \tau}{\hbar}
\sum_{n\mathbf{k}}
\partial_{k_x'}
\left[
f_{n\mathbf{k}'}
\bra{ \psi_{n\mathbf{k}'}}
T^{J_z}
\ket{\psi_{n\mathbf{k}'}}
\right]
\\
&=&
-
\left\langle 
T^{J_z}
\right\rangle^\mathrm{intra}.
\end{eqnarray}
\end{subequations}
\end{widetext}
Therefore, both interband and intraband responses for $T^{J_z}$ vanishes by the symmetries. By the procedure for different components of the torque, we arrive at the conclusion that the presence of $\mathcal{T}\mathcal{M}_x$ and $\mathcal{T}\mathcal{M}_y$ symmetries allows only $\langle T^{J_y} \rangle^\mathrm{inter}$ and $\langle T^{J_x} \rangle^\mathrm{intra}$ to be nonzero.

\section{Intraband Response}
\label{app:intraband_response}

In Fig. \ref{fig:Fe_intra}, intraband contributions appearing in Eq. \eqref{eq:equations_of_motion_intra} are plotted for each layer of Fe/W(110). We confirm that the sum of the current influx and torques vanishes for the intraband contributions, respectively for the orbital and spin, which confirms Eq. \eqref{eq:equations_of_motion_intra}. For the orbital [Fig. \ref{fig:Fe_intra}(a)], we find that $\langle \Phi [Q_z^{L_x}] \rangle^\mathrm{intra}$ tends to cancel with $\langle T_\mathrm{CF}^{L_x} \rangle^\mathrm{intra}$ and $\langle T_\mathrm{SO}^\mathrm{L_x} \rangle^\mathrm{intra}$ is small. 
Meanwhile, for the spin, not only $\langle \Phi [Q_z^{S_x}] \rangle^\mathrm{intra}$ and $\langle T_\mathrm{XC}^{S_x} \rangle^\mathrm{intra}$ but also $\langle T_\mathrm{SO}^{S_x} \rangle^\mathrm{intra}$ are of comparable magnitudes, which is distinct from the interband response [Fig. \ref{fig:Fe_inter}(b)]. However, near the Fe layers, $\langle T_\mathrm{SO}^{S_x} \rangle^\mathrm{intra}$ is small, and $\langle T_\mathrm{XC}^{S_x} \rangle^\mathrm{intra}$ tends to cancel with $\langle \Phi [Q_z^{S_x}] \rangle^\mathrm{intra}$. We attribute this behavior to small spin-orbit correlation in Fe [Fig. \ref{fig:structure}(c)], and quenching of the orbital moment. 
Fermi energy dependence plots in Fig. \ref{fig:Fe_intra} also show the cancellation behaviors between the the orbital current influx and crystal field torque, and between the spin current influx and and the exchange torque. Although the spin-orbital torque is not particularly small in general, only near the true Fermi energy it is suppressed. Therefore, the fieldlike torque originates in the spin current injection (spin torque mechanism).

In Ni/W(110), for the orbital, $\langle \Phi [Q_z^{L_x}] \rangle^\mathrm{intra}$ and $\langle T_\mathrm{CF}^{L_x} \rangle^\mathrm{intra}$ cancel each other, with small magnitude of $\langle T_\mathrm{SO}^\mathrm{L_x} \rangle^\mathrm{intra}$ [Fig. \ref{fig:Ni_intra}(a)]. For the spin, on the other hand, as well as $\langle \Phi [Q_z^{S_x}] \rangle^\mathrm{intra}$, $\langle T_\mathrm{SO}^{S_x} \rangle^\mathrm{intra}$ contributes to $\langle T_\mathrm{XC}^{S_x} \rangle^\mathrm{intra}$, in comparable magnitudes [Fig. \ref{fig:Ni_intra}(b)]. This is due to pronounced spin-orbit correlation of Ni at the Fermi energy [Fig. \ref{fig:structure}(d)]. The Fermi energy dependence plots in Figs. \ref{fig:Ni_intra}(c) and \ref{fig:Ni_intra}(d) also show that the spin-orbital torque is nonnegligible at the Fermi energy. Therefore, in Ni/W(110), the fieldlike torque is a combined effect of the spin injection and the spin-orbit coupling. Such behavior has also been observed in Pt/Co \cite{Mahfouzi2020}.

\begin{figure}[t!]
\includegraphics[angle=0, width=0.45\textwidth]{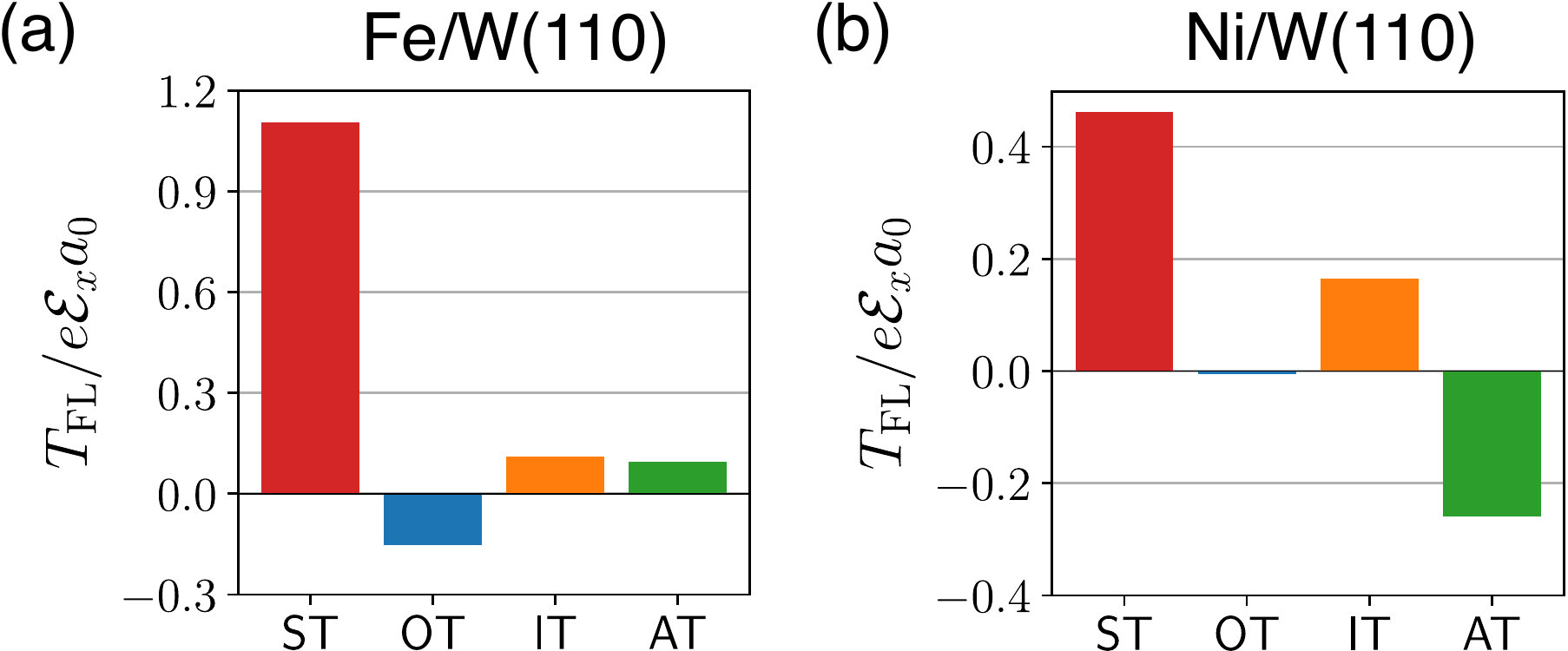}
\caption{
\label{fig:disentanglement_fieldlike torque}
Disentanglement of the fieldlike torque into the spin torque, orbital torque, interfacial torque, and anomalous torque in (a) Fe/W(11) and (b) Ni/W(110). In both systems, the spin torque is most dominant mechanism. We note that the anomalous torque is not negligible in Ni/W(110).
}
\end{figure}

To clarify microscopic mechanisms of different origins, we disentangle the fieldlike torque into the spin torque, orbital torque, interfacial torque, and interfacial torque, analogously to Fig. \ref{fig:disentanglement}. For Fe/W(110) [Fig. \ref{fig:disentanglement_fieldlike torque}(a)], we find that the spin torque is the most dominant contribution, as expected. On the other hand, for Ni/W(110) [Fig. \ref{fig:disentanglement_fieldlike torque}(b)], not only the spin torque but also the anomalous torque significantly contributes. This is due to pronounced spin-orbit correlation in Ni. Meanwhile, we also find that the interfacial torque is not negligible.

\section{Tight-binding Representation of the Continuity Equation}
\label{app:tight-binding_respresentation}

Here, we derive a tight-binding representation of the current influx and torque appearing in the continuity equation [Eq. \eqref{eq:continuity_equations}]. To do this, we first define $P_z$ as a projection operator onto a set of MLWFs located near a layer whose index is $z$. Then, for the spin operator $\mathbf{S}$, we define
\begin{eqnarray}
\mathbf{S}(z)
=
\frac{1}{2}
\left[
\mathbf{S} P_z
+
P_z\mathbf{S}
\right]
\end{eqnarray}
as the spin operator at $z$, such that 
\begin{eqnarray}
\mathbf{S}
=
\sum_{z} \mathbf{S}(z).
\end{eqnarray}
The Heisenberg equation of motion for $\mathbf{S}(z)$ is written as
\begin{subequations}
\begin{eqnarray}
\frac{d\mathbf{S}(z)}{dt}
&=&
\frac{1}{i\hbar}
\left[
\mathbf{S} (z), \mathcal{H}
\right]
\\
&=&
\frac{1}{2i\hbar}
\left[
\mathbf{S}P_z, P_z \mathbf{S}, \mathcal{H}
\right]
\\
&=&
\frac{1}{2i\hbar}
\left\{
[\mathbf{S},\mathcal{H}] P_z + \mathbf{S} [P_z, \mathcal{H}]
\right.
\nonumber
\\
& &
\ \ \ \ \ \ \ \ \ 
\left.
+
[P_z, \mathcal{H}] \mathbf{S} + P_z [\mathbf{S},\mathcal{H}]
\right\}
\\
&=&
T^\mathbf{S} (z)
+
\Phi [j^{\mathbf{S}}] (z).
\end{eqnarray}
\end{subequations}
We define local torque operator at $z$ by
\begin{subequations}
\begin{eqnarray}
T^\mathbf{S} (z)
&=&
\frac{1}{2i\hbar}
\left\{
P_z [\mathbf{S},\mathcal{H}]
+
[\mathbf{S},\mathcal{H}] P_z
\right\}
\\
&=&
\frac{1}{2}
\left[
T^\mathbf{S} P_z + P_z T^\mathbf{S}
\right],
\end{eqnarray}
\end{subequations}
where 
\begin{eqnarray}
T^\mathbf{S}
=
\frac{1}{i\hbar}
\left[ 
\mathbf{S}, \mathcal{H}
\right]
\end{eqnarray}
is the total torque operator, and we define 
\begin{eqnarray}
\Phi [j^{\mathbf{S}}] (z)
=
\frac{1}{2i\hbar}
\Big \{
[P_z,\mathcal{H}] \mathbf{S}
+
\mathbf{S} [P_z, \mathcal{H}]
\Big \}
\label{eq:spin_current_influx_app}
\end{eqnarray}
the spin current influx at $z$.

Although $\Phi [j^{\mathbf{S}}] (z)$ may not seem intuitive, it corresponds to an usual definition of the spin current influx. To demonstrate this point, we consider the case where $P = \ket{\mathbf{r}}\bra{\mathbf{r}}$ and $\mathcal{H}=-\hbar^2 \boldsymbol{\nabla}_\mathbf{r}^2/2m$, where $\ket{\mathbf{r}}$ is an eigenket for the position operator $\mathbf{r}$. Then $\Phi [j^{\mathbf{S}}]$ becomes
\begin{eqnarray}
\Phi [j^{\mathbf{S}}]
&=&
\frac{1}{2i\hbar}
\Big\{
\ket{\mathbf{r}} \bra{\mathbf{r}}
\mathcal{H} \mathbf{S}
-
\mathcal{H} 
\ket{\mathbf{r}} \bra{\mathbf{r}} 
\mathbf{S}
\nonumber
\\
& &
\ \ \ \ \ \ \ \ \ 
+
\mathbf{S}
\ket{\mathbf{r}} \bra{\mathbf{r}} 
\mathcal{H}
-
\mathbf{S}
\mathcal{H}
\ket{\mathbf{r}} \bra{\mathbf{r}} 
\Big \}.
\end{eqnarray}
Thus, a matrix element between states $\phi$ and $\psi$ is written as
\begin{subequations}
\begin{eqnarray}
\bra{\phi}
\Phi [j^{\mathbf{S}}]
\ket{\psi}
&=&
\frac{i\hbar}{2m}
\Big\{ 
\phi^* (\mathbf{r})
\mathbf{S}
\left[
\boldsymbol{\nabla}_\mathbf{r}^2
\psi (\mathbf{r})
\right]
\nonumber
\\
& &
\ \ \ \ \ \ \ \ 
-
\left[
\boldsymbol{\nabla}_\mathbf{r}^2
\phi^* (\mathbf{r})
\right]
\mathbf{S}
\psi (\mathbf{r})
\Big\}
\\
&=&
-\boldsymbol{\nabla}_\mathbf{r} \cdot 
\bra{\phi}
\mathbf{j}^{\mathbf{S}}
\ket{\psi}
,
\end{eqnarray}
\end{subequations}
where 
\begin{eqnarray}
\bra{\phi}
\mathbf{j}^{\mathbf{S}}
\ket{\psi}
&=&
-\frac{i\hbar}{2m}
\Big\{ 
\phi^* (\mathbf{r})
\mathbf{S}
\left[
\boldsymbol{\nabla}_\mathbf{r}
\psi (\mathbf{r})
\right]
\nonumber
+
\left[
\boldsymbol{\nabla}_\mathbf{r}
\phi^* (\mathbf{r})
\right]
\mathbf{S}
\psi (\mathbf{r})
\Big\}. 
\nonumber
\\
\label{eq:spin_influx_app}
\end{eqnarray}
From Eq. \eqref{eq:spin_influx_app}, we find that this is consistent with usual definition of the spin current $\mathbf{j}^\mathbf{S} = \mathbf{S} \otimes (\mathbf{p}/m)$. Therefore, Eq. \eqref{eq:spin_current_influx_app} can be understood as an operator of the spin current influx to the subspace defined by the projection $P_z$.

\section{Disentangling Different Contributions of the Current-Induced Torque}
\label{app:disentanglement}

To disentangle different contributions of the torque (Figs. \ref{fig:disentanglement} and \ref{fig:disentanglement_fieldlike torque}), we utilize a property that upon changing the sign of the spin-orbit coupling constant the orbital torque and anomalous torque flip their signs while the signs of the spin torque and interfacial torque remains invariant. That is, the total exchange torque is decomposed as the sum of the contribution driven by the spin-orbit coupling in the nonmagnet and the contribution driven by the spin-orbit coupling in the ferromagnet:
\begin{eqnarray}
\left \langle 
T_\mathrm{XC}^\mathbf{S}
\right\rangle^\mathrm{tot}
=
\left \langle 
T_\mathrm{XC}^\mathbf{S}
\right\rangle^\mathrm{NM-SOC}
+
\left \langle 
T_\mathrm{XC}^\mathbf{S}
\right\rangle^\mathrm{FM-SOC}.
\end{eqnarray}
In an auxiliary system where the sign of the spin-orbit coupling is flipped in the ferromagnet atoms, the exchange torque becomes
\begin{eqnarray}
\left \langle 
T_\mathrm{XC}^\mathbf{S}
\right\rangle^\mathrm{aux}
=
\left \langle 
T_\mathrm{XC}^\mathbf{S}
\right\rangle^\mathrm{NM-SOC}
-
\left \langle 
T_\mathrm{XC}^\mathbf{S}
\right\rangle^\mathrm{FM-SOC}.
\end{eqnarray}
Thus, the nonmagnet-spin-orbit coupling contribution is written as
\begin{eqnarray}
\left \langle 
T_\mathrm{XC}^\mathbf{S}
\right\rangle^\mathrm{NM-SOC}
=
\frac{1}{2}
\left[
\left \langle 
T_\mathrm{XC}^\mathbf{S}
\right\rangle^\mathrm{tot}
+
\left \langle 
T_\mathrm{XC}^\mathbf{S}
\right\rangle^\mathrm{aux}
\right],
\end{eqnarray}
and the ferromagnet-spin-orbit coupling contribution is written as
\begin{eqnarray}
\left \langle 
T_\mathrm{XC}^\mathbf{S}
\right\rangle^\mathrm{FM-SOC}
=
\frac{1}{2}
\left[
\left \langle 
T_\mathrm{XC}^\mathbf{S}
\right\rangle^\mathrm{tot}
-
\left \langle 
T_\mathrm{XC}^\mathbf{S}
\right\rangle^\mathrm{aux}
\right].
\end{eqnarray}
Then, by applying the electric field only in the nonmagnet or ferromagnet layers by Eq. \eqref{eq:vel_projection}, we can separately evaluate the spin torque, orbital torque, anomalous torque, and interfacial torque.

\bibliography{references}

\begin{thebibliography}{80}%
\makeatletter
\providecommand \@ifxundefined [1]{%
 \@ifx{#1\undefined}
}%
\providecommand \@ifnum [1]{%
 \ifnum #1\expandafter \@firstoftwo
 \else \expandafter \@secondoftwo
 \fi
}%
\providecommand \@ifx [1]{%
 \ifx #1\expandafter \@firstoftwo
 \else \expandafter \@secondoftwo
 \fi
}%
\providecommand \natexlab [1]{#1}%
\providecommand \enquote  [1]{``#1''}%
\providecommand \bibnamefont  [1]{#1}%
\providecommand \bibfnamefont [1]{#1}%
\providecommand \citenamefont [1]{#1}%
\providecommand \href@noop [0]{\@secondoftwo}%
\providecommand \href [0]{\begingroup \@sanitize@url \@href}%
\providecommand \@href[1]{\@@startlink{#1}\@@href}%
\providecommand \@@href[1]{\endgroup#1\@@endlink}%
\providecommand \@sanitize@url [0]{\catcode `\\12\catcode `\$12\catcode
  `\&12\catcode `\#12\catcode `\^12\catcode `\_12\catcode `\%12\relax}%
\providecommand \@@startlink[1]{}%
\providecommand \@@endlink[0]{}%
\providecommand \url  [0]{\begingroup\@sanitize@url \@url }%
\providecommand \@url [1]{\endgroup\@href {#1}{\urlprefix }}%
\providecommand \urlprefix  [0]{URL }%
\providecommand \Eprint [0]{\href }%
\providecommand \doibase [0]{http://dx.doi.org/}%
\providecommand \selectlanguage [0]{\@gobble}%
\providecommand \bibinfo  [0]{\@secondoftwo}%
\providecommand \bibfield  [0]{\@secondoftwo}%
\providecommand \translation [1]{[#1]}%
\providecommand \BibitemOpen [0]{}%
\providecommand \bibitemStop [0]{}%
\providecommand \bibitemNoStop [0]{.\EOS\space}%
\providecommand \EOS [0]{\spacefactor3000\relax}%
\providecommand \BibitemShut  [1]{\csname bibitem#1\endcsname}%
\let\auto@bib@innerbib\@empty
\bibitem [{\citenamefont {Hellman}\ \emph {et~al.}(2017)\citenamefont
  {Hellman}, \citenamefont {Hoffmann}, \citenamefont {Tserkovnyak},
  \citenamefont {Beach}, \citenamefont {Fullerton}, \citenamefont {Leighton},
  \citenamefont {MacDonald}, \citenamefont {Ralph}, \citenamefont {Arena},
  \citenamefont {D\"urr}, \citenamefont {Fischer}, \citenamefont {Grollier},
  \citenamefont {Heremans}, \citenamefont {Jungwirth}, \citenamefont {Kimel},
  \citenamefont {Koopmans}, \citenamefont {Krivorotov}, \citenamefont {May},
  \citenamefont {Petford-Long}, \citenamefont {Rondinelli}, \citenamefont
  {Samarth}, \citenamefont {Schuller}, \citenamefont {Slavin}, \citenamefont
  {Stiles}, \citenamefont {Tchernyshyov}, \citenamefont {Thiaville},\ and\
  \citenamefont {Zink}}]{Hellman2017}%
  \BibitemOpen
  \bibfield  {author} {\bibinfo {author} {\bibfnamefont {Frances}\ \bibnamefont
  {Hellman}}, \bibinfo {author} {\bibfnamefont {Axel}\ \bibnamefont
  {Hoffmann}}, \bibinfo {author} {\bibfnamefont {Yaroslav}\ \bibnamefont
  {Tserkovnyak}}, \bibinfo {author} {\bibfnamefont {Geoffrey S.~D.}\
  \bibnamefont {Beach}}, \bibinfo {author} {\bibfnamefont {Eric~E.}\
  \bibnamefont {Fullerton}}, \bibinfo {author} {\bibfnamefont {Chris}\
  \bibnamefont {Leighton}}, \bibinfo {author} {\bibfnamefont {Allan~H.}\
  \bibnamefont {MacDonald}}, \bibinfo {author} {\bibfnamefont {Daniel~C.}\
  \bibnamefont {Ralph}}, \bibinfo {author} {\bibfnamefont {Dario~A.}\
  \bibnamefont {Arena}}, \bibinfo {author} {\bibfnamefont {Hermann~A.}\
  \bibnamefont {D\"urr}}, \bibinfo {author} {\bibfnamefont {Peter}\
  \bibnamefont {Fischer}}, \bibinfo {author} {\bibfnamefont {Julie}\
  \bibnamefont {Grollier}}, \bibinfo {author} {\bibfnamefont {Joseph~P.}\
  \bibnamefont {Heremans}}, \bibinfo {author} {\bibfnamefont {Tomas}\
  \bibnamefont {Jungwirth}}, \bibinfo {author} {\bibfnamefont {Alexey~V.}\
  \bibnamefont {Kimel}}, \bibinfo {author} {\bibfnamefont {Bert}\ \bibnamefont
  {Koopmans}}, \bibinfo {author} {\bibfnamefont {Ilya~N.}\ \bibnamefont
  {Krivorotov}}, \bibinfo {author} {\bibfnamefont {Steven~J.}\ \bibnamefont
  {May}}, \bibinfo {author} {\bibfnamefont {Amanda~K.}\ \bibnamefont
  {Petford-Long}}, \bibinfo {author} {\bibfnamefont {James~M.}\ \bibnamefont
  {Rondinelli}}, \bibinfo {author} {\bibfnamefont {Nitin}\ \bibnamefont
  {Samarth}}, \bibinfo {author} {\bibfnamefont {Ivan~K.}\ \bibnamefont
  {Schuller}}, \bibinfo {author} {\bibfnamefont {Andrei~N.}\ \bibnamefont
  {Slavin}}, \bibinfo {author} {\bibfnamefont {Mark~D.}\ \bibnamefont
  {Stiles}}, \bibinfo {author} {\bibfnamefont {Oleg}\ \bibnamefont
  {Tchernyshyov}}, \bibinfo {author} {\bibfnamefont {Andr\'e}\ \bibnamefont
  {Thiaville}}, \ and\ \bibinfo {author} {\bibfnamefont {Barry~L.}\
  \bibnamefont {Zink}},\ }\bibfield  {title} {\enquote {\bibinfo {title}
  {Interface-induced phenomena in magnetism},}\ }\href {\doibase
  10.1103/RevModPhys.89.025006} {\bibfield  {journal} {\bibinfo  {journal}
  {Rev. Mod. Phys.}\ }\textbf {\bibinfo {volume} {89}},\ \bibinfo {pages}
  {025006} (\bibinfo {year} {2017})}\BibitemShut {NoStop}%
\bibitem [{\citenamefont {Gambardella}\ and\ \citenamefont
  {Miron}(2011)}]{Gambardella2011}%
  \BibitemOpen
  \bibfield  {author} {\bibinfo {author} {\bibfnamefont {Pietro}\ \bibnamefont
  {Gambardella}}\ and\ \bibinfo {author} {\bibfnamefont {Ioan~Mihai}\
  \bibnamefont {Miron}},\ }\bibfield  {title} {\enquote {\bibinfo {title}
  {Current-induced spin-orbit torques},}\ }\href {\doibase
  10.1098/rsta.2010.0336} {\bibfield  {journal} {\bibinfo  {journal}
  {Philosophical Transactions of the Royal Society A: Mathematical, Physical
  and Engineering Sciences}\ }\textbf {\bibinfo {volume} {369}},\ \bibinfo
  {pages} {3175--3197} (\bibinfo {year} {2011})}\BibitemShut {NoStop}%
\bibitem [{\citenamefont {Manchon}\ \emph {et~al.}(2019)\citenamefont
  {Manchon}, \citenamefont {\ifmmode~\check{Z}\else \v{Z}\fi{}elezn\'y},
  \citenamefont {Miron}, \citenamefont {Jungwirth}, \citenamefont {Sinova},
  \citenamefont {Thiaville}, \citenamefont {Garello},\ and\ \citenamefont
  {Gambardella}}]{Manchon2019}%
  \BibitemOpen
  \bibfield  {author} {\bibinfo {author} {\bibfnamefont {A.}~\bibnamefont
  {Manchon}}, \bibinfo {author} {\bibfnamefont {J.}~\bibnamefont
  {\ifmmode~\check{Z}\else \v{Z}\fi{}elezn\'y}}, \bibinfo {author}
  {\bibfnamefont {I.~M.}\ \bibnamefont {Miron}}, \bibinfo {author}
  {\bibfnamefont {T.}~\bibnamefont {Jungwirth}}, \bibinfo {author}
  {\bibfnamefont {J.}~\bibnamefont {Sinova}}, \bibinfo {author} {\bibfnamefont
  {A.}~\bibnamefont {Thiaville}}, \bibinfo {author} {\bibfnamefont
  {K.}~\bibnamefont {Garello}}, \ and\ \bibinfo {author} {\bibfnamefont
  {P.}~\bibnamefont {Gambardella}},\ }\bibfield  {title} {\enquote {\bibinfo
  {title} {Current-induced spin-orbit torques in ferromagnetic and
  antiferromagnetic systems},}\ }\href {\doibase 10.1103/RevModPhys.91.035004}
  {\bibfield  {journal} {\bibinfo  {journal} {Rev. Mod. Phys.}\ }\textbf
  {\bibinfo {volume} {91}},\ \bibinfo {pages} {035004} (\bibinfo {year}
  {2019})}\BibitemShut {NoStop}%
\bibitem [{\citenamefont {Miron}\ \emph
  {et~al.}(2011{\natexlab{a}})\citenamefont {Miron}, \citenamefont {Garello},
  \citenamefont {Gaudin}, \citenamefont {Zermatten}, \citenamefont {Costache},
  \citenamefont {Auffret}, \citenamefont {Bandiera}, \citenamefont {Rodmacq},
  \citenamefont {Schuhl},\ and\ \citenamefont {Gambardella}}]{Miron2011b}%
  \BibitemOpen
  \bibfield  {author} {\bibinfo {author} {\bibfnamefont {Ioan~Mihai}\
  \bibnamefont {Miron}}, \bibinfo {author} {\bibfnamefont {Kevin}\ \bibnamefont
  {Garello}}, \bibinfo {author} {\bibfnamefont {Gilles}\ \bibnamefont
  {Gaudin}}, \bibinfo {author} {\bibfnamefont {Pierre-Jean}\ \bibnamefont
  {Zermatten}}, \bibinfo {author} {\bibfnamefont {Marius~V.}\ \bibnamefont
  {Costache}}, \bibinfo {author} {\bibfnamefont {St{\'e}phane}\ \bibnamefont
  {Auffret}}, \bibinfo {author} {\bibfnamefont {S{\'e}bastien}\ \bibnamefont
  {Bandiera}}, \bibinfo {author} {\bibfnamefont {Bernard}\ \bibnamefont
  {Rodmacq}}, \bibinfo {author} {\bibfnamefont {Alain}\ \bibnamefont {Schuhl}},
  \ and\ \bibinfo {author} {\bibfnamefont {Pietro}\ \bibnamefont
  {Gambardella}},\ }\bibfield  {title} {\enquote {\bibinfo {title}
  {Perpendicular switching of a single ferromagnetic layer induced by in-plane
  current injection},}\ }\href {http://dx.doi.org/10.1038/nature10309}
  {\bibfield  {journal} {\bibinfo  {journal} {Nature (London)}\ }\textbf
  {\bibinfo {volume} {476}},\ \bibinfo {pages} {189} (\bibinfo {year}
  {2011}{\natexlab{a}})}\BibitemShut {NoStop}%
\bibitem [{\citenamefont {Liu}\ \emph {et~al.}(2012{\natexlab{a}})\citenamefont
  {Liu}, \citenamefont {Lee}, \citenamefont {Gudmundsen}, \citenamefont
  {Ralph},\ and\ \citenamefont {Buhrman}}]{Liu2012a}%
  \BibitemOpen
  \bibfield  {author} {\bibinfo {author} {\bibfnamefont {Luqiao}\ \bibnamefont
  {Liu}}, \bibinfo {author} {\bibfnamefont {O.~J.}\ \bibnamefont {Lee}},
  \bibinfo {author} {\bibfnamefont {T.~J.}\ \bibnamefont {Gudmundsen}},
  \bibinfo {author} {\bibfnamefont {D.~C.}\ \bibnamefont {Ralph}}, \ and\
  \bibinfo {author} {\bibfnamefont {R.~A.}\ \bibnamefont {Buhrman}},\
  }\bibfield  {title} {\enquote {\bibinfo {title} {{Current-Induced Switching
  of Perpendicularly Magnetized Magnetic Layers Using Spin Torque from the Spin
  Hall Effect}},}\ }\href {\doibase 10.1103/PhysRevLett.109.096602} {\bibfield
  {journal} {\bibinfo  {journal} {Phys. Rev. Lett.}\ }\textbf {\bibinfo
  {volume} {109}},\ \bibinfo {pages} {096602} (\bibinfo {year}
  {2012}{\natexlab{a}})}\BibitemShut {NoStop}%
\bibitem [{\citenamefont {Liu}\ \emph {et~al.}(2012{\natexlab{b}})\citenamefont
  {Liu}, \citenamefont {Pai}, \citenamefont {Li}, \citenamefont {Tseng},
  \citenamefont {Ralph},\ and\ \citenamefont {Buhrman}}]{Liu2012b}%
  \BibitemOpen
  \bibfield  {author} {\bibinfo {author} {\bibfnamefont {Luqiao}\ \bibnamefont
  {Liu}}, \bibinfo {author} {\bibfnamefont {Chi-Feng}\ \bibnamefont {Pai}},
  \bibinfo {author} {\bibfnamefont {Y.}~\bibnamefont {Li}}, \bibinfo {author}
  {\bibfnamefont {H.~W.}\ \bibnamefont {Tseng}}, \bibinfo {author}
  {\bibfnamefont {D.~C.}\ \bibnamefont {Ralph}}, \ and\ \bibinfo {author}
  {\bibfnamefont {R.~A.}\ \bibnamefont {Buhrman}},\ }\bibfield  {title}
  {\enquote {\bibinfo {title} {{Spin-Torque Switching with the Giant Spin Hall
  Effect of Tantalum}},}\ }\href {\doibase 10.1126/science.1218197} {\bibfield
  {journal} {\bibinfo  {journal} {Science}\ }\textbf {\bibinfo {volume}
  {336}},\ \bibinfo {pages} {555--558} (\bibinfo {year}
  {2012}{\natexlab{b}})}\BibitemShut {NoStop}%
\bibitem [{\citenamefont {Yu}\ \emph {et~al.}(2014)\citenamefont {Yu},
  \citenamefont {Upadhyaya}, \citenamefont {Fan}, \citenamefont {Alzate},
  \citenamefont {Jiang}, \citenamefont {Wong}, \citenamefont {Takei},
  \citenamefont {Bender}, \citenamefont {Chang}, \citenamefont {Jiang},
  \citenamefont {Lang}, \citenamefont {Tang}, \citenamefont {Wang},
  \citenamefont {Tserkovnyak}, \citenamefont {Amiri},\ and\ \citenamefont
  {Wang}}]{Yu2014}%
  \BibitemOpen
  \bibfield  {author} {\bibinfo {author} {\bibfnamefont {Guoqiang}\
  \bibnamefont {Yu}}, \bibinfo {author} {\bibfnamefont {Pramey}\ \bibnamefont
  {Upadhyaya}}, \bibinfo {author} {\bibfnamefont {Yabin}\ \bibnamefont {Fan}},
  \bibinfo {author} {\bibfnamefont {Juan~G.}\ \bibnamefont {Alzate}}, \bibinfo
  {author} {\bibfnamefont {Wanjun}\ \bibnamefont {Jiang}}, \bibinfo {author}
  {\bibfnamefont {Kin~L.}\ \bibnamefont {Wong}}, \bibinfo {author}
  {\bibfnamefont {So}~\bibnamefont {Takei}}, \bibinfo {author} {\bibfnamefont
  {Scott~A.}\ \bibnamefont {Bender}}, \bibinfo {author} {\bibfnamefont {Li-Te}\
  \bibnamefont {Chang}}, \bibinfo {author} {\bibfnamefont {Ying}\ \bibnamefont
  {Jiang}}, \bibinfo {author} {\bibfnamefont {Murong}\ \bibnamefont {Lang}},
  \bibinfo {author} {\bibfnamefont {Jianshi}\ \bibnamefont {Tang}}, \bibinfo
  {author} {\bibfnamefont {Yong}\ \bibnamefont {Wang}}, \bibinfo {author}
  {\bibfnamefont {Yaroslav}\ \bibnamefont {Tserkovnyak}}, \bibinfo {author}
  {\bibfnamefont {Pedram~Khalili}\ \bibnamefont {Amiri}}, \ and\ \bibinfo
  {author} {\bibfnamefont {Kang~L.}\ \bibnamefont {Wang}},\ }\bibfield  {title}
  {\enquote {\bibinfo {title} {Switching of perpendicular magnetization by
  spin-orbit torques in the absence of external magnetic fields},}\ }\href
  {https://doi.org/10.1038/nnano.2014.94} {\bibfield  {journal} {\bibinfo
  {journal} {Nat. Nanotechnol.}\ }\textbf {\bibinfo {volume} {9}},\ \bibinfo
  {pages} {548} (\bibinfo {year} {2014})}\BibitemShut {NoStop}%
\bibitem [{\citenamefont {Baumgartner}\ \emph {et~al.}(2017)\citenamefont
  {Baumgartner}, \citenamefont {Garello}, \citenamefont {Mendil}, \citenamefont
  {Avci}, \citenamefont {Grimaldi}, \citenamefont {Murer}, \citenamefont
  {Feng}, \citenamefont {Gabureac}, \citenamefont {Stamm}, \citenamefont
  {Acremann}, \citenamefont {Finizio}, \citenamefont {Wintz}, \citenamefont
  {Raabe},\ and\ \citenamefont {Gambardella}}]{Baumgartner2017}%
  \BibitemOpen
  \bibfield  {author} {\bibinfo {author} {\bibfnamefont {Manuel}\ \bibnamefont
  {Baumgartner}}, \bibinfo {author} {\bibfnamefont {Kevin}\ \bibnamefont
  {Garello}}, \bibinfo {author} {\bibfnamefont {Johannes}\ \bibnamefont
  {Mendil}}, \bibinfo {author} {\bibfnamefont {Can~Onur}\ \bibnamefont {Avci}},
  \bibinfo {author} {\bibfnamefont {Eva}\ \bibnamefont {Grimaldi}}, \bibinfo
  {author} {\bibfnamefont {Christoph}\ \bibnamefont {Murer}}, \bibinfo {author}
  {\bibfnamefont {Junxiao}\ \bibnamefont {Feng}}, \bibinfo {author}
  {\bibfnamefont {Mihai}\ \bibnamefont {Gabureac}}, \bibinfo {author}
  {\bibfnamefont {Christian}\ \bibnamefont {Stamm}}, \bibinfo {author}
  {\bibfnamefont {Yves}\ \bibnamefont {Acremann}}, \bibinfo {author}
  {\bibfnamefont {Simone}\ \bibnamefont {Finizio}}, \bibinfo {author}
  {\bibfnamefont {Sebastian}\ \bibnamefont {Wintz}}, \bibinfo {author}
  {\bibfnamefont {J{\"o}rg}\ \bibnamefont {Raabe}}, \ and\ \bibinfo {author}
  {\bibfnamefont {Pietro}\ \bibnamefont {Gambardella}},\ }\bibfield  {title}
  {\enquote {\bibinfo {title} {Spatially and time-resolved magnetization
  dynamics driven by spin-orbit torques},}\ }\href {\doibase
  10.1038/nnano.2017.151} {\bibfield  {journal} {\bibinfo  {journal} {Nature
  Nanotechnology}\ }\textbf {\bibinfo {volume} {12}},\ \bibinfo {pages}
  {980--986} (\bibinfo {year} {2017})}\BibitemShut {NoStop}%
\bibitem [{\citenamefont {Miron}\ \emph
  {et~al.}(2011{\natexlab{b}})\citenamefont {Miron}, \citenamefont {Moore},
  \citenamefont {Szambolics}, \citenamefont {Buda-Prejbeanu}, \citenamefont
  {Auffret}, \citenamefont {Rodmacq}, \citenamefont {Pizzini}, \citenamefont
  {Vogel}, \citenamefont {Bonfim}, \citenamefont {Schuhl},\ and\ \citenamefont
  {Gaudin}}]{Miron2011a}%
  \BibitemOpen
  \bibfield  {author} {\bibinfo {author} {\bibfnamefont {Ioan~Mihai}\
  \bibnamefont {Miron}}, \bibinfo {author} {\bibfnamefont {Thomas}\
  \bibnamefont {Moore}}, \bibinfo {author} {\bibfnamefont {Helga}\ \bibnamefont
  {Szambolics}}, \bibinfo {author} {\bibfnamefont {Liliana~Daniela}\
  \bibnamefont {Buda-Prejbeanu}}, \bibinfo {author} {\bibfnamefont
  {St{\'e}phane}\ \bibnamefont {Auffret}}, \bibinfo {author} {\bibfnamefont
  {Bernard}\ \bibnamefont {Rodmacq}}, \bibinfo {author} {\bibfnamefont
  {Stefania}\ \bibnamefont {Pizzini}}, \bibinfo {author} {\bibfnamefont {Jan}\
  \bibnamefont {Vogel}}, \bibinfo {author} {\bibfnamefont {Marlio}\
  \bibnamefont {Bonfim}}, \bibinfo {author} {\bibfnamefont {Alain}\
  \bibnamefont {Schuhl}}, \ and\ \bibinfo {author} {\bibfnamefont {Gilles}\
  \bibnamefont {Gaudin}},\ }\bibfield  {title} {\enquote {\bibinfo {title}
  {{Fast current-induced domain-wall motion controlled by the Rashba
  effect}},}\ }\href {https://doi.org/10.1038/nmat3020} {\bibfield  {journal}
  {\bibinfo  {journal} {Mat. Mater.}\ }\textbf {\bibinfo {volume} {10}},\
  \bibinfo {pages} {419} (\bibinfo {year} {2011}{\natexlab{b}})}\BibitemShut
  {NoStop}%
\bibitem [{\citenamefont {Ryu}\ \emph {et~al.}(2013)\citenamefont {Ryu},
  \citenamefont {Thomas}, \citenamefont {Yang},\ and\ \citenamefont
  {Parkin}}]{Ryu2013}%
  \BibitemOpen
  \bibfield  {author} {\bibinfo {author} {\bibfnamefont {Kwang-Su}\
  \bibnamefont {Ryu}}, \bibinfo {author} {\bibfnamefont {Luc}\ \bibnamefont
  {Thomas}}, \bibinfo {author} {\bibfnamefont {See-Hun}\ \bibnamefont {Yang}},
  \ and\ \bibinfo {author} {\bibfnamefont {Stuart}\ \bibnamefont {Parkin}},\
  }\bibfield  {title} {\enquote {\bibinfo {title} {Chiral spin torque at
  magnetic domain walls},}\ }\href {https://doi.org/10.1038/nnano.2013.102}
  {\bibfield  {journal} {\bibinfo  {journal} {Nat. Nanotechnol.}\ }\textbf
  {\bibinfo {volume} {8}},\ \bibinfo {pages} {527} (\bibinfo {year}
  {2013})}\BibitemShut {NoStop}%
\bibitem [{\citenamefont {Emori}\ \emph {et~al.}(2013)\citenamefont {Emori},
  \citenamefont {Bauer}, \citenamefont {Ahn}, \citenamefont {Martinez},\ and\
  \citenamefont {Beach}}]{Emori2013}%
  \BibitemOpen
  \bibfield  {author} {\bibinfo {author} {\bibfnamefont {Satoru}\ \bibnamefont
  {Emori}}, \bibinfo {author} {\bibfnamefont {Uwe}\ \bibnamefont {Bauer}},
  \bibinfo {author} {\bibfnamefont {Sung-Min}\ \bibnamefont {Ahn}}, \bibinfo
  {author} {\bibfnamefont {Eduardo}\ \bibnamefont {Martinez}}, \ and\ \bibinfo
  {author} {\bibfnamefont {Geoffrey S.~D.}\ \bibnamefont {Beach}},\ }\bibfield
  {title} {\enquote {\bibinfo {title} {Current-driven dynamics of chiral
  ferromagnetic domain walls},}\ }\href {https://doi.org/10.1038/nmat3675}
  {\bibfield  {journal} {\bibinfo  {journal} {Nat. Mater.}\ }\textbf {\bibinfo
  {volume} {12}},\ \bibinfo {pages} {611} (\bibinfo {year} {2013})}\BibitemShut
  {NoStop}%
\bibitem [{\citenamefont {Martinez}\ \emph {et~al.}(2013)\citenamefont
  {Martinez}, \citenamefont {Emori},\ and\ \citenamefont
  {Beach}}]{Martinez2013}%
  \BibitemOpen
  \bibfield  {author} {\bibinfo {author} {\bibfnamefont {Eduardo}\ \bibnamefont
  {Martinez}}, \bibinfo {author} {\bibfnamefont {Satoru}\ \bibnamefont
  {Emori}}, \ and\ \bibinfo {author} {\bibfnamefont {Geoffrey S.~D.}\
  \bibnamefont {Beach}},\ }\bibfield  {title} {\enquote {\bibinfo {title}
  {{Current-driven domain wall motion along high perpendicular anisotropy
  multilayers: The role of the Rashba field, the spin Hall effect, and the
  Dzyaloshinskii-Moriya interaction}},}\ }\href
  {https://doi.org/10.1063/1.4818723} {\bibfield  {journal} {\bibinfo
  {journal} {Appl. Phys. Lett.}\ }\textbf {\bibinfo {volume} {103}},\ \bibinfo
  {pages} {072406} (\bibinfo {year} {2013})}\BibitemShut {NoStop}%
\bibitem [{\citenamefont {Stiles}\ and\ \citenamefont
  {Zangwill}(2002)}]{Stiles2002}%
  \BibitemOpen
  \bibfield  {author} {\bibinfo {author} {\bibfnamefont {M.~D.}\ \bibnamefont
  {Stiles}}\ and\ \bibinfo {author} {\bibfnamefont {A.}~\bibnamefont
  {Zangwill}},\ }\bibfield  {title} {\enquote {\bibinfo {title} {Anatomy of
  spin-transfer torque},}\ }\href {\doibase 10.1103/PhysRevB.66.014407}
  {\bibfield  {journal} {\bibinfo  {journal} {Phys. Rev. B}\ }\textbf {\bibinfo
  {volume} {66}},\ \bibinfo {pages} {014407} (\bibinfo {year}
  {2002})}\BibitemShut {NoStop}%
\bibitem [{\citenamefont {Ralph}\ and\ \citenamefont
  {Stiles}(2008)}]{Ralph2008}%
  \BibitemOpen
  \bibfield  {author} {\bibinfo {author} {\bibfnamefont {D.C.}\ \bibnamefont
  {Ralph}}\ and\ \bibinfo {author} {\bibfnamefont {M.D.}\ \bibnamefont
  {Stiles}},\ }\bibfield  {title} {\enquote {\bibinfo {title} {Spin transfer
  torques},}\ }\href {\doibase https://doi.org/10.1016/j.jmmm.2007.12.019}
  {\bibfield  {journal} {\bibinfo  {journal} {Journal of Magnetism and Magnetic
  Materials}\ }\textbf {\bibinfo {volume} {320}},\ \bibinfo {pages} {1190 --
  1216} (\bibinfo {year} {2008})}\BibitemShut {NoStop}%
\bibitem [{\citenamefont {Manchon}\ and\ \citenamefont
  {Zhang}(2009)}]{Manchon2009}%
  \BibitemOpen
  \bibfield  {author} {\bibinfo {author} {\bibfnamefont {A.}~\bibnamefont
  {Manchon}}\ and\ \bibinfo {author} {\bibfnamefont {S.}~\bibnamefont
  {Zhang}},\ }\bibfield  {title} {\enquote {\bibinfo {title} {Theory of spin
  torque due to spin-orbit coupling},}\ }\href {\doibase
  10.1103/PhysRevB.79.094422} {\bibfield  {journal} {\bibinfo  {journal} {Phys.
  Rev. B}\ }\textbf {\bibinfo {volume} {79}},\ \bibinfo {pages} {094422}
  (\bibinfo {year} {2009})}\BibitemShut {NoStop}%
\bibitem [{\citenamefont {Garate}\ and\ \citenamefont
  {MacDonald}(2009)}]{Garate2009}%
  \BibitemOpen
  \bibfield  {author} {\bibinfo {author} {\bibfnamefont {Ion}\ \bibnamefont
  {Garate}}\ and\ \bibinfo {author} {\bibfnamefont {A.~H.}\ \bibnamefont
  {MacDonald}},\ }\bibfield  {title} {\enquote {\bibinfo {title} {Influence of
  a transport current on magnetic anisotropy in gyrotropic ferromagnets},}\
  }\href {\doibase 10.1103/PhysRevB.80.134403} {\bibfield  {journal} {\bibinfo
  {journal} {Phys. Rev. B}\ }\textbf {\bibinfo {volume} {80}},\ \bibinfo
  {pages} {134403} (\bibinfo {year} {2009})}\BibitemShut {NoStop}%
\bibitem [{\citenamefont {Kim}\ \emph {et~al.}(2012)\citenamefont {Kim},
  \citenamefont {Seo}, \citenamefont {Ryu}, \citenamefont {Lee},\ and\
  \citenamefont {Lee}}]{Kim2012}%
  \BibitemOpen
  \bibfield  {author} {\bibinfo {author} {\bibfnamefont {Kyoung-Whan}\
  \bibnamefont {Kim}}, \bibinfo {author} {\bibfnamefont {Soo-Man}\ \bibnamefont
  {Seo}}, \bibinfo {author} {\bibfnamefont {Jisu}\ \bibnamefont {Ryu}},
  \bibinfo {author} {\bibfnamefont {Kyung-Jin}\ \bibnamefont {Lee}}, \ and\
  \bibinfo {author} {\bibfnamefont {Hyun-Woo}\ \bibnamefont {Lee}},\ }\bibfield
   {title} {\enquote {\bibinfo {title} {{Magnetization dynamics induced by
  in-plane currents in ultrathin magnetic nanostructures with Rashba spin-orbit
  coupling}},}\ }\href {\doibase 10.1103/PhysRevB.85.180404} {\bibfield
  {journal} {\bibinfo  {journal} {Phys. Rev. B}\ }\textbf {\bibinfo {volume}
  {85}},\ \bibinfo {pages} {180404(R)} (\bibinfo {year} {2012})}\BibitemShut
  {NoStop}%
\bibitem [{\citenamefont {Haney}\ \emph
  {et~al.}(2013{\natexlab{a}})\citenamefont {Haney}, \citenamefont {Lee},
  \citenamefont {Lee}, \citenamefont {Manchon},\ and\ \citenamefont
  {Stiles}}]{Haney2013a}%
  \BibitemOpen
  \bibfield  {author} {\bibinfo {author} {\bibfnamefont {Paul~M.}\ \bibnamefont
  {Haney}}, \bibinfo {author} {\bibfnamefont {Hyun-Woo}\ \bibnamefont {Lee}},
  \bibinfo {author} {\bibfnamefont {Kyung-Jin}\ \bibnamefont {Lee}}, \bibinfo
  {author} {\bibfnamefont {Aur\'elien}\ \bibnamefont {Manchon}}, \ and\
  \bibinfo {author} {\bibfnamefont {M.~D.}\ \bibnamefont {Stiles}},\ }\bibfield
   {title} {\enquote {\bibinfo {title} {Current induced torques and interfacial
  spin-orbit coupling: Semiclassical modeling},}\ }\href {\doibase
  10.1103/PhysRevB.87.174411} {\bibfield  {journal} {\bibinfo  {journal} {Phys.
  Rev. B}\ }\textbf {\bibinfo {volume} {87}},\ \bibinfo {pages} {174411}
  (\bibinfo {year} {2013}{\natexlab{a}})}\BibitemShut {NoStop}%
\bibitem [{\citenamefont {Haney}\ \emph
  {et~al.}(2013{\natexlab{b}})\citenamefont {Haney}, \citenamefont {Lee},
  \citenamefont {Lee}, \citenamefont {Manchon},\ and\ \citenamefont
  {Stiles}}]{Haney2013b}%
  \BibitemOpen
  \bibfield  {author} {\bibinfo {author} {\bibfnamefont {Paul~M.}\ \bibnamefont
  {Haney}}, \bibinfo {author} {\bibfnamefont {Hyun-Woo}\ \bibnamefont {Lee}},
  \bibinfo {author} {\bibfnamefont {Kyung-Jin}\ \bibnamefont {Lee}}, \bibinfo
  {author} {\bibfnamefont {Aur\'elien}\ \bibnamefont {Manchon}}, \ and\
  \bibinfo {author} {\bibfnamefont {M.~D.}\ \bibnamefont {Stiles}},\ }\bibfield
   {title} {\enquote {\bibinfo {title} {Current-induced torques and interfacial
  spin-orbit coupling},}\ }\href {\doibase 10.1103/PhysRevB.88.214417}
  {\bibfield  {journal} {\bibinfo  {journal} {Phys. Rev. B}\ }\textbf {\bibinfo
  {volume} {88}},\ \bibinfo {pages} {214417} (\bibinfo {year}
  {2013}{\natexlab{b}})}\BibitemShut {NoStop}%
\bibitem [{\citenamefont {Amin}\ and\ \citenamefont
  {Stiles}(2016{\natexlab{a}})}]{Amin2016a}%
  \BibitemOpen
  \bibfield  {author} {\bibinfo {author} {\bibfnamefont {V.~P.}\ \bibnamefont
  {Amin}}\ and\ \bibinfo {author} {\bibfnamefont {M.~D.}\ \bibnamefont
  {Stiles}},\ }\bibfield  {title} {\enquote {\bibinfo {title} {Spin transport
  at interfaces with spin-orbit coupling: Formalism},}\ }\href {\doibase
  10.1103/PhysRevB.94.104419} {\bibfield  {journal} {\bibinfo  {journal} {Phys.
  Rev. B}\ }\textbf {\bibinfo {volume} {94}},\ \bibinfo {pages} {104419}
  (\bibinfo {year} {2016}{\natexlab{a}})}\BibitemShut {NoStop}%
\bibitem [{\citenamefont {Amin}\ and\ \citenamefont
  {Stiles}(2016{\natexlab{b}})}]{Amin2016b}%
  \BibitemOpen
  \bibfield  {author} {\bibinfo {author} {\bibfnamefont {V.~P.}\ \bibnamefont
  {Amin}}\ and\ \bibinfo {author} {\bibfnamefont {M.~D.}\ \bibnamefont
  {Stiles}},\ }\bibfield  {title} {\enquote {\bibinfo {title} {Spin transport
  at interfaces with spin-orbit coupling: Phenomenology},}\ }\href {\doibase
  10.1103/PhysRevB.94.104420} {\bibfield  {journal} {\bibinfo  {journal} {Phys.
  Rev. B}\ }\textbf {\bibinfo {volume} {94}},\ \bibinfo {pages} {104420}
  (\bibinfo {year} {2016}{\natexlab{b}})}\BibitemShut {NoStop}%
\bibitem [{\citenamefont {Kim}\ \emph {et~al.}(2017)\citenamefont {Kim},
  \citenamefont {Lee}, \citenamefont {Sinova}, \citenamefont {Lee},\ and\
  \citenamefont {Stiles}}]{Kim2017}%
  \BibitemOpen
  \bibfield  {author} {\bibinfo {author} {\bibfnamefont {Kyoung-Whan}\
  \bibnamefont {Kim}}, \bibinfo {author} {\bibfnamefont {Kyung-Jin}\
  \bibnamefont {Lee}}, \bibinfo {author} {\bibfnamefont {Jairo}\ \bibnamefont
  {Sinova}}, \bibinfo {author} {\bibfnamefont {Hyun-Woo}\ \bibnamefont {Lee}},
  \ and\ \bibinfo {author} {\bibfnamefont {M.~D.}\ \bibnamefont {Stiles}},\
  }\bibfield  {title} {\enquote {\bibinfo {title} {Spin-orbit torques from
  interfacial spin-orbit coupling for various interfaces},}\ }\href {\doibase
  10.1103/PhysRevB.96.104438} {\bibfield  {journal} {\bibinfo  {journal} {Phys.
  Rev. B}\ }\textbf {\bibinfo {volume} {96}},\ \bibinfo {pages} {104438}
  (\bibinfo {year} {2017})}\BibitemShut {NoStop}%
\bibitem [{\citenamefont {Amin}\ \emph {et~al.}(2018)\citenamefont {Amin},
  \citenamefont {Zemen},\ and\ \citenamefont {Stiles}}]{Amin2018}%
  \BibitemOpen
  \bibfield  {author} {\bibinfo {author} {\bibfnamefont {V.~P.}\ \bibnamefont
  {Amin}}, \bibinfo {author} {\bibfnamefont {J.}~\bibnamefont {Zemen}}, \ and\
  \bibinfo {author} {\bibfnamefont {M.~D.}\ \bibnamefont {Stiles}},\ }\bibfield
   {title} {\enquote {\bibinfo {title} {{Interface-Generated Spin Currents}},}\
  }\href {\doibase 10.1103/PhysRevLett.121.136805} {\bibfield  {journal}
  {\bibinfo  {journal} {Phys. Rev. Lett.}\ }\textbf {\bibinfo {volume} {121}},\
  \bibinfo {pages} {136805} (\bibinfo {year} {2018})}\BibitemShut {NoStop}%
\bibitem [{\citenamefont {Rashba}(1960)}]{Rashba1960}%
  \BibitemOpen
  \bibfield  {author} {\bibinfo {author} {\bibfnamefont {E.}~\bibnamefont
  {Rashba}},\ }\bibfield  {title} {\enquote {\bibinfo {title} {Properties of
  semiconductors with an extremum loop. 1. cyclotron and combinational
  resonance in a magnetic field perpendicular to the plane of the loop.}}\
  }\href@noop {} {\bibfield  {journal} {\bibinfo  {journal} {Sov. Phys. Solid
  State}\ }\textbf {\bibinfo {volume} {2}},\ \bibinfo {pages} {1109–1122}
  (\bibinfo {year} {1960})}\BibitemShut {NoStop}%
\bibitem [{\citenamefont {Manchon}\ \emph {et~al.}(2015)\citenamefont
  {Manchon}, \citenamefont {Koo}, \citenamefont {Nitta}, \citenamefont
  {Frolov},\ and\ \citenamefont {Duine}}]{Manchon2015}%
  \BibitemOpen
  \bibfield  {author} {\bibinfo {author} {\bibfnamefont {A.}~\bibnamefont
  {Manchon}}, \bibinfo {author} {\bibfnamefont {H.~C.}\ \bibnamefont {Koo}},
  \bibinfo {author} {\bibfnamefont {J.}~\bibnamefont {Nitta}}, \bibinfo
  {author} {\bibfnamefont {S.~M.}\ \bibnamefont {Frolov}}, \ and\ \bibinfo
  {author} {\bibfnamefont {R.~A.}\ \bibnamefont {Duine}},\ }\bibfield  {title}
  {\enquote {\bibinfo {title} {{New perspectives for Rashba spin-orbit
  coupling}},}\ }\href {https://doi.org/10.1038/nmat4360} {\bibfield  {journal}
  {\bibinfo  {journal} {Nat. Mater.}\ }\textbf {\bibinfo {volume} {14}},\
  \bibinfo {pages} {871} (\bibinfo {year} {2015})}\BibitemShut {NoStop}%
\bibitem [{\citenamefont {Bercioux}\ and\ \citenamefont
  {Lucignano}(2015)}]{Bercioux2015}%
  \BibitemOpen
  \bibfield  {author} {\bibinfo {author} {\bibfnamefont {Dario}\ \bibnamefont
  {Bercioux}}\ and\ \bibinfo {author} {\bibfnamefont {Procolo}\ \bibnamefont
  {Lucignano}},\ }\bibfield  {title} {\enquote {\bibinfo {title} {{Quantum
  transport in Rashba spin{\textendash}orbit materials: a review}},}\ }\href
  {\doibase 10.1088/0034-4885/78/10/106001} {\bibfield  {journal} {\bibinfo
  {journal} {Reports on Progress in Physics}\ }\textbf {\bibinfo {volume}
  {78}},\ \bibinfo {pages} {106001} (\bibinfo {year} {2015})}\BibitemShut
  {NoStop}%
\bibitem [{\citenamefont {Amin}\ \emph {et~al.}(2019)\citenamefont {Amin},
  \citenamefont {Li}, \citenamefont {Stiles},\ and\ \citenamefont
  {Haney}}]{Amin2019}%
  \BibitemOpen
  \bibfield  {author} {\bibinfo {author} {\bibfnamefont {V.~P.}\ \bibnamefont
  {Amin}}, \bibinfo {author} {\bibfnamefont {Junwen}\ \bibnamefont {Li}},
  \bibinfo {author} {\bibfnamefont {M.~D.}\ \bibnamefont {Stiles}}, \ and\
  \bibinfo {author} {\bibfnamefont {P.~M.}\ \bibnamefont {Haney}},\ }\bibfield
  {title} {\enquote {\bibinfo {title} {Intrinsic spin currents in
  ferromagnets},}\ }\href {\doibase 10.1103/PhysRevB.99.220405} {\bibfield
  {journal} {\bibinfo  {journal} {Phys. Rev. B}\ }\textbf {\bibinfo {volume}
  {99}},\ \bibinfo {pages} {220405(R)} (\bibinfo {year} {2019})}\BibitemShut
  {NoStop}%
\bibitem [{\citenamefont {Wang}\ \emph {et~al.}(2019)\citenamefont {Wang},
  \citenamefont {Wang}, \citenamefont {Amin}, \citenamefont {Wang},
  \citenamefont {Radhakrishnan}, \citenamefont {Davidson}, \citenamefont
  {Allen}, \citenamefont {Silva}, \citenamefont {Ohldag}, \citenamefont
  {Balzar}, \citenamefont {Zink}, \citenamefont {Haney}, \citenamefont {Xiao},
  \citenamefont {Cahill}, \citenamefont {Lorenz},\ and\ \citenamefont
  {Fan}}]{Wang2019}%
  \BibitemOpen
  \bibfield  {author} {\bibinfo {author} {\bibfnamefont {Wenrui}\ \bibnamefont
  {Wang}}, \bibinfo {author} {\bibfnamefont {Tao}\ \bibnamefont {Wang}},
  \bibinfo {author} {\bibfnamefont {Vivek~P.}\ \bibnamefont {Amin}}, \bibinfo
  {author} {\bibfnamefont {Yang}\ \bibnamefont {Wang}}, \bibinfo {author}
  {\bibfnamefont {Anil}\ \bibnamefont {Radhakrishnan}}, \bibinfo {author}
  {\bibfnamefont {Angie}\ \bibnamefont {Davidson}}, \bibinfo {author}
  {\bibfnamefont {Shane~R.}\ \bibnamefont {Allen}}, \bibinfo {author}
  {\bibfnamefont {T.~J.}\ \bibnamefont {Silva}}, \bibinfo {author}
  {\bibfnamefont {Hendrik}\ \bibnamefont {Ohldag}}, \bibinfo {author}
  {\bibfnamefont {Davor}\ \bibnamefont {Balzar}}, \bibinfo {author}
  {\bibfnamefont {Barry~L.}\ \bibnamefont {Zink}}, \bibinfo {author}
  {\bibfnamefont {Paul~M.}\ \bibnamefont {Haney}}, \bibinfo {author}
  {\bibfnamefont {John~Q.}\ \bibnamefont {Xiao}}, \bibinfo {author}
  {\bibfnamefont {David~G.}\ \bibnamefont {Cahill}}, \bibinfo {author}
  {\bibfnamefont {Virginia~O.}\ \bibnamefont {Lorenz}}, \ and\ \bibinfo
  {author} {\bibfnamefont {Xin}\ \bibnamefont {Fan}},\ }\bibfield  {title}
  {\enquote {\bibinfo {title} {Anomalous spin-orbit torques in magnetic
  single-layer films},}\ }\href {\doibase 10.1038/s41565-019-0504-0} {\bibfield
   {journal} {\bibinfo  {journal} {Nature Nanotechnology}\ }\textbf {\bibinfo
  {volume} {14}},\ \bibinfo {pages} {819--824} (\bibinfo {year}
  {2019})}\BibitemShut {NoStop}%
\bibitem [{\citenamefont {Davidson}\ \emph {et~al.}(2020)\citenamefont
  {Davidson}, \citenamefont {Amin}, \citenamefont {Aljuaid}, \citenamefont
  {Haney},\ and\ \citenamefont {Fan}}]{Davison2020}%
  \BibitemOpen
  \bibfield  {author} {\bibinfo {author} {\bibfnamefont {Angie}\ \bibnamefont
  {Davidson}}, \bibinfo {author} {\bibfnamefont {Vivek~P.}\ \bibnamefont
  {Amin}}, \bibinfo {author} {\bibfnamefont {Wafa~S.}\ \bibnamefont {Aljuaid}},
  \bibinfo {author} {\bibfnamefont {Paul~M.}\ \bibnamefont {Haney}}, \ and\
  \bibinfo {author} {\bibfnamefont {Xin}\ \bibnamefont {Fan}},\ }\bibfield
  {title} {\enquote {\bibinfo {title} {Perspectives of electrically generated
  spin currents in ferromagnetic materials},}\ }\href {\doibase
  https://doi.org/10.1016/j.physleta.2019.126228} {\bibfield  {journal}
  {\bibinfo  {journal} {Physics Letters A}\ }\textbf {\bibinfo {volume}
  {384}},\ \bibinfo {pages} {126228} (\bibinfo {year} {2020})}\BibitemShut
  {NoStop}%
\bibitem [{\citenamefont {Go}\ and\ \citenamefont {Lee}(2020)}]{Go2020}%
  \BibitemOpen
  \bibfield  {author} {\bibinfo {author} {\bibfnamefont {Dongwook}\
  \bibnamefont {Go}}\ and\ \bibinfo {author} {\bibfnamefont {Hyun-Woo}\
  \bibnamefont {Lee}},\ }\bibfield  {title} {\enquote {\bibinfo {title}
  {Orbital torque: Torque generation by orbital current injection},}\ }\href
  {\doibase 10.1103/PhysRevResearch.2.013177} {\bibfield  {journal} {\bibinfo
  {journal} {Phys. Rev. Research}\ }\textbf {\bibinfo {volume} {2}},\ \bibinfo
  {pages} {013177} (\bibinfo {year} {2020})}\BibitemShut {NoStop}%
\bibitem [{\citenamefont {Tanaka}\ \emph {et~al.}(2008)\citenamefont {Tanaka},
  \citenamefont {Kontani}, \citenamefont {Naito}, \citenamefont {Naito},
  \citenamefont {Hirashima}, \citenamefont {Yamada},\ and\ \citenamefont
  {Inoue}}]{Tanaka2008}%
  \BibitemOpen
  \bibfield  {author} {\bibinfo {author} {\bibfnamefont {T.}~\bibnamefont
  {Tanaka}}, \bibinfo {author} {\bibfnamefont {H.}~\bibnamefont {Kontani}},
  \bibinfo {author} {\bibfnamefont {M.}~\bibnamefont {Naito}}, \bibinfo
  {author} {\bibfnamefont {T.}~\bibnamefont {Naito}}, \bibinfo {author}
  {\bibfnamefont {D.~S.}\ \bibnamefont {Hirashima}}, \bibinfo {author}
  {\bibfnamefont {K.}~\bibnamefont {Yamada}}, \ and\ \bibinfo {author}
  {\bibfnamefont {J.}~\bibnamefont {Inoue}},\ }\bibfield  {title} {\enquote
  {\bibinfo {title} {{Intrinsic spin Hall effect and orbital Hall effect in
  $4d$ and $5d$ transition metals}},}\ }\href {\doibase
  10.1103/PhysRevB.77.165117} {\bibfield  {journal} {\bibinfo  {journal} {Phys.
  Rev. B}\ }\textbf {\bibinfo {volume} {77}},\ \bibinfo {pages} {165117}
  (\bibinfo {year} {2008})}\BibitemShut {NoStop}%
\bibitem [{\citenamefont {Kontani}\ \emph {et~al.}(2009)\citenamefont
  {Kontani}, \citenamefont {Tanaka}, \citenamefont {Hirashima}, \citenamefont
  {Yamada},\ and\ \citenamefont {Inoue}}]{Kontani2009}%
  \BibitemOpen
  \bibfield  {author} {\bibinfo {author} {\bibfnamefont {H.}~\bibnamefont
  {Kontani}}, \bibinfo {author} {\bibfnamefont {T.}~\bibnamefont {Tanaka}},
  \bibinfo {author} {\bibfnamefont {D.~S.}\ \bibnamefont {Hirashima}}, \bibinfo
  {author} {\bibfnamefont {K.}~\bibnamefont {Yamada}}, \ and\ \bibinfo {author}
  {\bibfnamefont {J.}~\bibnamefont {Inoue}},\ }\bibfield  {title} {\enquote
  {\bibinfo {title} {{Giant Orbital Hall Effect in Transition Metals: Origin of
  Large Spin and Anomalous Hall Effects}},}\ }\href {\doibase
  10.1103/PhysRevLett.102.016601} {\bibfield  {journal} {\bibinfo  {journal}
  {Phys. Rev. Lett.}\ }\textbf {\bibinfo {volume} {102}},\ \bibinfo {pages}
  {016601} (\bibinfo {year} {2009})}\BibitemShut {NoStop}%
\bibitem [{\citenamefont {Go}\ \emph {et~al.}(2018)\citenamefont {Go},
  \citenamefont {Jo}, \citenamefont {Kim},\ and\ \citenamefont {Lee}}]{Go2018}%
  \BibitemOpen
  \bibfield  {author} {\bibinfo {author} {\bibfnamefont {Dongwook}\
  \bibnamefont {Go}}, \bibinfo {author} {\bibfnamefont {Daegeun}\ \bibnamefont
  {Jo}}, \bibinfo {author} {\bibfnamefont {Changyoung}\ \bibnamefont {Kim}}, \
  and\ \bibinfo {author} {\bibfnamefont {Hyun-Woo}\ \bibnamefont {Lee}},\
  }\bibfield  {title} {\enquote {\bibinfo {title} {{Intrinsic Spin and Orbital
  Hall Effects from Orbital Texture}},}\ }\href {\doibase
  10.1103/PhysRevLett.121.086602} {\bibfield  {journal} {\bibinfo  {journal}
  {Phys. Rev. Lett.}\ }\textbf {\bibinfo {volume} {121}},\ \bibinfo {pages}
  {086602} (\bibinfo {year} {2018})}\BibitemShut {NoStop}%
\bibitem [{\citenamefont {Jo}\ \emph {et~al.}(2018)\citenamefont {Jo},
  \citenamefont {Go},\ and\ \citenamefont {Lee}}]{Jo2018}%
  \BibitemOpen
  \bibfield  {author} {\bibinfo {author} {\bibfnamefont {Daegeun}\ \bibnamefont
  {Jo}}, \bibinfo {author} {\bibfnamefont {Dongwook}\ \bibnamefont {Go}}, \
  and\ \bibinfo {author} {\bibfnamefont {Hyun-Woo}\ \bibnamefont {Lee}},\
  }\bibfield  {title} {\enquote {\bibinfo {title} {Gigantic intrinsic orbital
  hall effects in weakly spin-orbit coupled metals},}\ }\href {\doibase
  10.1103/PhysRevB.98.214405} {\bibfield  {journal} {\bibinfo  {journal} {Phys.
  Rev. B}\ }\textbf {\bibinfo {volume} {98}},\ \bibinfo {pages} {214405}
  (\bibinfo {year} {2018})}\BibitemShut {NoStop}%
\bibitem [{\citenamefont {Freimuth}\ \emph {et~al.}(2014)\citenamefont
  {Freimuth}, \citenamefont {Bl\"ugel},\ and\ \citenamefont
  {Mokrousov}}]{Freimuth2014}%
  \BibitemOpen
  \bibfield  {author} {\bibinfo {author} {\bibfnamefont {Frank}\ \bibnamefont
  {Freimuth}}, \bibinfo {author} {\bibfnamefont {Stefan}\ \bibnamefont
  {Bl\"ugel}}, \ and\ \bibinfo {author} {\bibfnamefont {Yuriy}\ \bibnamefont
  {Mokrousov}},\ }\bibfield  {title} {\enquote {\bibinfo {title} {{Spin-orbit
  torques in Co/Pt(111) and Mn/W(001) magnetic bilayers from first
  principles}},}\ }\href {\doibase 10.1103/PhysRevB.90.174423} {\bibfield
  {journal} {\bibinfo  {journal} {Phys. Rev. B}\ }\textbf {\bibinfo {volume}
  {90}},\ \bibinfo {pages} {174423} (\bibinfo {year} {2014})}\BibitemShut
  {NoStop}%
\bibitem [{\citenamefont {Freimuth}\ \emph {et~al.}(2015)\citenamefont
  {Freimuth}, \citenamefont {Bl\"ugel},\ and\ \citenamefont
  {Mokrousov}}]{Freimuth2015}%
  \BibitemOpen
  \bibfield  {author} {\bibinfo {author} {\bibfnamefont {Frank}\ \bibnamefont
  {Freimuth}}, \bibinfo {author} {\bibfnamefont {Stefan}\ \bibnamefont
  {Bl\"ugel}}, \ and\ \bibinfo {author} {\bibfnamefont {Yuriy}\ \bibnamefont
  {Mokrousov}},\ }\bibfield  {title} {\enquote {\bibinfo {title} {Direct and
  inverse spin-orbit torques},}\ }\href {\doibase 10.1103/PhysRevB.92.064415}
  {\bibfield  {journal} {\bibinfo  {journal} {Phys. Rev. B}\ }\textbf {\bibinfo
  {volume} {92}},\ \bibinfo {pages} {064415} (\bibinfo {year}
  {2015})}\BibitemShut {NoStop}%
\bibitem [{\citenamefont {G\'eranton}\ \emph {et~al.}(2015)\citenamefont
  {G\'eranton}, \citenamefont {Freimuth}, \citenamefont {Bl\"ugel},\ and\
  \citenamefont {Mokrousov}}]{Geraton2015}%
  \BibitemOpen
  \bibfield  {author} {\bibinfo {author} {\bibfnamefont {Guillaume}\
  \bibnamefont {G\'eranton}}, \bibinfo {author} {\bibfnamefont {Frank}\
  \bibnamefont {Freimuth}}, \bibinfo {author} {\bibfnamefont {Stefan}\
  \bibnamefont {Bl\"ugel}}, \ and\ \bibinfo {author} {\bibfnamefont {Yuriy}\
  \bibnamefont {Mokrousov}},\ }\bibfield  {title} {\enquote {\bibinfo {title}
  {Spin-orbit torques in $l{1}_{0}\ensuremath{-}\mathrm{FePt}/\mathrm{Pt}$ thin
  films driven by electrical and thermal currents},}\ }\href {\doibase
  10.1103/PhysRevB.91.014417} {\bibfield  {journal} {\bibinfo  {journal} {Phys.
  Rev. B}\ }\textbf {\bibinfo {volume} {91}},\ \bibinfo {pages} {014417}
  (\bibinfo {year} {2015})}\BibitemShut {NoStop}%
\bibitem [{\citenamefont {Mahfouzi}\ and\ \citenamefont
  {Kioussis}(2018)}]{Mahfouzi2018}%
  \BibitemOpen
  \bibfield  {author} {\bibinfo {author} {\bibfnamefont {Farzad}\ \bibnamefont
  {Mahfouzi}}\ and\ \bibinfo {author} {\bibfnamefont {Nicholas}\ \bibnamefont
  {Kioussis}},\ }\bibfield  {title} {\enquote {\bibinfo {title}
  {{First-principles study of the angular dependence of the spin-orbit torque
  in Pt/Co and Pd/Co bilayers}},}\ }\href {\doibase 10.1103/PhysRevB.97.224426}
  {\bibfield  {journal} {\bibinfo  {journal} {Phys. Rev. B}\ }\textbf {\bibinfo
  {volume} {97}},\ \bibinfo {pages} {224426} (\bibinfo {year}
  {2018})}\BibitemShut {NoStop}%
\bibitem [{\citenamefont {Belashchenko}\ \emph {et~al.}(2019)\citenamefont
  {Belashchenko}, \citenamefont {Kovalev},\ and\ \citenamefont {van
  Schilfgaarde}}]{Belashchenko2019}%
  \BibitemOpen
  \bibfield  {author} {\bibinfo {author} {\bibfnamefont {K.~D.}\ \bibnamefont
  {Belashchenko}}, \bibinfo {author} {\bibfnamefont {Alexey~A.}\ \bibnamefont
  {Kovalev}}, \ and\ \bibinfo {author} {\bibfnamefont {M.}~\bibnamefont {van
  Schilfgaarde}},\ }\bibfield  {title} {\enquote {\bibinfo {title}
  {{First-principles calculation of spin-orbit torque in a Co/Pt bilayer}},}\
  }\href {\doibase 10.1103/PhysRevMaterials.3.011401} {\bibfield  {journal}
  {\bibinfo  {journal} {Phys. Rev. Materials}\ }\textbf {\bibinfo {volume}
  {3}},\ \bibinfo {pages} {011401} (\bibinfo {year} {2019})}\BibitemShut
  {NoStop}%
\bibitem [{\citenamefont {Guimar{\~a}es}\ \emph {et~al.}(2020)\citenamefont
  {Guimar{\~a}es}, \citenamefont {Bouaziz}, \citenamefont {dos Santos~Dias},\
  and\ \citenamefont {Lounis}}]{Guimaraes2020}%
  \BibitemOpen
  \bibfield  {author} {\bibinfo {author} {\bibfnamefont {Filipe S.~M.}\
  \bibnamefont {Guimar{\~a}es}}, \bibinfo {author} {\bibfnamefont {Juba}\
  \bibnamefont {Bouaziz}}, \bibinfo {author} {\bibfnamefont {Manuel}\
  \bibnamefont {dos Santos~Dias}}, \ and\ \bibinfo {author} {\bibfnamefont
  {Samir}\ \bibnamefont {Lounis}},\ }\bibfield  {title} {\enquote {\bibinfo
  {title} {Spin-orbit torques and their associated effective fields from
  gigahertz to terahertz},}\ }\href {\doibase 10.1038/s42005-020-0282-x}
  {\bibfield  {journal} {\bibinfo  {journal} {Communications Physics}\ }\textbf
  {\bibinfo {volume} {3}},\ \bibinfo {pages} {19} (\bibinfo {year}
  {2020})}\BibitemShut {NoStop}%
\bibitem [{\citenamefont {Valenzuela}\ and\ \citenamefont
  {Tinkham}(2006)}]{Valenzuela2006}%
  \BibitemOpen
  \bibfield  {author} {\bibinfo {author} {\bibfnamefont {S.~O.}\ \bibnamefont
  {Valenzuela}}\ and\ \bibinfo {author} {\bibfnamefont {M.}~\bibnamefont
  {Tinkham}},\ }\bibfield  {title} {\enquote {\bibinfo {title} {{Direct
  electronic measurement of the spin Hall effect}},}\ }\href {\doibase
  10.1038/nature04937} {\bibfield  {journal} {\bibinfo  {journal} {Nature}\
  }\textbf {\bibinfo {volume} {442}},\ \bibinfo {pages} {176--179} (\bibinfo
  {year} {2006})}\BibitemShut {NoStop}%
\bibitem [{\citenamefont {Kimura}\ \emph {et~al.}(2007)\citenamefont {Kimura},
  \citenamefont {Otani}, \citenamefont {Sato}, \citenamefont {Takahashi},\ and\
  \citenamefont {Maekawa}}]{Kimura2007}%
  \BibitemOpen
  \bibfield  {author} {\bibinfo {author} {\bibfnamefont {T.}~\bibnamefont
  {Kimura}}, \bibinfo {author} {\bibfnamefont {Y.}~\bibnamefont {Otani}},
  \bibinfo {author} {\bibfnamefont {T.}~\bibnamefont {Sato}}, \bibinfo {author}
  {\bibfnamefont {S.}~\bibnamefont {Takahashi}}, \ and\ \bibinfo {author}
  {\bibfnamefont {S.}~\bibnamefont {Maekawa}},\ }\bibfield  {title} {\enquote
  {\bibinfo {title} {{Room-Temperature Reversible Spin Hall Effect}},}\ }\href
  {\doibase 10.1103/PhysRevLett.98.156601} {\bibfield  {journal} {\bibinfo
  {journal} {Phys. Rev. Lett.}\ }\textbf {\bibinfo {volume} {98}},\ \bibinfo
  {pages} {156601} (\bibinfo {year} {2007})}\BibitemShut {NoStop}%
\bibitem [{\citenamefont {Chen}\ \emph {et~al.}(2018)\citenamefont {Chen},
  \citenamefont {Niu},\ and\ \citenamefont {MacDonald}}]{Chen2018}%
  \BibitemOpen
  \bibfield  {author} {\bibinfo {author} {\bibfnamefont {Hua}\ \bibnamefont
  {Chen}}, \bibinfo {author} {\bibfnamefont {Qian}\ \bibnamefont {Niu}}, \ and\
  \bibinfo {author} {\bibfnamefont {Allan~H.}\ \bibnamefont {MacDonald}},\
  }\href@noop {} {\enquote {\bibinfo {title} {{Spin Hall effects without spin
  currents in magnetic insulators}},}\ } (\bibinfo {year} {2018}),\ \Eprint
  {http://arxiv.org/abs/arXiv:1803.01294} {arXiv:1803.01294} \BibitemShut
  {NoStop}%
\bibitem [{\citenamefont {Park}\ \emph {et~al.}(2011)\citenamefont {Park},
  \citenamefont {Kim}, \citenamefont {Yu}, \citenamefont {Han},\ and\
  \citenamefont {Kim}}]{Park2013a}%
  \BibitemOpen
  \bibfield  {author} {\bibinfo {author} {\bibfnamefont {Seung~Ryong}\
  \bibnamefont {Park}}, \bibinfo {author} {\bibfnamefont {Choong~H.}\
  \bibnamefont {Kim}}, \bibinfo {author} {\bibfnamefont {Jaejun}\ \bibnamefont
  {Yu}}, \bibinfo {author} {\bibfnamefont {Jung~Hoon}\ \bibnamefont {Han}}, \
  and\ \bibinfo {author} {\bibfnamefont {Changyoung}\ \bibnamefont {Kim}},\
  }\bibfield  {title} {\enquote {\bibinfo {title} {{Orbital-Angular-Momentum
  Based Origin of Rashba-Type Surface Band Splitting}},}\ }\href {\doibase
  10.1103/PhysRevLett.107.156803} {\bibfield  {journal} {\bibinfo  {journal}
  {Phys. Rev. Lett.}\ }\textbf {\bibinfo {volume} {107}},\ \bibinfo {pages}
  {156803} (\bibinfo {year} {2011})}\BibitemShut {NoStop}%
\bibitem [{\citenamefont {Park}\ \emph {et~al.}(2013)\citenamefont {Park},
  \citenamefont {Kim}, \citenamefont {Lee},\ and\ \citenamefont
  {Han}}]{Park2013b}%
  \BibitemOpen
  \bibfield  {author} {\bibinfo {author} {\bibfnamefont {Jin-Hong}\
  \bibnamefont {Park}}, \bibinfo {author} {\bibfnamefont {Choong~H.}\
  \bibnamefont {Kim}}, \bibinfo {author} {\bibfnamefont {Hyun-Woo}\
  \bibnamefont {Lee}}, \ and\ \bibinfo {author} {\bibfnamefont {Jung~Hoon}\
  \bibnamefont {Han}},\ }\bibfield  {title} {\enquote {\bibinfo {title}
  {Orbital chirality and rashba interaction in magnetic bands},}\ }\href
  {\doibase 10.1103/PhysRevB.87.041301} {\bibfield  {journal} {\bibinfo
  {journal} {Phys. Rev. B}\ }\textbf {\bibinfo {volume} {87}},\ \bibinfo
  {pages} {041301(R)} (\bibinfo {year} {2013})}\BibitemShut {NoStop}%
\bibitem [{\citenamefont {Go}\ \emph {et~al.}(2017)\citenamefont {Go},
  \citenamefont {Hanke}, \citenamefont {Buhl}, \citenamefont {Freimuth},
  \citenamefont {Bihlmayer}, \citenamefont {Lee}, \citenamefont {Mokrousov},\
  and\ \citenamefont {Bl{\"u}gel}}]{Go2017}%
  \BibitemOpen
  \bibfield  {author} {\bibinfo {author} {\bibfnamefont {Dongwook}\
  \bibnamefont {Go}}, \bibinfo {author} {\bibfnamefont {Jan-Philipp}\
  \bibnamefont {Hanke}}, \bibinfo {author} {\bibfnamefont {Patrick~M.}\
  \bibnamefont {Buhl}}, \bibinfo {author} {\bibfnamefont {Frank}\ \bibnamefont
  {Freimuth}}, \bibinfo {author} {\bibfnamefont {Gustav}\ \bibnamefont
  {Bihlmayer}}, \bibinfo {author} {\bibfnamefont {Hyun-Woo}\ \bibnamefont
  {Lee}}, \bibinfo {author} {\bibfnamefont {Yuriy}\ \bibnamefont {Mokrousov}},
  \ and\ \bibinfo {author} {\bibfnamefont {Stefan}\ \bibnamefont
  {Bl{\"u}gel}},\ }\bibfield  {title} {\enquote {\bibinfo {title} {Toward
  surface orbitronics: giant orbital magnetism from the orbital rashba effect
  at the surface of sp-metals},}\ }\href {\doibase 10.1038/srep46742}
  {\bibfield  {journal} {\bibinfo  {journal} {Scientific Reports}\ }\textbf
  {\bibinfo {volume} {7}},\ \bibinfo {pages} {46742} (\bibinfo {year}
  {2017})}\BibitemShut {NoStop}%
\bibitem [{\citenamefont {Salemi}\ \emph {et~al.}(2019)\citenamefont {Salemi},
  \citenamefont {Berritta}, \citenamefont {Nandy},\ and\ \citenamefont
  {Oppeneer}}]{Salemi2019}%
  \BibitemOpen
  \bibfield  {author} {\bibinfo {author} {\bibfnamefont {Leandro}\ \bibnamefont
  {Salemi}}, \bibinfo {author} {\bibfnamefont {Marco}\ \bibnamefont
  {Berritta}}, \bibinfo {author} {\bibfnamefont {Ashis~K.}\ \bibnamefont
  {Nandy}}, \ and\ \bibinfo {author} {\bibfnamefont {Peter~M.}\ \bibnamefont
  {Oppeneer}},\ }\bibfield  {title} {\enquote {\bibinfo {title} {Orbitally
  dominated rashba-edelstein effect in noncentrosymmetric antiferromagnets},}\
  }\href {\doibase 10.1038/s41467-019-13367-z} {\bibfield  {journal} {\bibinfo
  {journal} {Nature Communications}\ }\textbf {\bibinfo {volume} {10}},\
  \bibinfo {pages} {5381} (\bibinfo {year} {2019})}\BibitemShut {NoStop}%
\bibitem [{\citenamefont {Canonico}\ \emph {et~al.}(2020)\citenamefont
  {Canonico}, \citenamefont {Cysne}, \citenamefont {Molina-Sanchez},
  \citenamefont {Muniz},\ and\ \citenamefont {Rappoport}}]{Canonico2020}%
  \BibitemOpen
  \bibfield  {author} {\bibinfo {author} {\bibfnamefont {Luis~M.}\ \bibnamefont
  {Canonico}}, \bibinfo {author} {\bibfnamefont {Tarik~P.}\ \bibnamefont
  {Cysne}}, \bibinfo {author} {\bibfnamefont {Alejandro}\ \bibnamefont
  {Molina-Sanchez}}, \bibinfo {author} {\bibfnamefont {R.~B.}\ \bibnamefont
  {Muniz}}, \ and\ \bibinfo {author} {\bibfnamefont {Tatiana~G.}\ \bibnamefont
  {Rappoport}},\ }\href@noop {} {\enquote {\bibinfo {title} {Orbital hall
  insulating phase in transition metal dichalcogenide monolayers},}\ }
  (\bibinfo {year} {2020}),\ \Eprint {http://arxiv.org/abs/arXiv:2001.03592}
  {arXiv:2001.03592} \BibitemShut {NoStop}%
\bibitem [{\citenamefont {Haney}\ and\ \citenamefont
  {Stiles}(2010)}]{Haney2010}%
  \BibitemOpen
  \bibfield  {author} {\bibinfo {author} {\bibfnamefont {Paul~M.}\ \bibnamefont
  {Haney}}\ and\ \bibinfo {author} {\bibfnamefont {M.~D.}\ \bibnamefont
  {Stiles}},\ }\bibfield  {title} {\enquote {\bibinfo {title} {{Current-Induced
  Torques in the Presence of Spin-Orbit Coupling}},}\ }\href {\doibase
  10.1103/PhysRevLett.105.126602} {\bibfield  {journal} {\bibinfo  {journal}
  {Phys. Rev. Lett.}\ }\textbf {\bibinfo {volume} {105}},\ \bibinfo {pages}
  {126602} (\bibinfo {year} {2010})}\BibitemShut {NoStop}%
\bibitem [{\citenamefont {Hoffmann}\ \emph {et~al.}(2015)\citenamefont
  {Hoffmann}, \citenamefont {Weischenberg}, \citenamefont {Dup\'e},
  \citenamefont {Freimuth}, \citenamefont {Ferriani}, \citenamefont
  {Mokrousov},\ and\ \citenamefont {Heinze}}]{Hoffmann2015}%
  \BibitemOpen
  \bibfield  {author} {\bibinfo {author} {\bibfnamefont {M.}~\bibnamefont
  {Hoffmann}}, \bibinfo {author} {\bibfnamefont {J.}~\bibnamefont
  {Weischenberg}}, \bibinfo {author} {\bibfnamefont {B.}~\bibnamefont
  {Dup\'e}}, \bibinfo {author} {\bibfnamefont {F.}~\bibnamefont {Freimuth}},
  \bibinfo {author} {\bibfnamefont {P.}~\bibnamefont {Ferriani}}, \bibinfo
  {author} {\bibfnamefont {Y.}~\bibnamefont {Mokrousov}}, \ and\ \bibinfo
  {author} {\bibfnamefont {S.}~\bibnamefont {Heinze}},\ }\bibfield  {title}
  {\enquote {\bibinfo {title} {Topological orbital magnetization and emergent
  hall effect of an atomic-scale spin lattice at a surface},}\ }\href {\doibase
  10.1103/PhysRevB.92.020401} {\bibfield  {journal} {\bibinfo  {journal} {Phys.
  Rev. B}\ }\textbf {\bibinfo {volume} {92}},\ \bibinfo {pages} {020401(R)}
  (\bibinfo {year} {2015})}\BibitemShut {NoStop}%
\bibitem [{\citenamefont {Hanke}\ \emph {et~al.}(2017)\citenamefont {Hanke},
  \citenamefont {Freimuth}, \citenamefont {Bl{\"u}gel},\ and\ \citenamefont
  {Mokrousov}}]{Hanke2017}%
  \BibitemOpen
  \bibfield  {author} {\bibinfo {author} {\bibfnamefont {Jan-Philipp}\
  \bibnamefont {Hanke}}, \bibinfo {author} {\bibfnamefont {Frank}\ \bibnamefont
  {Freimuth}}, \bibinfo {author} {\bibfnamefont {Stefan}\ \bibnamefont
  {Bl{\"u}gel}}, \ and\ \bibinfo {author} {\bibfnamefont {Yuriy}\ \bibnamefont
  {Mokrousov}},\ }\bibfield  {title} {\enquote {\bibinfo {title} {{Prototypical
  topological orbital ferromagnet $\gamma$-FeMn}},}\ }\href {\doibase
  10.1038/srep41078} {\bibfield  {journal} {\bibinfo  {journal} {Scientific
  Reports}\ }\textbf {\bibinfo {volume} {7}},\ \bibinfo {pages} {41078}
  (\bibinfo {year} {2017})}\BibitemShut {NoStop}%
\bibitem [{\citenamefont {dos Santos~Dias}\ \emph {et~al.}(2016)\citenamefont
  {dos Santos~Dias}, \citenamefont {Bouaziz}, \citenamefont {Bouhassoune},
  \citenamefont {Bl{\"u}gel},\ and\ \citenamefont
  {Lounis}}]{dosSantosDias2016}%
  \BibitemOpen
  \bibfield  {author} {\bibinfo {author} {\bibfnamefont {Manuel}\ \bibnamefont
  {dos Santos~Dias}}, \bibinfo {author} {\bibfnamefont {Juba}\ \bibnamefont
  {Bouaziz}}, \bibinfo {author} {\bibfnamefont {Mohammed}\ \bibnamefont
  {Bouhassoune}}, \bibinfo {author} {\bibfnamefont {Stefan}\ \bibnamefont
  {Bl{\"u}gel}}, \ and\ \bibinfo {author} {\bibfnamefont {Samir}\ \bibnamefont
  {Lounis}},\ }\bibfield  {title} {\enquote {\bibinfo {title} {Chirality-driven
  orbital magnetic moments as a new probe for topological magnetic
  structures},}\ }\href {\doibase 10.1038/ncomms13613} {\bibfield  {journal}
  {\bibinfo  {journal} {Nature Communications}\ }\textbf {\bibinfo {volume}
  {7}},\ \bibinfo {pages} {13613} (\bibinfo {year} {2016})}\BibitemShut
  {NoStop}%
\bibitem [{\citenamefont {Lux}\ \emph {et~al.}(2018)\citenamefont {Lux},
  \citenamefont {Freimuth}, \citenamefont {Bl{\"u}gel},\ and\ \citenamefont
  {Mokrousov}}]{Lux2018}%
  \BibitemOpen
  \bibfield  {author} {\bibinfo {author} {\bibfnamefont {Fabian~R.}\
  \bibnamefont {Lux}}, \bibinfo {author} {\bibfnamefont {Frank}\ \bibnamefont
  {Freimuth}}, \bibinfo {author} {\bibfnamefont {Stefan}\ \bibnamefont
  {Bl{\"u}gel}}, \ and\ \bibinfo {author} {\bibfnamefont {Yuriy}\ \bibnamefont
  {Mokrousov}},\ }\bibfield  {title} {\enquote {\bibinfo {title} {Engineering
  chiral and topological orbital magnetism of domain walls and skyrmions},}\
  }\href {\doibase 10.1038/s42005-018-0055-y} {\bibfield  {journal} {\bibinfo
  {journal} {Communications Physics}\ }\textbf {\bibinfo {volume} {1}},\
  \bibinfo {pages} {60} (\bibinfo {year} {2018})}\BibitemShut {NoStop}%
\bibitem [{\citenamefont {Grytsiuk}\ \emph {et~al.}(2020)\citenamefont
  {Grytsiuk}, \citenamefont {Hanke}, \citenamefont {Hoffmann}, \citenamefont
  {Bouaziz}, \citenamefont {Gomonay}, \citenamefont {Bihlmayer}, \citenamefont
  {Lounis}, \citenamefont {Mokrousov},\ and\ \citenamefont
  {Bl{\"u}gel}}]{Grytsiuk2020}%
  \BibitemOpen
  \bibfield  {author} {\bibinfo {author} {\bibfnamefont {S.}~\bibnamefont
  {Grytsiuk}}, \bibinfo {author} {\bibfnamefont {J.-P.}\ \bibnamefont {Hanke}},
  \bibinfo {author} {\bibfnamefont {M.}~\bibnamefont {Hoffmann}}, \bibinfo
  {author} {\bibfnamefont {J.}~\bibnamefont {Bouaziz}}, \bibinfo {author}
  {\bibfnamefont {O.}~\bibnamefont {Gomonay}}, \bibinfo {author} {\bibfnamefont
  {G.}~\bibnamefont {Bihlmayer}}, \bibinfo {author} {\bibfnamefont
  {S.}~\bibnamefont {Lounis}}, \bibinfo {author} {\bibfnamefont
  {Y.}~\bibnamefont {Mokrousov}}, \ and\ \bibinfo {author} {\bibfnamefont
  {S.}~\bibnamefont {Bl{\"u}gel}},\ }\bibfield  {title} {\enquote {\bibinfo
  {title} {Topological-chiral magnetic interactions driven by emergent orbital
  magnetism},}\ }\href {\doibase 10.1038/s41467-019-14030-3} {\bibfield
  {journal} {\bibinfo  {journal} {Nature Communications}\ }\textbf {\bibinfo
  {volume} {11}},\ \bibinfo {pages} {511} (\bibinfo {year} {2020})}\BibitemShut
  {NoStop}%
\bibitem [{\citenamefont {chuan Zhang}\ \emph {et~al.}(2019)\citenamefont
  {chuan Zhang}, \citenamefont {Lux}, \citenamefont {Hanke}, \citenamefont
  {Buhl}, \citenamefont {Grytsiuk}, \citenamefont {Bl\"ugel},\ and\
  \citenamefont {Mokrousov}}]{Zhang2019}%
  \BibitemOpen
  \bibfield  {author} {\bibinfo {author} {\bibfnamefont {Li}~\bibnamefont
  {chuan Zhang}}, \bibinfo {author} {\bibfnamefont {Fabian~R.}\ \bibnamefont
  {Lux}}, \bibinfo {author} {\bibfnamefont {Jan-Philipp}\ \bibnamefont
  {Hanke}}, \bibinfo {author} {\bibfnamefont {Patrick~M.}\ \bibnamefont
  {Buhl}}, \bibinfo {author} {\bibfnamefont {Sergii}\ \bibnamefont {Grytsiuk}},
  \bibinfo {author} {\bibfnamefont {Stefan}\ \bibnamefont {Bl\"ugel}}, \ and\
  \bibinfo {author} {\bibfnamefont {Yuriy}\ \bibnamefont {Mokrousov}},\
  }\href@noop {} {\enquote {\bibinfo {title} {{Orbital Nernst Effect of
  Magnons}},}\ } (\bibinfo {year} {2019}),\ \Eprint
  {http://arxiv.org/abs/arXiv:1910.03317} {arXiv:1910.03317} \BibitemShut
  {NoStop}%
\bibitem [{\citenamefont {Hanke}\ \emph {et~al.}(2016)\citenamefont {Hanke},
  \citenamefont {Freimuth}, \citenamefont {Nandy}, \citenamefont {Zhang},
  \citenamefont {Bl\"ugel},\ and\ \citenamefont {Mokrousov}}]{Hanke2016}%
  \BibitemOpen
  \bibfield  {author} {\bibinfo {author} {\bibfnamefont {J.-P.}\ \bibnamefont
  {Hanke}}, \bibinfo {author} {\bibfnamefont {F.}~\bibnamefont {Freimuth}},
  \bibinfo {author} {\bibfnamefont {A.~K.}\ \bibnamefont {Nandy}}, \bibinfo
  {author} {\bibfnamefont {H.}~\bibnamefont {Zhang}}, \bibinfo {author}
  {\bibfnamefont {S.}~\bibnamefont {Bl\"ugel}}, \ and\ \bibinfo {author}
  {\bibfnamefont {Y.}~\bibnamefont {Mokrousov}},\ }\bibfield  {title} {\enquote
  {\bibinfo {title} {{Role of Berry phase theory for describing orbital
  magnetism: From magnetic heterostructures to topological orbital
  ferromagnets}},}\ }\href {\doibase 10.1103/PhysRevB.94.121114} {\bibfield
  {journal} {\bibinfo  {journal} {Phys. Rev. B}\ }\textbf {\bibinfo {volume}
  {94}},\ \bibinfo {pages} {121114(R)} (\bibinfo {year} {2016})}\BibitemShut
  {NoStop}%
\bibitem [{\citenamefont {Thonhauser}\ \emph {et~al.}(2005)\citenamefont
  {Thonhauser}, \citenamefont {Ceresoli}, \citenamefont {Vanderbilt},\ and\
  \citenamefont {Resta}}]{Thonhauser2005}%
  \BibitemOpen
  \bibfield  {author} {\bibinfo {author} {\bibfnamefont {T.}~\bibnamefont
  {Thonhauser}}, \bibinfo {author} {\bibfnamefont {Davide}\ \bibnamefont
  {Ceresoli}}, \bibinfo {author} {\bibfnamefont {David}\ \bibnamefont
  {Vanderbilt}}, \ and\ \bibinfo {author} {\bibfnamefont {R.}~\bibnamefont
  {Resta}},\ }\bibfield  {title} {\enquote {\bibinfo {title} {{Orbital
  Magnetization in Periodic Insulators}},}\ }\href {\doibase
  10.1103/PhysRevLett.95.137205} {\bibfield  {journal} {\bibinfo  {journal}
  {Phys. Rev. Lett.}\ }\textbf {\bibinfo {volume} {95}},\ \bibinfo {pages}
  {137205} (\bibinfo {year} {2005})}\BibitemShut {NoStop}%
\bibitem [{\citenamefont {Ceresoli}\ \emph {et~al.}(2006)\citenamefont
  {Ceresoli}, \citenamefont {Thonhauser}, \citenamefont {Vanderbilt},\ and\
  \citenamefont {Resta}}]{Ceresoli2006}%
  \BibitemOpen
  \bibfield  {author} {\bibinfo {author} {\bibfnamefont {Davide}\ \bibnamefont
  {Ceresoli}}, \bibinfo {author} {\bibfnamefont {T.}~\bibnamefont
  {Thonhauser}}, \bibinfo {author} {\bibfnamefont {David}\ \bibnamefont
  {Vanderbilt}}, \ and\ \bibinfo {author} {\bibfnamefont {R.}~\bibnamefont
  {Resta}},\ }\bibfield  {title} {\enquote {\bibinfo {title} {{Orbital
  magnetization in crystalline solids: Multi-band insulators, Chern insulators,
  and metals}},}\ }\href {\doibase 10.1103/PhysRevB.74.024408} {\bibfield
  {journal} {\bibinfo  {journal} {Phys. Rev. B}\ }\textbf {\bibinfo {volume}
  {74}},\ \bibinfo {pages} {024408} (\bibinfo {year} {2006})}\BibitemShut
  {NoStop}%
\bibitem [{\citenamefont {Shi}\ \emph {et~al.}(2007)\citenamefont {Shi},
  \citenamefont {Vignale}, \citenamefont {Xiao},\ and\ \citenamefont
  {Niu}}]{Shi2007}%
  \BibitemOpen
  \bibfield  {author} {\bibinfo {author} {\bibfnamefont {Junren}\ \bibnamefont
  {Shi}}, \bibinfo {author} {\bibfnamefont {G.}~\bibnamefont {Vignale}},
  \bibinfo {author} {\bibfnamefont {Di}~\bibnamefont {Xiao}}, \ and\ \bibinfo
  {author} {\bibfnamefont {Qian}\ \bibnamefont {Niu}},\ }\bibfield  {title}
  {\enquote {\bibinfo {title} {{Quantum Theory of Orbital Magnetization and Its
  Generalization to Interacting Systems}},}\ }\href {\doibase
  10.1103/PhysRevLett.99.197202} {\bibfield  {journal} {\bibinfo  {journal}
  {Phys. Rev. Lett.}\ }\textbf {\bibinfo {volume} {99}},\ \bibinfo {pages}
  {197202} (\bibinfo {year} {2007})}\BibitemShut {NoStop}%
\bibitem [{\citenamefont {Mathon}(2001)}]{Mathon2001}%
  \BibitemOpen
  \bibfield  {author} {\bibinfo {author} {\bibfnamefont {J.}~\bibnamefont
  {Mathon}},\ }\enquote {\bibinfo {title} {Phenomenological theory of giant
  magnetoresistance},}\ in\ \href {\doibase 10.1007/3-540-45258-3_4} {\emph
  {\bibinfo {booktitle} {Spin Electronics}}},\ \bibinfo {editor} {edited by\
  \bibinfo {editor} {\bibfnamefont {Michael}\ \bibnamefont {Ziese}}\ and\
  \bibinfo {editor} {\bibfnamefont {Martin~J.}\ \bibnamefont {Thornton}}}\
  (\bibinfo  {publisher} {Springer Berlin Heidelberg},\ \bibinfo {address}
  {Berlin, Heidelberg},\ \bibinfo {year} {2001})\ pp.\ \bibinfo {pages}
  {71--88}\BibitemShut {NoStop}%
\bibitem [{\citenamefont {Mahan}(2000)}]{Mahan2000}%
  \BibitemOpen
  \bibfield  {author} {\bibinfo {author} {\bibfnamefont {Gerald~D.}\
  \bibnamefont {Mahan}},\ }\enquote {\bibinfo {title} {dc conductivities},}\
  in\ \href {\doibase 10.1007/978-1-4757-5714-9_8} {\emph {\bibinfo {booktitle}
  {Many-Particle Physics}}}\ (\bibinfo  {publisher} {Springer US},\ \bibinfo
  {address} {Boston, MA},\ \bibinfo {year} {2000})\ pp.\ \bibinfo {pages}
  {499--577}\BibitemShut {NoStop}%
\bibitem [{\citenamefont {Ghosh}\ and\ \citenamefont
  {Manchon}(2018)}]{Ghosh2018}%
  \BibitemOpen
  \bibfield  {author} {\bibinfo {author} {\bibfnamefont {S.}~\bibnamefont
  {Ghosh}}\ and\ \bibinfo {author} {\bibfnamefont {A.}~\bibnamefont
  {Manchon}},\ }\bibfield  {title} {\enquote {\bibinfo {title} {Spin-orbit
  torque in a three-dimensional topological insulator--ferromagnet
  heterostructure: Crossover between bulk and surface transport},}\ }\href
  {\doibase 10.1103/PhysRevB.97.134402} {\bibfield  {journal} {\bibinfo
  {journal} {Phys. Rev. B}\ }\textbf {\bibinfo {volume} {97}},\ \bibinfo
  {pages} {134402} (\bibinfo {year} {2018})}\BibitemShut {NoStop}%
\bibitem [{\citenamefont {Manchon}\ \emph {et~al.}(2020)\citenamefont
  {Manchon}, \citenamefont {Ghosh}, \citenamefont {Barreteau},\ and\
  \citenamefont {Manchon}}]{Manchon2020}%
  \BibitemOpen
  \bibfield  {author} {\bibinfo {author} {\bibfnamefont {G.}~\bibnamefont
  {Manchon}}, \bibinfo {author} {\bibfnamefont {S.}~\bibnamefont {Ghosh}},
  \bibinfo {author} {\bibfnamefont {C.}~\bibnamefont {Barreteau}}, \ and\
  \bibinfo {author} {\bibfnamefont {A.}~\bibnamefont {Manchon}},\ }\href@noop
  {} {\enquote {\bibinfo {title} {Semi-realistic tight-binding model for
  spin-orbit torques},}\ } (\bibinfo {year} {2020}),\ \Eprint
  {http://arxiv.org/abs/arXiv:2002.05533} {arXiv:2002.05533} \BibitemShut
  {NoStop}%
\bibitem [{\citenamefont {Hoffmann}\ \emph {et~al.}(2017)\citenamefont
  {Hoffmann}, \citenamefont {Zimmermann}, \citenamefont {M{\"u}ller},
  \citenamefont {Sch{\"u}rhoff}, \citenamefont {Kiselev}, \citenamefont
  {Melcher},\ and\ \citenamefont {Bl{\"u}gel}}]{Hoffmann2017}%
  \BibitemOpen
  \bibfield  {author} {\bibinfo {author} {\bibfnamefont {Markus}\ \bibnamefont
  {Hoffmann}}, \bibinfo {author} {\bibfnamefont {Bernd}\ \bibnamefont
  {Zimmermann}}, \bibinfo {author} {\bibfnamefont {Gideon~P.}\ \bibnamefont
  {M{\"u}ller}}, \bibinfo {author} {\bibfnamefont {Daniel}\ \bibnamefont
  {Sch{\"u}rhoff}}, \bibinfo {author} {\bibfnamefont {Nikolai~S.}\ \bibnamefont
  {Kiselev}}, \bibinfo {author} {\bibfnamefont {Christof}\ \bibnamefont
  {Melcher}}, \ and\ \bibinfo {author} {\bibfnamefont {Stefan}\ \bibnamefont
  {Bl{\"u}gel}},\ }\bibfield  {title} {\enquote {\bibinfo {title}
  {{Antiskyrmions stabilized at interfaces by anisotropic Dzyaloshinskii-Moriya
  interactions}},}\ }\href {\doibase 10.1038/s41467-017-00313-0} {\bibfield
  {journal} {\bibinfo  {journal} {Nature Communications}\ }\textbf {\bibinfo
  {volume} {8}},\ \bibinfo {pages} {308} (\bibinfo {year} {2017})}\BibitemShut
  {NoStop}%
\bibitem [{\citenamefont {Wang}\ \emph {et~al.}(2016)\citenamefont {Wang},
  \citenamefont {Wesselink}, \citenamefont {Liu}, \citenamefont {Yuan},
  \citenamefont {Xia},\ and\ \citenamefont {Kelly}}]{Wang2016}%
  \BibitemOpen
  \bibfield  {author} {\bibinfo {author} {\bibfnamefont {Lei}\ \bibnamefont
  {Wang}}, \bibinfo {author} {\bibfnamefont {R.~J.~H.}\ \bibnamefont
  {Wesselink}}, \bibinfo {author} {\bibfnamefont {Yi}~\bibnamefont {Liu}},
  \bibinfo {author} {\bibfnamefont {Zhe}\ \bibnamefont {Yuan}}, \bibinfo
  {author} {\bibfnamefont {Ke}~\bibnamefont {Xia}}, \ and\ \bibinfo {author}
  {\bibfnamefont {Paul~J.}\ \bibnamefont {Kelly}},\ }\bibfield  {title}
  {\enquote {\bibinfo {title} {{Giant Room Temperature Interface Spin Hall and
  Inverse Spin Hall Effects}},}\ }\href {\doibase
  10.1103/PhysRevLett.116.196602} {\bibfield  {journal} {\bibinfo  {journal}
  {Phys. Rev. Lett.}\ }\textbf {\bibinfo {volume} {116}},\ \bibinfo {pages}
  {196602} (\bibinfo {year} {2016})}\BibitemShut {NoStop}%
\bibitem [{\citenamefont {Kim}\ \emph {et~al.}(2013)\citenamefont {Kim},
  \citenamefont {Sinha}, \citenamefont {Hayashi}, \citenamefont {Yamanouchi},
  \citenamefont {Fukami}, \citenamefont {Suzuki}, \citenamefont {Mitani},\ and\
  \citenamefont {Ohno}}]{Kim2013}%
  \BibitemOpen
  \bibfield  {author} {\bibinfo {author} {\bibfnamefont {Junyeon}\ \bibnamefont
  {Kim}}, \bibinfo {author} {\bibfnamefont {Jaivardhan}\ \bibnamefont {Sinha}},
  \bibinfo {author} {\bibfnamefont {Masamitsu}\ \bibnamefont {Hayashi}},
  \bibinfo {author} {\bibfnamefont {Michihiko}\ \bibnamefont {Yamanouchi}},
  \bibinfo {author} {\bibfnamefont {Shunsuke}\ \bibnamefont {Fukami}}, \bibinfo
  {author} {\bibfnamefont {Tetsuhiro}\ \bibnamefont {Suzuki}}, \bibinfo
  {author} {\bibfnamefont {Seiji}\ \bibnamefont {Mitani}}, \ and\ \bibinfo
  {author} {\bibfnamefont {Hideo}\ \bibnamefont {Ohno}},\ }\bibfield  {title}
  {\enquote {\bibinfo {title} {{Layer thickness dependence of the
  current-induced effective field vector in Ta/CoFeB/MgO}},}\ }\href {\doibase
  10.1038/nmat3522} {\bibfield  {journal} {\bibinfo  {journal} {Nature
  Materials}\ }\textbf {\bibinfo {volume} {12}},\ \bibinfo {pages} {240--245}
  (\bibinfo {year} {2013})}\BibitemShut {NoStop}%
\bibitem [{\citenamefont {Ramaswamy}\ \emph {et~al.}(2016)\citenamefont
  {Ramaswamy}, \citenamefont {Qiu}, \citenamefont {Dutta}, \citenamefont
  {Pollard},\ and\ \citenamefont {Yang}}]{Ramaswamy2016}%
  \BibitemOpen
  \bibfield  {author} {\bibinfo {author} {\bibfnamefont {Rajagopalan}\
  \bibnamefont {Ramaswamy}}, \bibinfo {author} {\bibfnamefont {Xuepeng}\
  \bibnamefont {Qiu}}, \bibinfo {author} {\bibfnamefont {Tanmay}\ \bibnamefont
  {Dutta}}, \bibinfo {author} {\bibfnamefont {Shawn~David}\ \bibnamefont
  {Pollard}}, \ and\ \bibinfo {author} {\bibfnamefont {Hyunsoo}\ \bibnamefont
  {Yang}},\ }\bibfield  {title} {\enquote {\bibinfo {title} {{Hf thickness
  dependence of spin-orbit torques in Hf/CoFeB/MgO heterostructures}},}\ }\href
  {\doibase 10.1063/1.4951674} {\bibfield  {journal} {\bibinfo  {journal}
  {Applied Physics Letters}\ }\textbf {\bibinfo {volume} {108}},\ \bibinfo
  {pages} {202406} (\bibinfo {year} {2016})}\BibitemShut {NoStop}%
\bibitem [{\citenamefont {Zheng}\ \emph {et~al.}(2020)\citenamefont {Zheng},
  \citenamefont {Guo}, \citenamefont {Jo}, \citenamefont {Go}, \citenamefont
  {Wang}, \citenamefont {Chen}, \citenamefont {Yin}, \citenamefont {Wang},
  \citenamefont {Yu}, \citenamefont {He}, \citenamefont {Lee}, \citenamefont
  {Teng},\ and\ \citenamefont {Zhu}}]{Zheng2020}%
  \BibitemOpen
  \bibfield  {author} {\bibinfo {author} {\bibfnamefont {Z.~C.}\ \bibnamefont
  {Zheng}}, \bibinfo {author} {\bibfnamefont {Q.~X.}\ \bibnamefont {Guo}},
  \bibinfo {author} {\bibfnamefont {D.}~\bibnamefont {Jo}}, \bibinfo {author}
  {\bibfnamefont {D.}~\bibnamefont {Go}}, \bibinfo {author} {\bibfnamefont
  {L.~H.}\ \bibnamefont {Wang}}, \bibinfo {author} {\bibfnamefont {H.~C.}\
  \bibnamefont {Chen}}, \bibinfo {author} {\bibfnamefont {W.}~\bibnamefont
  {Yin}}, \bibinfo {author} {\bibfnamefont {X.~M.}\ \bibnamefont {Wang}},
  \bibinfo {author} {\bibfnamefont {G.~H.}\ \bibnamefont {Yu}}, \bibinfo
  {author} {\bibfnamefont {W.}~\bibnamefont {He}}, \bibinfo {author}
  {\bibfnamefont {H.-W.}\ \bibnamefont {Lee}}, \bibinfo {author} {\bibfnamefont
  {J.}~\bibnamefont {Teng}}, \ and\ \bibinfo {author} {\bibfnamefont
  {T.}~\bibnamefont {Zhu}},\ }\bibfield  {title} {\enquote {\bibinfo {title}
  {{Magnetization switching driven by current-induced torque from weakly
  spin-orbit coupled Zr}},}\ }\href {\doibase 10.1103/PhysRevResearch.2.013127}
  {\bibfield  {journal} {\bibinfo  {journal} {Phys. Rev. Research}\ }\textbf
  {\bibinfo {volume} {2}},\ \bibinfo {pages} {013127} (\bibinfo {year}
  {2020})}\BibitemShut {NoStop}%
\bibitem [{\citenamefont {Guo}\ \emph {et~al.}(2008)\citenamefont {Guo},
  \citenamefont {Murakami}, \citenamefont {Chen},\ and\ \citenamefont
  {Nagaosa}}]{Guo2008}%
  \BibitemOpen
  \bibfield  {author} {\bibinfo {author} {\bibfnamefont {G.~Y.}\ \bibnamefont
  {Guo}}, \bibinfo {author} {\bibfnamefont {S.}~\bibnamefont {Murakami}},
  \bibinfo {author} {\bibfnamefont {T.-W.}\ \bibnamefont {Chen}}, \ and\
  \bibinfo {author} {\bibfnamefont {N.}~\bibnamefont {Nagaosa}},\ }\bibfield
  {title} {\enquote {\bibinfo {title} {{Intrinsic Spin Hall Effect in Platinum:
  First-Principles Calculations}},}\ }\href {\doibase
  10.1103/PhysRevLett.100.096401} {\bibfield  {journal} {\bibinfo  {journal}
  {Phys. Rev. Lett.}\ }\textbf {\bibinfo {volume} {100}},\ \bibinfo {pages}
  {096401} (\bibinfo {year} {2008})}\BibitemShut {NoStop}%
\bibitem [{\citenamefont {Mahfouzi}\ \emph {et~al.}(2020)\citenamefont
  {Mahfouzi}, \citenamefont {Mishra}, \citenamefont {Chang}, \citenamefont
  {Yang},\ and\ \citenamefont {Kioussis}}]{Mahfouzi2020}%
  \BibitemOpen
  \bibfield  {author} {\bibinfo {author} {\bibfnamefont {Farzad}\ \bibnamefont
  {Mahfouzi}}, \bibinfo {author} {\bibfnamefont {Rahul}\ \bibnamefont
  {Mishra}}, \bibinfo {author} {\bibfnamefont {Po-Hao}\ \bibnamefont {Chang}},
  \bibinfo {author} {\bibfnamefont {Hyunsoo}\ \bibnamefont {Yang}}, \ and\
  \bibinfo {author} {\bibfnamefont {Nicholas}\ \bibnamefont {Kioussis}},\
  }\bibfield  {title} {\enquote {\bibinfo {title} {Microscopic origin of
  spin-orbit torque in ferromagnetic heterostructures: A first-principles
  approach},}\ }\href {\doibase 10.1103/PhysRevB.101.060405} {\bibfield
  {journal} {\bibinfo  {journal} {Phys. Rev. B}\ }\textbf {\bibinfo {volume}
  {101}},\ \bibinfo {pages} {060405(R)} (\bibinfo {year} {2020})}\BibitemShut
  {NoStop}%
\bibitem [{\citenamefont {Fan}\ \emph {et~al.}(2014)\citenamefont {Fan},
  \citenamefont {Celik}, \citenamefont {Wu}, \citenamefont {Ni}, \citenamefont
  {Lee}, \citenamefont {Lorenz},\ and\ \citenamefont {Xiao}}]{Fan2014}%
  \BibitemOpen
  \bibfield  {author} {\bibinfo {author} {\bibfnamefont {Xin}\ \bibnamefont
  {Fan}}, \bibinfo {author} {\bibfnamefont {Halise}\ \bibnamefont {Celik}},
  \bibinfo {author} {\bibfnamefont {Jun}\ \bibnamefont {Wu}}, \bibinfo {author}
  {\bibfnamefont {Chaoying}\ \bibnamefont {Ni}}, \bibinfo {author}
  {\bibfnamefont {Kyung-Jin}\ \bibnamefont {Lee}}, \bibinfo {author}
  {\bibfnamefont {Virginia~O.}\ \bibnamefont {Lorenz}}, \ and\ \bibinfo
  {author} {\bibfnamefont {John~Q.}\ \bibnamefont {Xiao}},\ }\bibfield  {title}
  {\enquote {\bibinfo {title} {Quantifying interface and bulk contributions to
  spin-orbit torque in magnetic bilayers},}\ }\href {\doibase
  10.1038/ncomms4042} {\bibfield  {journal} {\bibinfo  {journal} {Nature
  Communications}\ }\textbf {\bibinfo {volume} {5}},\ \bibinfo {pages} {3042}
  (\bibinfo {year} {2014})}\BibitemShut {NoStop}%
\bibitem [{\citenamefont {Hayashi}\ \emph {et~al.}(2020)\citenamefont
  {Hayashi}, \citenamefont {Musha}, \citenamefont {Sakimura},\ and\
  \citenamefont {Ando}}]{Hiroki2020}%
  \BibitemOpen
  \bibfield  {author} {\bibinfo {author} {\bibfnamefont {Hiroki}\ \bibnamefont
  {Hayashi}}, \bibinfo {author} {\bibfnamefont {Akira}\ \bibnamefont {Musha}},
  \bibinfo {author} {\bibfnamefont {Hiroto}\ \bibnamefont {Sakimura}}, \ and\
  \bibinfo {author} {\bibfnamefont {Kazuya}\ \bibnamefont {Ando}},\ }\href@noop
  {} {\enquote {\bibinfo {title} {{Spin-orbit torques originating from bulk and
  interface in Pt-based structures}},}\ } (\bibinfo {year} {2020}),\ \Eprint
  {http://arxiv.org/abs/arXiv:2003.07271} {arXiv:2003.07271} \BibitemShut
  {NoStop}%
\bibitem [{\citenamefont {Zhu}\ \emph {et~al.}(2019)\citenamefont {Zhu},
  \citenamefont {Ralph},\ and\ \citenamefont {Buhrman}}]{Zhu2019}%
  \BibitemOpen
  \bibfield  {author} {\bibinfo {author} {\bibfnamefont {Lijun}\ \bibnamefont
  {Zhu}}, \bibinfo {author} {\bibfnamefont {D.~C.}\ \bibnamefont {Ralph}}, \
  and\ \bibinfo {author} {\bibfnamefont {R.~A.}\ \bibnamefont {Buhrman}},\
  }\bibfield  {title} {\enquote {\bibinfo {title} {{Spin-Orbit Torques in
  Heavy-Metal--Ferromagnet Bilayers with Varying Strengths of Interfacial
  Spin-Orbit Coupling}},}\ }\href {\doibase 10.1103/PhysRevLett.122.077201}
  {\bibfield  {journal} {\bibinfo  {journal} {Phys. Rev. Lett.}\ }\textbf
  {\bibinfo {volume} {122}},\ \bibinfo {pages} {077201} (\bibinfo {year}
  {2019})}\BibitemShut {NoStop}%
\bibitem [{\citenamefont {Tokura}\ and\ \citenamefont
  {Nagaosa}(2000)}]{Tokura2000}%
  \BibitemOpen
  \bibfield  {author} {\bibinfo {author} {\bibfnamefont {Y.}~\bibnamefont
  {Tokura}}\ and\ \bibinfo {author} {\bibfnamefont {N.}~\bibnamefont
  {Nagaosa}},\ }\bibfield  {title} {\enquote {\bibinfo {title} {Orbital physics
  in transition-metal oxides},}\ }\href {\doibase 10.1126/science.288.5465.462}
  {\bibfield  {journal} {\bibinfo  {journal} {Science}\ }\textbf {\bibinfo
  {volume} {288}},\ \bibinfo {pages} {462--468} (\bibinfo {year}
  {2000})}\BibitemShut {NoStop}%
\bibitem [{\citenamefont {Mochizuki}\ and\ \citenamefont
  {Imada}(2004)}]{Mochizuki2004}%
  \BibitemOpen
  \bibfield  {author} {\bibinfo {author} {\bibfnamefont {Masahito}\
  \bibnamefont {Mochizuki}}\ and\ \bibinfo {author} {\bibfnamefont {Masatoshi}\
  \bibnamefont {Imada}},\ }\bibfield  {title} {\enquote {\bibinfo {title}
  {Orbital physics in the perovskite ti oxides},}\ }\href {\doibase
  10.1088/1367-2630/6/1/154} {\bibfield  {journal} {\bibinfo  {journal} {New
  Journal of Physics}\ }\textbf {\bibinfo {volume} {6}},\ \bibinfo {pages}
  {154--154} (\bibinfo {year} {2004})}\BibitemShut {NoStop}%
\bibitem [{\citenamefont {Ole{\'{s}}}(2012)}]{Andrzej2012}%
  \BibitemOpen
  \bibfield  {author} {\bibinfo {author} {\bibfnamefont {Andrzej~M}\
  \bibnamefont {Ole{\'{s}}}},\ }\bibfield  {title} {\enquote {\bibinfo {title}
  {Fingerprints of spin{\textendash}orbital entanglement in transition metal
  oxides},}\ }\href {\doibase 10.1088/0953-8984/24/31/313201} {\bibfield
  {journal} {\bibinfo  {journal} {Journal of Physics: Condensed Matter}\
  }\textbf {\bibinfo {volume} {24}},\ \bibinfo {pages} {313201} (\bibinfo
  {year} {2012})}\BibitemShut {NoStop}%
\bibitem [{\citenamefont {Wimmer}\ \emph {et~al.}(1981)\citenamefont {Wimmer},
  \citenamefont {Krakauer}, \citenamefont {Weinert},\ and\ \citenamefont
  {Freeman}}]{Wimmer1981}%
  \BibitemOpen
  \bibfield  {author} {\bibinfo {author} {\bibfnamefont {E.}~\bibnamefont
  {Wimmer}}, \bibinfo {author} {\bibfnamefont {H.}~\bibnamefont {Krakauer}},
  \bibinfo {author} {\bibfnamefont {M.}~\bibnamefont {Weinert}}, \ and\
  \bibinfo {author} {\bibfnamefont {A.~J.}\ \bibnamefont {Freeman}},\
  }\bibfield  {title} {\enquote {\bibinfo {title} {{Full-potential
  self-consistent linearized-augmented-plane-wave method for calculating the
  electronic structure of molecules and surfaces: ${\mathrm{O}}_{2}$
  molecule}},}\ }\href {\doibase 10.1103/PhysRevB.24.864} {\bibfield  {journal}
  {\bibinfo  {journal} {Phys. Rev. B}\ }\textbf {\bibinfo {volume} {24}},\
  \bibinfo {pages} {864--875} (\bibinfo {year} {1981})}\BibitemShut {NoStop}%
\bibitem [{FLE()}]{FLEUR}%
  \BibitemOpen
  \href@noop {} {}\bibinfo {howpublished}
  {\url{https://www.flapw.de}}\BibitemShut {NoStop}%
\bibitem [{\citenamefont {Perdew}\ \emph {et~al.}(1996)\citenamefont {Perdew},
  \citenamefont {Burke},\ and\ \citenamefont {Ernzerhof}}]{Perdew1996}%
  \BibitemOpen
  \bibfield  {author} {\bibinfo {author} {\bibfnamefont {John~P.}\ \bibnamefont
  {Perdew}}, \bibinfo {author} {\bibfnamefont {Kieron}\ \bibnamefont {Burke}},
  \ and\ \bibinfo {author} {\bibfnamefont {Matthias}\ \bibnamefont
  {Ernzerhof}},\ }\bibfield  {title} {\enquote {\bibinfo {title} {{Generalized
  Gradient Approximation Made Simple}},}\ }\href {\doibase
  10.1103/PhysRevLett.77.3865} {\bibfield  {journal} {\bibinfo  {journal}
  {Phys. Rev. Lett.}\ }\textbf {\bibinfo {volume} {77}},\ \bibinfo {pages}
  {3865--3868} (\bibinfo {year} {1996})}\BibitemShut {NoStop}%
\bibitem [{\citenamefont {Pizzi}\ \emph {et~al.}(2020)\citenamefont {Pizzi},
  \citenamefont {Vitale}, \citenamefont {Arita}, \citenamefont {Bl\"ugel},
  \citenamefont {Freimuth}, \citenamefont {G{\'{e}}ranton}, \citenamefont
  {Gibertini}, \citenamefont {Gresch}, \citenamefont {Johnson}, \citenamefont
  {Koretsune}, \citenamefont {Iba{\~{n}}ez-Azpiroz}, \citenamefont {Lee},
  \citenamefont {Lihm}, \citenamefont {Marchand}, \citenamefont {Marrazzo},
  \citenamefont {Mokrousov}, \citenamefont {Mustafa}, \citenamefont {Nohara},
  \citenamefont {Nomura}, \citenamefont {Paulatto}, \citenamefont
  {Ponc{\'{e}}}, \citenamefont {Ponweiser}, \citenamefont {Qiao}, \citenamefont
  {Th\"ole}, \citenamefont {Tsirkin}, \citenamefont {Wierzbowska},
  \citenamefont {Marzari}, \citenamefont {Vanderbilt}, \citenamefont {Souza},
  \citenamefont {Mostofi},\ and\ \citenamefont {Yates}}]{Pizzi2020}%
  \BibitemOpen
  \bibfield  {author} {\bibinfo {author} {\bibfnamefont {Giovanni}\
  \bibnamefont {Pizzi}}, \bibinfo {author} {\bibfnamefont {Valerio}\
  \bibnamefont {Vitale}}, \bibinfo {author} {\bibfnamefont {Ryotaro}\
  \bibnamefont {Arita}}, \bibinfo {author} {\bibfnamefont {Stefan}\
  \bibnamefont {Bl\"ugel}}, \bibinfo {author} {\bibfnamefont {Frank}\
  \bibnamefont {Freimuth}}, \bibinfo {author} {\bibfnamefont {Guillaume}\
  \bibnamefont {G{\'{e}}ranton}}, \bibinfo {author} {\bibfnamefont {Marco}\
  \bibnamefont {Gibertini}}, \bibinfo {author} {\bibfnamefont {Dominik}\
  \bibnamefont {Gresch}}, \bibinfo {author} {\bibfnamefont {Charles}\
  \bibnamefont {Johnson}}, \bibinfo {author} {\bibfnamefont {Takashi}\
  \bibnamefont {Koretsune}}, \bibinfo {author} {\bibfnamefont {Julen}\
  \bibnamefont {Iba{\~{n}}ez-Azpiroz}}, \bibinfo {author} {\bibfnamefont
  {Hyungjun}\ \bibnamefont {Lee}}, \bibinfo {author} {\bibfnamefont {Jae-Mo}\
  \bibnamefont {Lihm}}, \bibinfo {author} {\bibfnamefont {Daniel}\ \bibnamefont
  {Marchand}}, \bibinfo {author} {\bibfnamefont {Antimo}\ \bibnamefont
  {Marrazzo}}, \bibinfo {author} {\bibfnamefont {Yuriy}\ \bibnamefont
  {Mokrousov}}, \bibinfo {author} {\bibfnamefont {Jamal~I}\ \bibnamefont
  {Mustafa}}, \bibinfo {author} {\bibfnamefont {Yoshiro}\ \bibnamefont
  {Nohara}}, \bibinfo {author} {\bibfnamefont {Yusuke}\ \bibnamefont {Nomura}},
  \bibinfo {author} {\bibfnamefont {Lorenzo}\ \bibnamefont {Paulatto}},
  \bibinfo {author} {\bibfnamefont {Samuel}\ \bibnamefont {Ponc{\'{e}}}},
  \bibinfo {author} {\bibfnamefont {Thomas}\ \bibnamefont {Ponweiser}},
  \bibinfo {author} {\bibfnamefont {Junfeng}\ \bibnamefont {Qiao}}, \bibinfo
  {author} {\bibfnamefont {Florian}\ \bibnamefont {Th\"ole}}, \bibinfo {author}
  {\bibfnamefont {Stepan~S}\ \bibnamefont {Tsirkin}}, \bibinfo {author}
  {\bibfnamefont {Ma{\l}gorzata}\ \bibnamefont {Wierzbowska}}, \bibinfo
  {author} {\bibfnamefont {Nicola}\ \bibnamefont {Marzari}}, \bibinfo {author}
  {\bibfnamefont {David}\ \bibnamefont {Vanderbilt}}, \bibinfo {author}
  {\bibfnamefont {Ivo}\ \bibnamefont {Souza}}, \bibinfo {author} {\bibfnamefont
  {Arash~A}\ \bibnamefont {Mostofi}}, \ and\ \bibinfo {author} {\bibfnamefont
  {Jonathan~R}\ \bibnamefont {Yates}},\ }\bibfield  {title} {\enquote {\bibinfo
  {title} {Wannier90 as a community code: new features and applications},}\
  }\href {\doibase 10.1088/1361-648x/ab51ff} {\bibfield  {journal} {\bibinfo
  {journal} {Journal of Physics: Condensed Matter}\ }\textbf {\bibinfo {volume}
  {32}},\ \bibinfo {pages} {165902} (\bibinfo {year} {2020})}\BibitemShut
  {NoStop}%
\end{thebibliography}%

\end{document}